\documentclass[aps,prl,notitlepage,superscriptaddress,twocolumn,footinbib]{revtex4-2}
\usepackage{amsmath,amsfonts,amssymb,amsthm,epsfig,array}
\usepackage{bbold}
\usepackage{dsfont}
\usepackage{slashed}
\usepackage{graphics}
\usepackage{float}
\usepackage{verbatim}
\usepackage{color}
\usepackage[dvipsnames]{xcolor}
\usepackage[hidelinks,colorlinks,linkcolor=blue,
citecolor=blue,urlcolor=blue]{hyperref}
\usepackage[titletoc,title]{appendix}
\usepackage{multirow}
\usepackage{bibentry}
\usepackage{tabularx}
\usepackage[mathscr]{euscript}
\usepackage{mathtools}
\usepackage{physics}
\usepackage{siunitx}
\usepackage{cleveref}
\DeclareRobustCommand{\App}[1]{App.~\ref{#1}}

\newcommand{\bsl}[1]{\boldsymbol{#1}}

\DeclareRobustCommand{\App}[1]{App.~\ref{#1}}

\DeclareRobustCommand{\Fig}[1]{Fig.~\ref{#1}}

\DeclareRobustCommand{\Eq}[1]{Eq.~\ref{#1}}

\newcommand{\bea}{\begin{equation} \begin{aligned}}
\newcommand{\eea}{\end{aligned} \end{equation} }
\DeclareRobustCommand{\Eq}[1]{Eq.~\ref{#1}}
\newcommand{\la}{\lambda}
\newcommand{\be}{\beta}

\newcommand{\al}{\alpha}
\newcommand{\bpm}{\begin{pmatrix}}
\newcommand{\epm}{\end{pmatrix}}
\newcommand{\eps}{\epsilon}

\renewcommand{\th}{\theta}

\newcommand{\lp}{\left(}
\newcommand{\rp}{\right)}

\newcommand{\mbf}[1]{\mathbf{#1}}

\renewcommand{\Tr}{\text{Tr }}

\usepackage{braket}

\begin{document}

\title{The ``Moir\'e Capacitor Effect" and Stabilization of Fractional Chern Insulators in Rhombohedral Graphene Superlattices}

\author{Nicolas Regnault}
\affiliation{Center for Computational Quantum Physics, Flatiron Institute, 162 5th Avenue, New York, NY 10010, USA}
\affiliation{Laboratoire de Physique de l'Ecole normale sup\'{e}rieure, ENS, Universit\'{e} PSL, CNRS, Sorbonne Universit\'{e}, Universit\'{e} Paris-Diderot, Sorbonne Paris Cit\'{e}, 75005 Paris, France}
\affiliation{Department of Physics, Princeton University, Princeton, New Jersey 08544, USA}

\author{Heqiu Li}
\affiliation{Donostia International Physics Center, P. Manuel de Lardizabal 4, 20018 Donostia-San Sebastian, Spain}

\author{Yves H. Kwan}
\affiliation{Princeton Center for Theoretical Science, Princeton University, Princeton, NJ 08544}
\affiliation{Department of Physics, University of Texas at Dallas, Richardson, Texas 75080, USA.}

\author{B. Andrei Bernevig}
\email{bernevig@princeton.edu}
\affiliation{Department of Physics, Princeton University, Princeton, New Jersey 08544, USA}
\affiliation{Donostia International Physics Center, P. Manuel de Lardizabal 4, 20018 Donostia-San Sebastian, Spain}
\affiliation{IKERBASQUE, Basque Foundation for Science, Bilbao, Spain}

\author{Jonah Herzog-Arbeitman}
\email{jonahha@mit.edu}
\affiliation{Department of Physics, Princeton University, Princeton, New Jersey 08544, USA}
\affiliation{Department of Physics, Massachusetts Institute of Technology, Cambridge MA 02139, USA}

\begin{abstract}
While the necessity of a moir\'e potential for fractional Chern insulators (FCIs) in rhombohedral graphene-hBN superlattices, first predicted in Ref.~\cite{PhysRevB.112.075110}, is now grounded in experiments, a theory of its origin and importance remains at large. We present a mechanism---the moir\'e capacitor effect---that enhances the moiré potential by electrostatically imprinting the valence charge density onto the conduction bands. We derive the analytical form of this term and reveal its crucial role in stabilizing a parent state with Chern number $C=1$ at filling $\nu=1$. We propose a parent state theory which posits that stability of the Chern insulator and flatness of its hole excitations are necessary for obtaining FCIs upon doping.
We then perform multi-band exact diagonalization calculations to confirm the emergence of FCIs at $\nu = 2/3$ in the presence of the moir\'e capacitor effect. 
Our FCI state is stabilized by inter-band fluctuations, unlike in the moire-free case which collapses with band-mixing.
We provide the first consistent theory for this state in aligned samples and explain its absence in unaligned ones. 
\end{abstract}

\maketitle

\textbf{Introduction.} The discovery of FCIs in moiré materials~\cite{cai2023signatures,zeng2023thermodynamic,Park2023observation,Xu2023observation,75gl-jzl6,Lu2024fractional,xie2024even,2025Natur.639..342C,2025PhRvX..15a1045W,2025arXiv250406972X,park2026observation,2025arXiv251015309H,2025arXiv250601485X,2025Natur.637.1090L,2025arXiv251221609Z,2025arXiv250501767L,2025arXiv250720647U,2025arXiv251203622L,li2026fractional} has revealed departures from the fractional quantum Hall (FQH) phase diagram. This stems from the band structures of these materials~\cite{ju2024fractional,bernevig2025fractional}, which deviate from the flat dispersion and uniform, ideal quantum geometry of the well-isolated lowest Landau level, as in twisted MoTe$_2$ for example\cite{Wu2019TIintTMD,PhysRevResearch.3.L032070,Crepel2022FCITMD,PhysRevLett.131.136502,Xiao2023tMoTe2abinitio,Morales_MacDonald_2023,Zhang2023MoTe2b,Yu2024MFCI0,MFCI1,2023PhRvB.108h5117R,Xu2024maximally,Zaletel2023tMoTe2FCI,Abouelkomsan2024mixing,fu2021flat,Li2024tmdtbg,PhysRevLett.133.186602-2024}. Nevertheless, early theory~\cite{neupert,regnaultbernevig,sheng,Sun2011,yang2012arbitrary,Tang11} on lattice models demonstrates that FCIs can appear despite such deviations. 

\begin{figure*}
\centering
\includegraphics[width=0.98\linewidth]{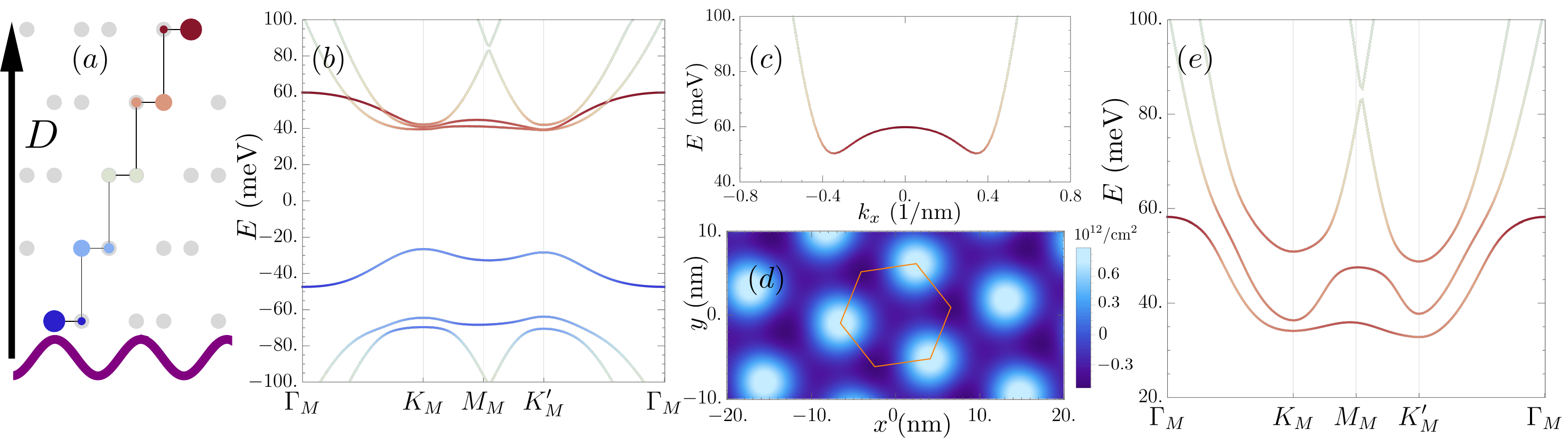}
\caption{$(a)$ Structure of R5G, including two sublattice-polarized surface states. $(b)$ Non-interacting moir\'e band structure at $V=30$meV and $\theta=0.77^\circ$, showing strong (weak) moir\'e Umklapp in the valence (conduction) bands. States are colored by layer polarization according to panel $(a)$. The conduction bands are essentially folded from pristine R5G $(c)$, while the valence bands exhibit moiré-modulated density $(d)$.  The filled valence bands  generate Hartree terms (the moir\'e capacitor effect) which imprint on the conduction bands $(e)$, causing moir\'e Umklapp.}
\label{fig_nonint_effective}
\end{figure*}

FCIs in $L$-layer rhombohedral graphene twisted on hBN (R$L$G/hBN)~\cite{herzog2024MFCI2,kwan2023MFCI3,dong2023anomalous,zhou2023fractional,dong2023theory,dong2024stabilityahc,guo2023theory,PhysRevB.112.075110,2025arXiv250919764L} diverge fundamentally from the Landau level setting. When a displacement field polarizes the doped electrons away from the hBN (Fig.~\ref{fig_nonint_effective}(a)), the lowest non-interacting conduction band has $|C|=L$  but is essentially gapless since moir\'e Umklapp is suppressed (Fig.~\ref{fig_nonint_effective}b). Nevertheless, experiments for $L=4,5,6$ observe the Jain sequence~\cite{Lu2024fractional,ju2024fractional,2025Natur.639..342C,2025Natur.637.1090L,2025arXiv251015309H,2026arXiv260606450B} (in particular a robust $\nu=2/3$ FCI) over a range of twist angles, which disappears in the absence of hBN 
alignment~\cite{2024Sci...384..647H,2025Natur.643..654H,2025arXiv250815909D,2025arXiv250924672K,2025arXiv250405129Q,2025arXiv251116578K,2025arXiv251117423G,2026arXiv260530316S,2026arXiv260705039N,2026arXiv260605356K,2026NatPh..22..862K,2026Natur.653..384L,2026arXiv260728425O,2026arXiv260400113H,2026arXiv260715014Z}. 
Intense scrutiny revealed that both the Chern insulator \cite{kwan2023MFCI3,2026PhRvL.136p6503D,2025PNAS..12215532D} and the FCI \cite{PhysRevB.112.075110} are unstable in the high-displacement field regime, challenging moir\'e-free theories \cite{zhou2023fractional,dong2023theory,huang2024selfconsistent,kwan2023MFCI3}. Hence, the role of the hBN has not been understood theoretically, and controlled, microscopic calculations have yet to produce FCIs in moir\'e systems.

In this work, we show that the \emph{valence} electrons proximate to the hBN are bound into the moir\'e potential and create a moiré-periodic charge density, which electrostatically imprints on the conduction bands. This mechanism, dubbed the ``moiré capacitor effect," underlies the stability of a suitably flat parent $C=1$ band, which we verify can host $\nu=2/3$ FCIs using multi-band exact diagonalization (ED) calculations. Our findings propose a solution to the so far elusive physics of R$L$G/hBN, and provide a candidate framework for the emergence of FCIs in other systems far from the Landau level limit.

\textbf{Moir\'e Capacitor Effect.} We begin by deriving an effective Hamiltonian $H_\text{con}$ that faithfully describes the low-energy physics of the conduction electrons in R$L$G/hBN. 
The continuum Hamiltonian  is $H = H_0 + H_{\text{hBN}} + H_{\text{int}}$, where $H_0$ describes pristine R$L$G and $H_{\text{int}}$ is the Coulomb interaction defined below . We first neglect the moir\'e term $H_{\text{hBN}}$ arising from hybridization with the hBN. At low energies for $|v_F k| \leq t_1/(1+ \frac{1}{L})$, where $v_F$ is the graphene velocity and $t_1$ is the nearest-neighbor interlayer hopping, $H_0$ consists of `surface' states that are weakly dispersing and polarized to the top and bottom graphene layers (see \App{app:exacteigenstates} for details).
$H_0$ includes the inversion-breaking displacement field $D$, which opens a gap at neutrality and generates Berry curvature in the surface bands in each valley. The layer-resolved Coulomb interaction is~\cite{PhysRevB.108.195148}
\bea
\label{eq:Hamint}
H_{\text{int}} &= \frac{1}{2 \Omega_{tot}} \sum_{\mbf{q} ll'}V_{ll'}(\mbf{q})\delta \rho_{\mbf{q},l} \delta \rho_{-\mbf{q},l'}, \\
\delta \rho_{\mbf{q},l} &=  \sum_{\mbf{k},\sigma \eta s} (c^\dag_{\mbf{k}+\mbf{q}, l \sigma \eta s}c_{\mbf{k}, l \sigma \eta s} - \frac{1}{2}\delta_{\mbf{q},0}) 
\eea
where $\Omega_{tot}$ is the total area, $\sigma,\eta,s$ denote sublattice, valley and spin, and $V_{ll'}(\mbf{q})$ is the interaction between electrons on layers $l,l'$ with $\mbf{q} \in \mathbb{R}^2$. 
The density operator $\delta \rho_{\mbf{q},l}$, expressed in terms of plane wave operators, is measured relative to charge neutrality, which is necessary when taking the thermodynamic limit with Coulomb interactions~\cite{giuliani2008quantum}. For graphene-like systems, however, the charge density of the valence electrons is unbounded in the Dirac approximation~\cite{RevModPhys.81.109}.
The Dirac sea picture breaks down for momenta $|\mathbf{k}|$ approaching the inverse lattice spacing, which motivates a cutoff $|\mbf{k}| \leq \Lambda$~\cite{PhysRevB.84.085446,PhysRevLett.118.266801} to make \Eq{eq:Hamint} well-defined. For monolayer graphene~\cite{Gonzalez1994HoneycombRG,Gonzalez1999MarginalFermiLiquid,DasSarma2007ManyBodyGraphene,Barlas2007ChiralityCorrelations,Hwang2008QuasiparticleSpectralFunction,Polini2007PseudochiralFermiLiquid}, certain observables famously depend on $\Lambda$, such as the effective Fermi velocity which is renormalized by Coulomb interactions. This can be shown analytically within Hartree-Fock (HF) at charge neutrality~\cite{Borghi2009FermiVelocityEnhancement}, and the renormalized dispersion is consistent with recent experiments \cite{2026NanoL..26.4046L} using quantum twisting microscopy (QTM)~\cite{2023Natur.614..682I}.

In R$L$G, the Fock potential of the filled valence states similarly renormalizes the single-particle hoppings through Coulomb interactions. These effective hoppings have previously been extracted by fitting to transport experiments \cite{2021Natur.598..429Z,PhysRevB.111.075103} and have been used in previous theoretical work \cite{dong2023anomalous,zhou2023fractional,dong2023theory,guo2023theory,huang2024selfconsistent}. Therefore, we simply adopt these effective hoppings to model the effect of Fock renormalization (see Eq.~\ref{eq:parameters_exp} in App.~\ref{app:AppA}). 
We benchmark this approach by comparing with self-consistent HF up to a large cutoff $\Lambda =6\,\text{nm}^{-1}$~\cite{2026NanoL..26.4046L} starting from bare Slater-Koster hoppings, and find reasonable agreement (see \App{app:fockcommentary}). 
In addition, the Coulomb interaction screens $D$ through Hartree terms. Intuitively, a positive $D$ pushes the valence electrons to the bottom layer ($l=0$), which is accompanied by a deficit of electron density on the top layer. The resulting charge imbalance forms a classical capacitor, whose internal field opposes $D$. The resulting screened field can be modeled by self-consistent Hartree calculations that include the classical (electrostatic) and quantum capacitance. Our calculations (see \App{appA:nonint}) show that for large $D\simeq 0.9\,\text{Vnm}^{-1}$ relevant for experiments, the solution can be approximated by an effective interlayer potential $V\simeq 30\,$meV generating a gap $\sim (L-1)V$ at charge neutrality.  In summary, once HF corrections arising from the valence bands are included (or modeled with modified hoppings/potentials), we can obtain a low-energy model by projecting into the conduction bands.  

We now consider the effect of the hBN. Hybridization with the aligned hBN adjacent to the bottom graphene layer generates a moir\'e-periodic coupling~\cite{PhysRevB.90.155406,2015NatCo...6.6308J,PhysRevB.96.085442}
\bea
V_M(\mbf{r}) = V_1 e^{i\psi}\sum_{j=1}^3 e^{i \mbf{g}_j\cdot\mbf{r}}T_j  + h.c., \ T_j= \bpm 1& \omega^{-j} \\ \omega^{j+1} &\omega \epm,
\eea
acting on the sublattices of layer $l=0$. $\mbf{g}_j$ are $C_3$-related moir\'e reciprocal vectors, $\omega = e^{2\pi i/3}$, and $V_1 e^{i\psi} = (21\,\text{meV}) e^{i 16^\circ}$ is the coupling amplitude in the two-center approximation~\cite{PhysRevB.90.155406}.   Encapsulation by hBN also induces a spatially-uniform potential on the outer layers $V_{tb} \approx 30$meV that is twist-independent and can be absorbed into $H_0$ (see \App{appA:nonint} for details). 
As illustrated by the band structure (Fig.~\ref{fig_nonint_effective}(b)), the valence bands develop large moir\'e gaps. By contrast, the conduction bands barely feel $V_M(\mathbf{r})$ as they are polarized towards the top, and are essentially the bands of R$L$G (\Fig{fig_nonint_effective}(c)) folded into the moir\'e Brillouin zone (mBZ) --- in contradiction with the indispensable role of hBN alignment in experiment.

To resolve this inconsistency, we must reconsider the effect of the valence bands in the presence of hBN alignment.  
The key physics arises from the moir\'e-modulated charge density of the occupied valence manifold (\Fig{fig_nonint_effective}(d)), which is \emph{convergent} in $\Lambda$ as proven in \App{app:hartreecommentary}.
It generates the electrostatic potential~\footnote{In \App{app:fockcommentary}, we demonstrate that the Fock terms generated by the hBN perturbation are less than $1\,$meV, and will hence be neglected.}
\bea
\label{eq:MCE}
\mathcal{V}_{\text{val}., l}(\mbf{r}) &= \sum_{\mbf{G} l'} e^{i \mbf{G} \cdot \mbf{r}} V_{ll'}(\mbf{G}) \rho^{\text{val.}}_{-\mbf{G},l'}, \\
\eea
where $\rho^{\text{val.}}_{-\mbf{G},l'}=\braket{\delta\rho_{-\mbf{G},l'}}/\Omega_{tot}$ is the density of the occupied valence bands at charge-neutrality. The $\mbf{G} \neq 0$ terms break the continuous translation symmetry in the conduction bands down to moir\'e translations. We term this the ``moir\'e capacitor effect" since these terms generalize the $\mbf{G}=0$ uniform, classical capacitor potential created by the valence bands without the moiré.

\begin{figure*}
\centering
\includegraphics[width=0.98\linewidth]{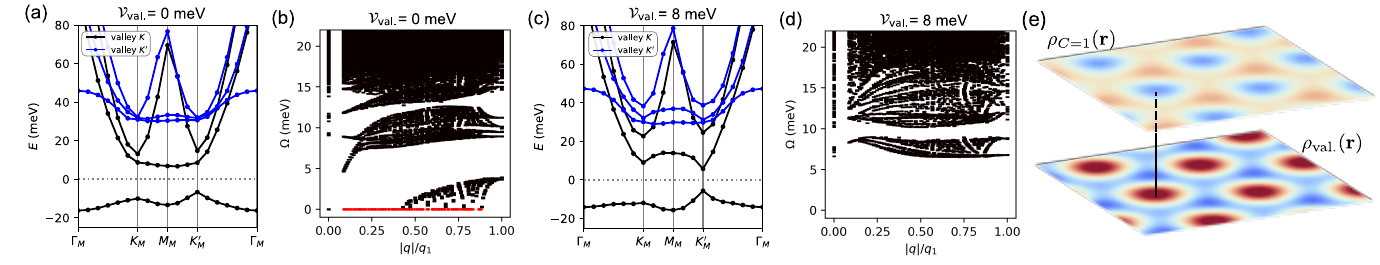}
\caption{$(a)$ HF band structure for the $\nu=1$, $C=1$ state in $\theta=0.77^\circ$ R5G/hBN with $\mathcal{V}_\text{val.}=0\,$meV on a $18\times 18$ system. Only bands in the polarized spin-$\uparrow$ sector are shown. Dotted line indicates Fermi level. $(b)$ TDHF spectrum as a function of momentum $|q|/q_1$. Results for multiple system sizes $N_x\times N_y$ with $N_x =N_y=8,9,\ldots,20$ are overlaid. Red dots indicate momenta with complex/negative TDHF eigenvalues, indicating a local instability.  $(c,d)$ Same as $(a,b)$ except with $\mathcal{V}_\text{val.}=8\,$meV. $(e)$ Top/bottom shows charge density variation of the occupied HF band/valence background with $\mathcal{V}_\text{val.}=8\,$meV. We use parameters $V=22\,$meV, $V_{tb}=32\,$meV, $\psi_\text{val.}=(\frac{4\pi}{3}+0.12)\,$rad, and keep the $3$ lowest conduction bands. Red/blue is $\pm0.5\times 10^{12}/\text{cm}^{2}$.}
\label{fig_maintext_bandstruct_colleigs}
\end{figure*} 

We derive explicit expressions for $\mathcal{V}_{\text{val}., l}(\mbf{r})$ when the neutrality gap $(L-1)V$ is much larger than the hybridization amplitude $V_1$ and interaction scale. Within first-order perturbation theory, the lowest Fourier harmonic of the valence charge background on the bottom layer is (see App.~\ref{app:hartreecommentary} for derivation)
\bea
\label{eq:perthry}
\rho^{\text{val.}}_{-\mbf{g}_1,0} \!&= \!\frac{4 V_{1} e^{i \psi}}{\Omega_{tot}} \!\! \sum_{\substack{\mbf{q}\in\mathds{R}^2 \\ m\in\text{val} \\ n\in\text{con}}} \! \frac{\text{Tr}_{l=0}[P_{m}(\mbf{q}) T_1 P_n(\mbf{q}-\mbf{g}_1)]}{\lambda_{m}(\mbf{q}) - \lambda_{n}(\mbf{q}-\mbf{g}_1)} - (m \!\leftrightarrow \! n)
\eea
where $\lambda_n(\mbf{q}),P_n(\mbf{q})$ are the energies and wavefunction projectors of the $n$th pristine R$L$G band, and the trace is restricted to the layer $l=0$ where the aligned hBN touches the graphene. 

Using the Coulomb interaction $V_{ll'}(\mbf{G}) = \frac{e^2}{2 \eps_\perp|\mbf{G}|} e^{-d|l-l'||\mbf{G}|}$, with interlayer spacing $d = 0.33\,$nm and perpendicular dielectric constant $\eps_\perp = 3$, we can determine the strength of $\mathcal{V}_{\text{val}., L-1}(\mbf{r})$ on the top layer. 
Henceforth focusing on $\theta=0.77^\circ$ R5G/hBN at $V=30\,$meV relevant for Ref.~\cite{Lu2024fractional}, we obtain an order of magnitude estimate by approximating the sum  by $((L-1)V)^{-1} \times \Omega_{mBZ} $, with mBZ area $\Omega_{mBZ} = \frac{\sqrt{3}}{2} |\mbf{g}_1|^2 $, yielding
\bea
\label{eq:Vcalestimate}
\mathcal{V}_{\text{val.}} \approx \frac{e^2}{2 \eps_\perp |\mbf{g}_1|} e^{- d (L-1) |\mbf{g}_1|}  \frac{4 V_1}{(L-1)V} \frac{\Omega_{mBZ}}{(2\pi)^2} = 12\,\text{meV},
\eea
close to the value $9\,$meV computed directly from \Eq{eq:perthry}. 

The conduction states are localized on the top layer, so we can replace the layer-dependent $\mathcal{V}_{\text{val}.,l}$ in \Eq{eq:MCE} by its value on the top layer $\mathcal{V}_{\text{val}.,L}$
  (\App{app:hartreecommentary}) and obtain 
\bea
\label{eq:scalarpot}
\mathcal{V}_{\text{val}.}(\mbf{r}) &= 2 \mathcal{V}_{\text{val.}}\sum_j \cos (\mbf{g}_j \cdot \mbf{r} + \psi_{\text{val.}})
\eea
where $\psi_{\text{val.}} = \frac{4\pi}{3}+0.12$ is numerically computed from \Eq{eq:perthry}. Note that a $C_6$-symmetric triangular lattice potential has $\psi_{\text{val.}} = \frac{2\pi n}{3}$. 
\Fig{fig_nonint_effective}(e) plots the conduction bands in the presence of $\mathcal{V}_\text{val.}(\mathbf{r})$, revealing that the three-fold degeneracies at the mBZ corners are strongly split (see \App{app:phasediagram} for analytical estimates).

To summarize, our analysis has distilled the effects of hBN down to the scalar potential \Eq{eq:scalarpot} depending on only two parameters, $\mathcal{V}_{\text{val.}}$ and $\psi_{\text{val.}}$, which is useful given the uncertainties in the bare moir\'e parameters. While $\mathcal{V}_{\text{val.}}$ becomes negligible for large enough displacement fields (see Eq.~\ref{eq:Vcalestimate}), a key prediction of our work is that the $\sim100\,$meV neutrality gap that we estimate for the FCI experiments is not sufficiently large to neglect the valence charge background. Instead, doped electrons  experience an approximately triangular $\sim 10\,$meV potential, leading to the reconstructed bands in \Fig{fig_nonint_effective}(e).

\textbf{Stabilization of the Chern Insulator.} We now explain the role of the moir\'e capacitor effect (Eq.~\ref{eq:scalarpot}) in stabilizing an interaction-induced $C=1$ insulator at $\nu=1$. We have derived the effective Hamiltonian describing the conduction bands
\bea
\bar{H}_{\text{con}} &= \bar{H}_0 + \bar{\mathcal{V}}_{\text{val}.} + : \bar{H}_{\text{int}}:
\eea
where bars denote projection to the conduction bands and the colon denotes conventional normal ordering against the charge-neutrality gap.   In the following, we retain the three low-energy bands in \Fig{fig_nonint_effective}(e)~\footnote{In App.~\ref{secapp:HD_TDHF_phase_diagrams}, we demonstrate that including additional conduction bands does not affect our conclusions.}. In the absence of moir\'e, these bands would be degenerate at the mBZ corners. 
Since the conduction electrons are localized towards the top layer, we now neglect the 3D portion of the interaction and instead use a 2D gate-screened interaction in $:\bar{H}_\text{int}:$ for simplicity (see \App{app:HF}).

Our HF calculations in \Fig{fig_maintext_bandstruct_colleigs}, which enforce moir\'e translation invariance, reveal a spin-valley polarized $C=1$ insulator, whose lowest band is significantly flatter than the single-particle bands of $\bar{H}_{\text{con}}$. This remains true for a range of $\mathcal{V}_{\text{val}.}$, including $\mathcal{V}_{\text{val}.} = 0$ (\Fig{fig_maintext_bandstruct_colleigs}(a)) where the hBN moir\'e vanishes and the breaking of continuous translation is spontaneous, leading to an anomalous Hall crystal \cite{dong2023anomalous,zhou2023fractional,dong2023theory,guo2023theory,huang2024selfconsistent}. However, the latter is actually not the ground state beyond restricted HF---it has a local instability that can be, and was early on, detected via the collective modes in time-dependent HF (TDHF)~\cite{kwan2023MFCI3}. Goldstone's theorem protects two gapless electronic phonon modes because of spontaneously-broken translation. However, as seen in \Fig{fig_maintext_bandstruct_colleigs}(b), one of the phonon branches has complex frequencies, with an imaginary velocity and finite-momentum instabilities throughout the mBZ.  

By contrast, the inclusion of a finite $\mathcal{V}_{\text{val}.}$ in \Fig{fig_maintext_bandstruct_colleigs}(c,d) cures this instability, yielding a neutral gap of $\Delta_{\text{TDHF}} = 6\,$meV, comparable to the $10\,$meV HF charge gap~\footnote{The true many-body gaps are expected to be smaller. Our $21$-site multi-band ED calculations in App.~C show a neutral gap $\sim3$\,meV. }. As illustrated in \Fig{fig_maintext_bandstruct_colleigs}(e), the density of the filled $C=1$ band is supported on a hexagonal lattice (with imbalanced sublattices since inversion is broken), which is electrostatically ``locked" to the valence charge background. The Hartree potential $\mathcal{V}_{\text{val.}}(\mbf{r})$ therefore favors the $C=1$ state -- it gaps and hence stabilizes the phonon branches. This interpretation relies on the valence background being triangular with a single maximum per unit cell for the honeycomb charge density to center around. This is true when $\psi_{\text{val.}} - \frac{2\pi n }{3} \in (-\frac{\pi}{6},\frac{\pi}{6})$, in agreement with our computed value.

We emphasize that this $C=1$ Chern band does not persist to the non-interacting limit~\footnote{Indeed as discussed in App.~\ref{app:phasediagram}, the lowest band of Fig.~\ref{fig_nonint_effective}(e) has a Chern number of either $C=0$ or $C=-1$, which differs from $C=1$ of the HF solution.} and is fundamentally interaction-induced. 
Hence, in contrast to the cyclotron gap in the Landau level setting, its charge and neutral gaps \emph{must} be comparable to the interaction strength. As such, upon hole-doping from $\nu=1$, there is no \emph{a priori} justification for projection into this single $C=1$ band --- a fundamental departure from the typical settings where FCIs have been established (although see Refs.~\cite{PhysRevLett.112.126806,2025PhRvL.134s6501Y,2025arXiv251215041Y,2025arXiv250504138L,c6zs-m1wk,PhysRevB.90.165101,LIU2024515}). Nevertheless, experiments in R$L$G/hBN show a robust $\nu=2/3$ FCI. To explain this, we now develop a parent state picture that generalizes the conventional criteria for FCIs in isolated Chern bands and allows us to predict where multi-band FCIs can occur.  

\begin{figure*}
\centering
\includegraphics[width=1\linewidth]{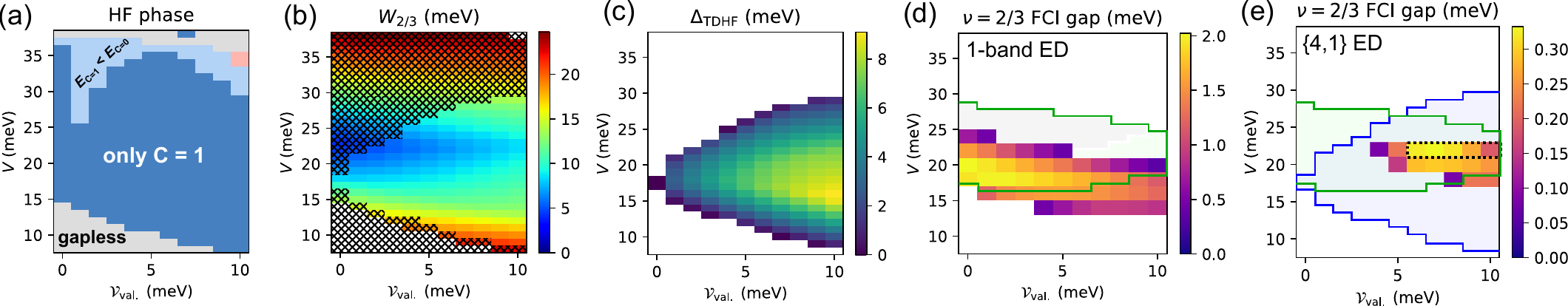}
\caption{$(a)$ HF phase diagram for $\theta=0.77^\circ$ R5G/hBN at $\nu=1$. The dark blue region labeled `only $C=1$' indicates where the HF ground state has $C=1$, and no $C=0$ state is obtained even as a metastable solution. The light blue region labeled `$E_{C=1}<E_{C=0}$' indicates where the HF ground state has $C=1$, but metastable $C=0$ solutions are also obtained. The small pink region at $\mathcal{V}_\text{val.}=10\,$meV indicates where the HF ground state has $C=0$, but metastable $C=1$ solutions are also obtained. The grey region labeled `gapless' denotes where the HF ground state does not have a positive indirect gap. System size is $12\times12$ and $3$ conduction bands are kept. $(b)$ Effective bandwidth $W_{2/3}$ (see text for definition) of the $C=1$ state, where the $C=1$ HF solution can be obtained. Note that the $C=1$ solution is not always the global HF ground state (see phase diagram in $(a)$). Cross-hatching corresponds to the white region in $(c)$.  $(c)$ Lowest TDHF collective mode eigenvalue $\Delta_\text{TDHF}$ of the $C=1$ state in $(a)$. White regions indicate where either the $C=1$ state yields negative/complex TDHF eigenvalues, or a $C=1$ HF state could not be obtained.
$(d)$ FCI gap at $\nu=2/3$ obtained from $21$-site ED calculations that project into the $C=1$ HF band. White regions indicate absence of FCI. Green border denotes where effective bandwidth $W_{\text{2/3}}$ is less than $10$\,meV.  
$(e)$ Same as $(d)$ except the ED incorporates fluctuations by allowing up to $4$ ($1$) particles in the second (third) band with orbital truncation 14,2 (see \App{app:ED_23}). Blue border denotes $\Delta_\text{TDHF}>0$ in $(c)$. 
}
\label{fig_maintext_HF_TDHF_phase_schematic}
\end{figure*}

\textbf{Parent State Principles For Multi-Band FCIs}. In \Fig{fig_maintext_HF_TDHF_phase_schematic}(a), we begin by examining the HF phase diagram as a function of interlayer potential $V$ and the electrostatic potential amplitude $\mathcal{V}_\text{val.}$, to determine where a $C=1$ HF solution exists at $\nu=1$. We then test two types of ED calculations at $\nu=2/3$ on a 21-site system for the same values of $V$ and $\mathcal{V}_{\text{val.}}$. In \Fig{fig_maintext_HF_TDHF_phase_schematic}(d), the ED calculation is projected to the $C=1$ HF band, while in \Fig{fig_maintext_HF_TDHF_phase_schematic}(e), we further include many-body fluctuations into the higher HF bands using techniques developed in Ref.~\cite{PhysRevB.112.075110} (to be discussed momentarily). The 1-band ED shows a wide region of FCIs whose many-body gap $\Delta_\text{FCI}\simeq 2\,$meV is largest at $\mathcal{V}_{\text{val.}}=0$.  Upon including band-mixing, the FCI phase shrinks and shifts such that it only appears for finite $\mathcal{V}_\text{val.}\geq 4\,\text{meV}.$   
The maximal many-body gap now occurs at $\mathcal{V}_{\text{val.}}=7\,$meV where we find a diminished $\Delta_\text{FCI}\simeq 0.3\,$meV. The dramatic impact of inter-band fluctuations has previously been uncovered in Ref.~\cite{PhysRevB.112.075110} \footnote{Note that the Hamiltonian studied by Ref.~\cite{PhysRevB.112.075110} did not show a converged FCI gap. In Ref.~\cite{PhysRevB.112.075110}, a finite cutoff of 4 momentum shells $\sim 1.3\,\text{nm}^{-1}$ was employed to compute the Hartree-Fock valence band terms. In this work, we account for the cutoff-divergence of the Fock term by using renormalized hoppings, and we account for the cutoff-convergent Hartree term analytically. This key difference is responsible for the appearance of FCIs in our finite size ED calculations.}. Physically, 1-band ED projects out low-energy collective modes which mix the $C=1$ band into the higher bands. At integer filling, these modes start at $\Delta_\text{TDHF}\lesssim 6\,$meV, which is \emph{smaller} than the $\sim 20$\,meV interaction scale. Without moir\'e, they are gapless and cannot be neglected. Hence band mixing cannot be ignored, and 1-band ED is uncontrolled and unreliable. 

The canonical setting for FCIs, motivated by the FQH problem, involves doping a $C=1$ band that is (\textbf{I}) flat and (\textbf{II}) isolated. To understand where multi-band FCIs in R$L$G/hBN may be favorable (\Fig{fig_maintext_HF_TDHF_phase_schematic}(e)), we generalize criteria (\textbf{I}) and (\textbf{II}). 
For (\textbf{I}), we introduce the notion of `effective' flatness $W_{\nu}$, which aims to capture the interaction-renormalized bandwidth relevant for FCI physics (see App.~\ref{secapp:HF_formalism} for a detailed discussion). Specifically, we first construct the effective Hamiltonian $H^{C=1}_{HF}(\nu)=\bar{H}_0+\bar{\mathcal{V}}_\text{val.} + \nu(H_H+H_F)$, where $H_H,H_F$ are the Hartree and Fock potentials of the $\nu=1$ Chern insulator (see also Ref.~\cite{huang2024selfconsistent} for a similar construction). This corresponds to the mean-field Hamiltonian obtained by uniformly scaling the $C=1$ HF density matrix. $W_\nu$ is defined as the bandwidth of the lowest HF band in $H^{C=1}_{HF}(\nu)$. Note that $W_1$ corresponds to the bandwidth of the occupied HF band in the $C=1$ state, which would \emph{appear} to be a good indicator for criterion (\textbf{I}). However, for FCI physics at $\nu=2/3$, we argue that $W_1$ overestimates the interaction-induced renormalization of the bandwidth, because the putative FCI contains only two-thirds of the electronic density of the parent $C=1$ state at $\nu=1$. Since the electronic occupation in an FCI is typically homogeneous in momentum space, this motivates $W_{2/3}$ as a more refined indicator for band flatness. We show $W_{2/3}$ in \Fig{fig_maintext_HF_TDHF_phase_schematic}(b), and we use green outlines in the ED plots of \Fig{fig_maintext_HF_TDHF_phase_schematic}(d,e) to indicate where $W_{2/3}$ is less than $10$\,meV (roughly half the interaction scale). We find that the FCI phase, both with and without band-mixing, predominantly lies within the region where $W_{2/3}$ is small. \Fig{fig:app_V_vs_V1_extended} in App.~\ref{secapp:HD_TDHF_phase_diagrams} demonstrates that $W_1$ is a much poorer predictor of FCI stability compared to $W_{2/3}$.

For (\textbf{II}), we have remarked that it is impossible for the excitation gap of the $C=1$ HF band to be larger than the interaction scale. However, it is natural to at least demand many-body stability of the $C=1$ integer-filling ground state.  Here, we use $\Delta_{\text{TDHF}} \geq 0$ as an indicator of local stability. This simple requirement is necessary for any parent state picture to hold, yet it rules out a significant portion of the phase diagram in \Fig{fig_maintext_HF_TDHF_phase_schematic}(c). Since stability favors large $\mathcal{V}_{\text{val}.}$ and flatness favors small $\mathcal{V}_{\text{val}.}$, this competition restricts FCIs to a narrow region, in agreement with multi-band ED (\Fig{fig_maintext_HF_TDHF_phase_schematic}(e)). 

\begin{figure}
\centering
\includegraphics[width=\columnwidth]{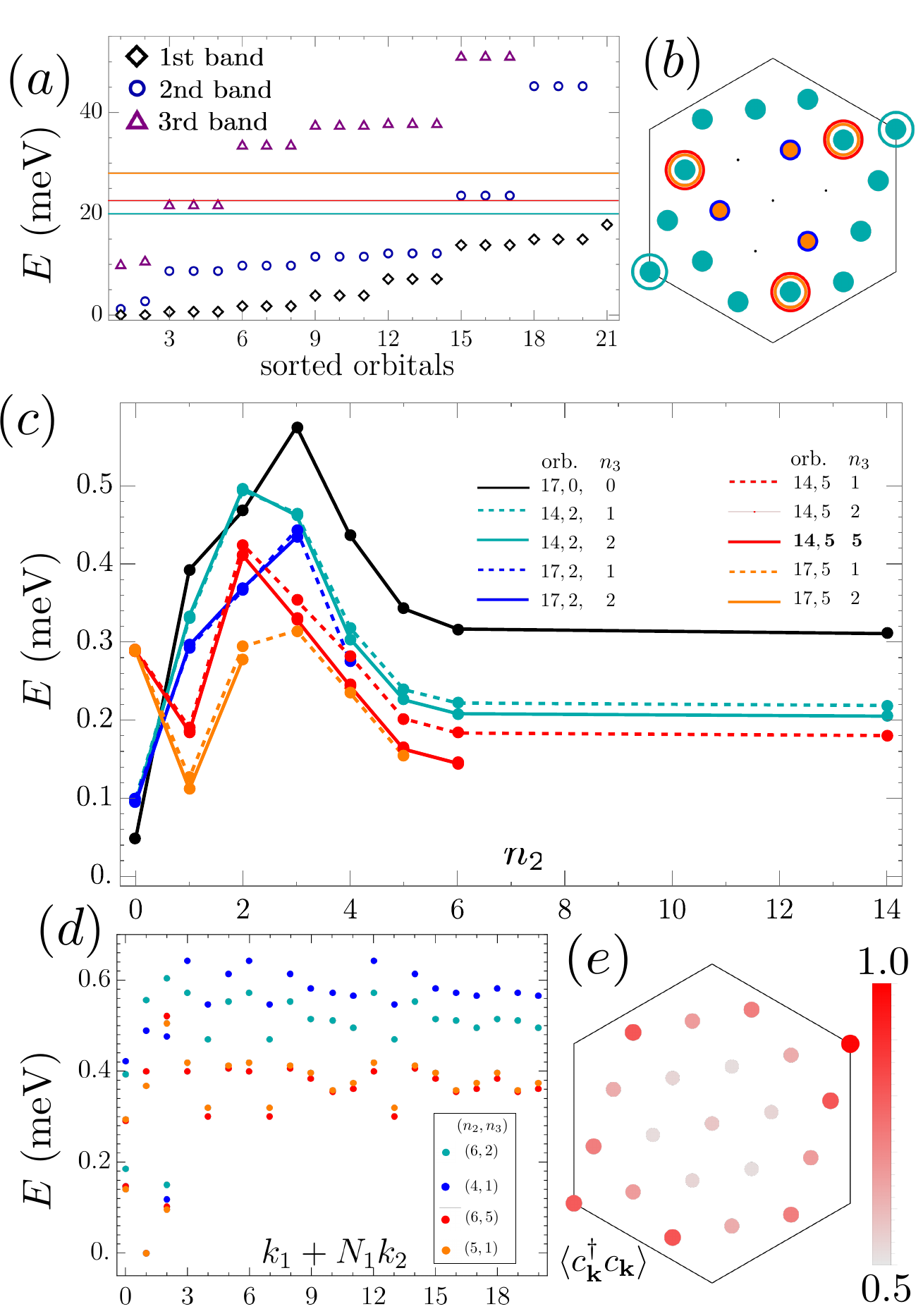}
\caption{Multi-band convergence of the $\nu=2/3$ FCI on 21 sites (using the parameters of \Fig{fig_maintext_bandstruct_colleigs}(c,d)). $(a)$ Single-particle levels of $\bar{H}_{\text{con}}$ sorted by energy. 
Orbitals are truncated based on the energy cutoff.
$(b)$ The 14,2 truncation is shown in cyan with solid (open) circles denoting orbitals in the second (third) band. The 17,2/14,5/17,5 truncation includes the additional blue/red/orange orbitals. $(c)$ FCI gap as a function of band mixing. $(d)$ ED spectra (colored as in $(b)$) show that the FCI ground states rapidly converge and the gap to the continuum is stabilized at large truncations. $(e)$ Momentum-resolved density $\braket{c^\dag_{\mathbf{k}}c_{\mathbf{k}}}$ in the  $C=1$ HF band exhibits increased weight near the mBZ edge, where mixing into the higher bands increases the occupation. The average occupation in the 1st/2nd/3rd band is $0.6/0.06/0.005$.}
\label{fig_EDbandmax}
\end{figure}

\textbf{Multi-band Exact Diagonalization.} We proceed to a detailed convergence study of the multi-band $\nu=2/3$ FCI. The presence of 3 low-energy bands, analogous to an extreme form of Landau level mixing in the FQH problem, prevents reliable single-band ED calculations, even when attempted with various optimization strategies~\cite{huang2024selfconsistent,zlqg-sj86}. 
Full 3-band ED is challenging because of the large Hilbert space dimension $\sim 10^{12}$ on 21 sites.
However, \Fig{fig_nonint_effective}(e) shows that some states in the higher bands (e.g. near $\Gamma_M$) have large kinetic energies \cite{bernevig2025berrytrashcanmodelinteracting}, suggesting that the $\nu=2/3$ ground state has only small support on them. 
To exploit this, we implement two truncation parameters. First, we impose energy cutoffs (\Fig{fig_EDbandmax}(a,b)) to restrict the $\mathbf{k}$-points to 14 or 17 (2 or 5) orbitals in the 2nd (3rd) band. 
Second, we impose a ``bandmax'' truncation that limits the maximum number of particles in the upper two bands to $n_2,n_3$ respectively (see \App{app:ED_23}.).
We work in the HF basis such that the parent $C=1$ band (the 1st band) is fully accessible for any truncation~\footnote{It is clear that the choice of single-particle basis does not affect the full 3-band ED calculation, and hence does not introduce bias in the truncated calculation if it is converged.}.

\Fig{fig_EDbandmax}(c) plots the FCI gap for Hilbert spaces up to $1\times 10^9$.
For $n_3=1,\ldots,5$, both 14,2 and 14,5 orbital truncations show convergence in $n_2$ by $n_2=6$. The 14,2 and 17,2 truncations show identical trends in $\Delta_\text{FCI}$ for accessible $n_2$. Finally, the 17,5 calculation exhibits nearly identical gaps for $n_3=1,2$, and behaves similarly to the 14,5 calculation.
Our results suggest that $\Delta_\text{FCI}$ converges to a finite $>0.1$meV gap (corresponding to a temperature $\sim 1\,$K) for 21 sites. While we are unable to access larger systems, \Fig{fig_EDbandmax}(c) points to the existence of a converged multi-band FCIs in the presence of the moir\'e capacitor effect. The finite-size splitting within the 3 FCI ground states are barely affected by the truncation, while the continuum of excitations is strongly affected (\Fig{fig_EDbandmax}d).
\Fig{fig_EDbandmax}e reveals significant variation in $\braket{c^\dag_{\mathbf{k}}c_{\mathbf{k}}}$ , showing peaks at the mBZ edge where band mixing is largest. Quantitatively, at $K_M,K_{M'}$ the density in the dispersive bands is roughly $50\%$ of the $C=1$ band's density --- far from the 1-band FCI limit. In fact, for $n_2 = n_3 = 0$ when band mixing is not included, the ground state shows signatures of a charge density wave and is \emph{not} an FCI at these parameters (see \App{app:ED_23} for comprehensive ED calculations). Hence the FCI state out-competes the charge density wave through band mixing. We conclude that, while the FCI is dominated by the $C=1$ HF band, its properties are qualitatively affected by the higher bands.

\textbf{Discussion.} We have established that the moir\'e capacitor effect---an intrinsic electrostatic term that imprints a moir\'e potential on the conduction bands---leads to the emergence of FCIs in RLG/hBN. Our extensive prior investigation (without moir\'e and with other normal orderings)~\cite{PhysRevB.112.075110, zlqg-sj86} failed to yield stable FCIs, so we conclude that, to the best of our knowledge, unbiased ED calculations can only support this scenario. 

While estimates of the important parameters $V_1$ and $\psi$ vary considerably, they may be constrained by experiment\footnote{Estimates of the strength of $V_1$ vary from $21$\,meV~\cite{PhysRevB.90.155406} computed in the two-center approximation, to $15$\,meV~\cite{2015NatCo...6.6308J} computed by interpolating aligned (non-moir\'e) ab initio calculations, to $5$\,meV \cite{MFCI1} computed using classical relaxation. Attempts to fit de Haas-van Alphen oscillations in Bernal graphene/hBN yield $V_1 \in (4,9)$\,meV \cite{2024Sci...383...42B} .}. For instance, QTM on monolayer graphene/hBN has directly imaged an electrostatic potential~\cite{2026Natur.650..875K}. We can analyze this via the moir\'e capacitor effect in \Eq{eq:perthry}, which remains valid (without a gap) in the limit of large $v_F$ where Umklapp is weak. We find $\rho^{\text{val.}}_{-\mbf{g}_1} =  V_1 e^{i \psi} \frac{|\mbf{g}_1|}{8v_F} \omega^*$ analytically (see \App{app:othersystems}) so that
\bea
\mathcal{V}_{\text{mono.}}(\mbf{r}) 
&= \frac{V_1}{8v_F}\frac{e^2}{\eps_\perp} e^{- d |\mbf{g}_1|} \sum_j \cos (\mbf{g}_j \cdot \mbf{r} + \psi - \frac{2\pi}{3}) \ . \\
\eea
For reasonable values of the dielectric ($\eps_\perp = 4.6 \eps_0$ \cite{2023Natur.614..682I}) and thickness $d = 1\,$nm, the min-to-max amplitude of $\mathcal{V}_{\text{mono.}}(\mbf{r})$ is $58\,$meV for $V_1 =21\,$meV, in good agreement with the $55-63$meV value observed in Ref.~\cite{2026Natur.650..875K}. The observed weak breaking of $C_6$ symmetry is best fit by $\psi \sim 6^\circ$, slightly smaller than the two-center value. 
As another example, STM studies on R3G/hBN~\cite{2025arXiv251009548S} observe density modulation on the top layer that cannot be explained by the bare moir\'e potential, but rather requires an effective potential acting on the top layer with reduced strength and opposite sign. This is qualitatively consistent with the moir\'e capacitor effect. We compute the corresponding electrostatic potential for the trilayer system and find it is nearly triangular (see \App{app:othersystems}) and predict it will induce a hexagonal charge distribution on the top layer. Finally, in R$5$G, Ref. \cite{2026arXiv260606450B} has measured charge density waves at fillings commensurate with the hBN moir\'e in the Wigner solid region at low density. This is good agreement with the moir\'e capacitor effect.

Moving forward, one can apply our framework to other fillings in the Jain sequence, including states with non-trivial spin structure, although a correct treatment of fluctuations due to band-mixing appears demanding. Alternatively, the moir\'e capacitor effect may be useful for probing states in moir\'e-less devices, such as the superconductor \cite{2025Natur.643..654H} or Wigner crystal~\cite{2026arXiv260400113H}, by studying their response to modulated potentials. Beyond R$L$G/hBN, the moir\'e capacitor effect may be operative in other classes of multilayer heterostructures. Finally, we expect that our generalized FCI criteria will find use in predicting FCIs in other multi-band systems, in concert with efficient many-body methods~\cite{Yu2024MFCI0,PhysRevB.112.075110,2024PhRvB.109l1107A,2024PhRvB.109x5125H,2025arXiv250605330G,2025arXiv251116641H,PhysRevB.111.205117,2026PhRvL.136f6504H,PhysRevLett.134.176503,2026arXiv260406316Z,2026arXiv260416847M}. 

\textbf{Note added.} While finalizing this manuscript, we became aware of Ref.~\cite{2026arXiv260620377Q}, whose Hartree-Fock results are in accordance with our previously presented work \cite{HerzogArbeitman2026MoireCapacitor,HerzogArbeitman2026MultibandFCI}. In addition, Ref. \cite{Wang2026TransMoire} reports STM measurements of a moir\'e-distant potential, corroborated by mean-field calculations. Both are in agreement with our theory of the moir\'e capacitor effect. Finally, we note that our theory is in qualitative agreement with the recent results of Ref.\cite{2026arXiv260708822D}, which also finds that a moir\'e potential -- of precisely the form derived here -- is required to stabilize an FCI in idealized models of R$L$G. 

\textbf{Acknowledgments.} This paper was not generated using any AI tools. We are grateful for stimulating discussions with Long Ju, Ray Ashoori, Allan MacDonald, Shahal Ilani, Leonid Levitov, Trithep Devakul, Jackson Butler, and Tonghang Han, Jiewen Xiao, and Jiabin Yu. B.A.B. was supported by the Gordon and Betty Moore Foundation through Grant No.~GBMF8685 towards the Princeton theory program, the Gordon and Betty Moore Foundation’s EPiQS Initiative (Grant No.~GBMF11070), the Global Collaborative Network Grant at Princeton University, the Simons Investigator Grant No.~404513, the BSF Israel US foundation No.~2018226, the NSF-MERSEC (Grant No.~MERSEC DMR 2011750), the Simons Collaboration on New Frontiers in Superconductivity (SFI-MPS-NFS-00006741-01), the Schmidt Fund at the Princeton University and the Princeton Catalysis Innitiative. J.H.-A. was supported by a Hertz Fellowship and the MIT Pappalardo Fellowship. H.L was partially supported by the European Research Council (ERC) under the European Union’s Horizon 2020 Research and Innovation Programme (Grant Agreement No. 101020833). The Flatiron Institute is a division of the Simons Foundation. Part of the simulations presented in this article were performed on computational resources managed and supported by Princeton Research Computing, a consortium of groups including the Princeton Institute for Computational Science and Engineering (PICSciE) and Research Computing at Princeton University.

\bibliography{re.bib}

\clearpage
\newpage

\appendix
\setcounter{secnumdepth}{3}
\onecolumngrid

\tableofcontents

\section{Moir\'e Hamiltonian}
\label{app:AppA}

We present the moir\'e Hamiltonian of rhombohedral graphene aligned with hBN in \App{appA:nonint} and derive an effective interacting Hamiltonian $H_\text{con}$ acting on the conduction bands in the presence of a displacement field in \App{app:MBhamiltonian}. The key result is that the valence bands, although separated by a $\sim 100$\,meV  gap from the conduction bands, cannot be neglected. Instead, our treatment shows that they generate two terms when they are integrated out in \App{app:moirePT}. One is a Fock term -- without significant moir\'e dependence -- which can be absorbed into renormalized single-particle hoppings and velocities (see \App{app:fockcommentary}). The other is a Hartree term which imprints the moir\'e-periodic valence electron density onto the conduction bands as an electrostatic potential. We derive an analytical form of this potential in \App{app:hartreecommentary}. Finally, we devote \App{app:exacteigenstates} and \App{app:phasediagram} to analytical derivations of the single-particle band structure and topology. In \App{app:exacteigenstates} we solve the minimal-hopping model of R$L$G using a transfer matrix approach, and in \App{app:phasediagram} we determine the topological phase diagram in the presence of the moir\'e capacitor effect. 

\subsection{Band Structure of Rhombohedral Graphene}
\label{appA:nonint}

\begin{figure*}
\centering
\includegraphics[width=0.95\linewidth]{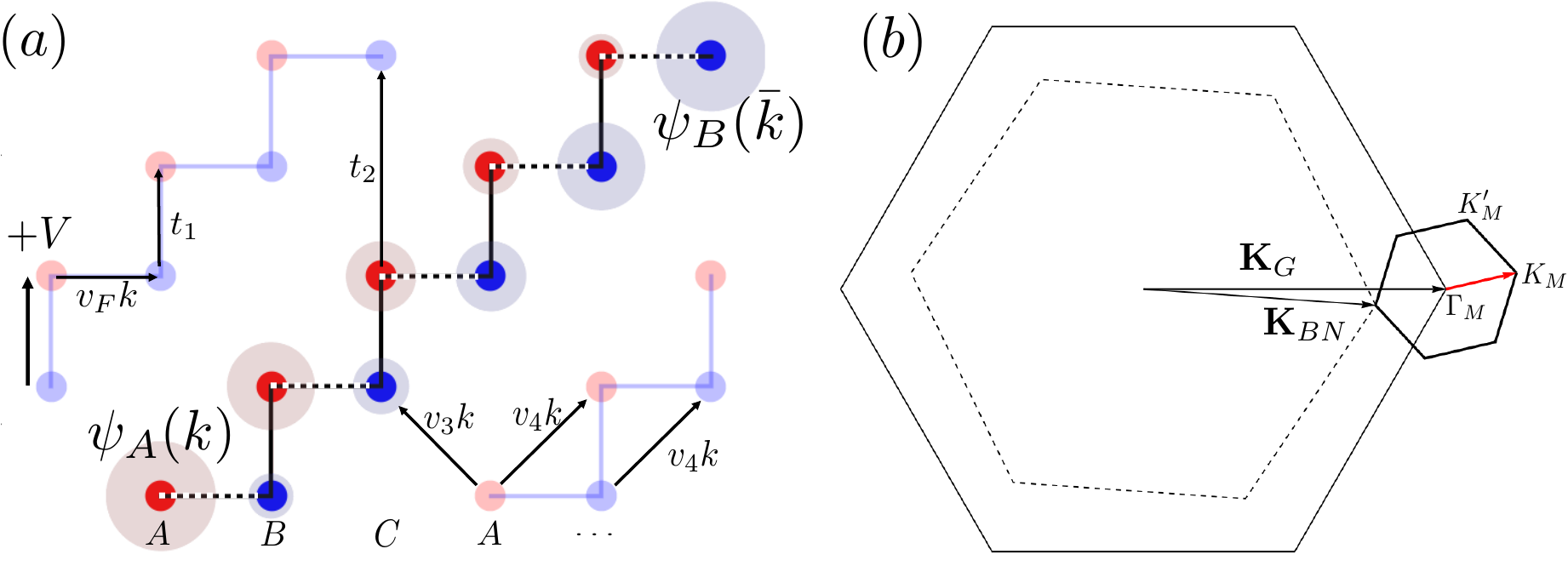}
\caption{(a) Lattice structure and hoppings of rhombohedral graphene. Parameter values may be found in \Eq{eq:parameters_bare} and \Eq{eq:parameters_exp}. 
(b) Moir\'e Brillouin zone (mBZ, thick) with graphene BZ (solid) and hBN BZ (dashed, not to scale). The red arrow is the moir\'e scattering vector $\mbf{q}_1$.}  
\label{fig_app_geometry}
\end{figure*}

We begin by presenting the low-energy continuum model for pristine non-interacting rhombohedral graphene. In the $K$ valley, the Hamiltonian can be written~\cite{PhysRevB.90.155406,Park2023RMGhBNChernFlatBands,PhysRevB.89.205414,herzog2024MFCI2}
\bea 
\label{eq:H_K}
H_{K}(\mbf{k}) &= \bpm
v_F\mbf{k} \cdot \pmb{\sigma}  & t^\dag(\mbf{k}) & t'^\dagger &   &\\
t(\mbf{k}) & \ddots & \ddots & t'^\dagger \\
t' & \ddots & v_F\mbf{k} \cdot \pmb{\sigma} & t^\dagger(\mbf{k})\\
& t' & t(\mbf{k})  & v_F\mbf{k} \cdot \pmb{\sigma}
\epm + H_{ISP} + H_D, 
\eea
which acts on the layer $\ell = 0,\dots,L-1$ and sublattice $\sigma = A,B$ tensor product basis, and $\bsl{\sigma}=(\sigma_x,\sigma_y)$ is a vector of Pauli matrices. The longer-range hoppings $t(\mathbf{k})$ and $t'$ are
\bea
t(\mbf{k}) = \bpm -v_4 (k_x+i k_y) & t_1 \\ -v_3 (k_x-i k_y) &  -v_4 (k_x+i k_y) \epm, \qquad  \qquad t' = \bpm 0 & 0 \\ t_2 & 0 \epm \ .
\eea
The parameters are as follows: $v_F$ is the Fermi velocity, $t_1,v_3,v_4$ describe hoppings between consecutive layers, and $t_2$ describes hopping between next-nearest layers. Here $\mbf{k} = - i \pmb{\nabla}$ is a continuum momentum measured from the Dirac momentum (the $K$ point of the graphene BZ). The model is only valid within a cutoff $|\mbf{k}| < \Lambda$ around the Dirac momentum. Although the cutoff $\Lambda$ is much larger than the moir\'e wavevector, it famously plays an important role in the interacting physics of Fermi velocity renormalization in graphene \cite{Gonzalez1994HoneycombRG,Gonzalez1999MarginalFermiLiquid,DasSarma2007ManyBodyGraphene,Barlas2007ChiralityCorrelations,Hwang2008QuasiparticleSpectralFunction,Polini2007PseudochiralFermiLiquid}. We will discuss this in detail later in App.~\ref{app:fockcommentary}. The model in the other valley $K' = -K$ can be obtained by time-reversal $H_{-K}(\mbf{k}) = H_{K}(-\mbf{k})^*$. 

\begin{figure*}
\centering
\includegraphics[width=0.95\linewidth]{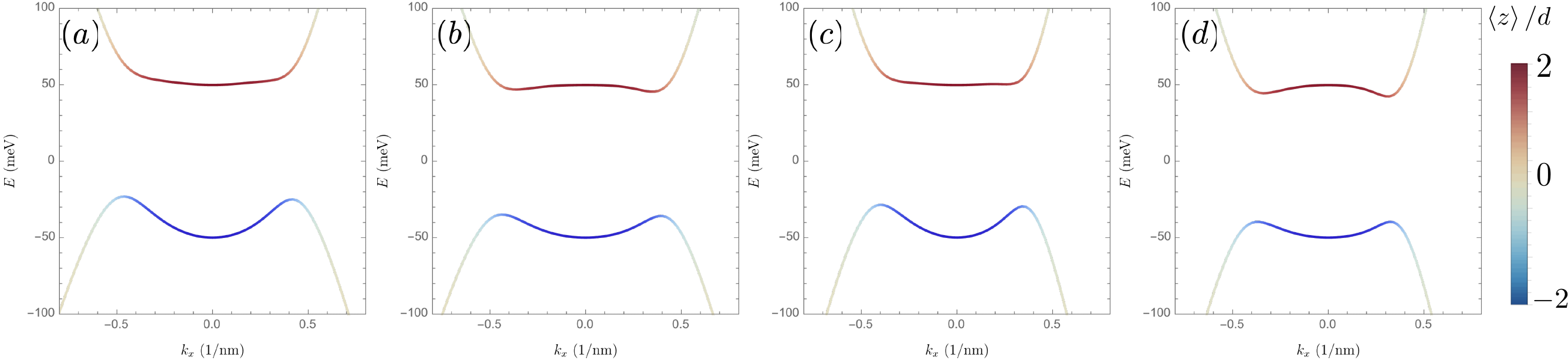}
\caption{Non-interacting band structures at $V=25$meV in the $K$ valley. $(a)$ and $(b)$ use the bare parameters in \Eq{eq:parameters_bare}, and $(b)$ includes the ISP term in \Eq{eq:ISP}. $(c)$ and $(d)$ use the effective parameters in \Eq{eq:parameters_exp}, and $(d)$ includes the ISP term in \Eq{eq:ISP}. The ISP term increases the energy at $\mbf{k}=0$, and the effective parameters lead to a steeper dispersion. The color coding of the bands reflects the layer polarization according to Fig.~\ref{fig_nonint_effective}$(a)$ of the Main Text (red/blue is top/bottom).}  
\label{fig_app_h0_bands}
\end{figure*}

The terms so far treat all layers equivalently, but for a finite stack there is no symmetry that requires this (e.g. the outer layers have different chemical environments than the inner layers). \emph{Ab initio} calculations reveal terms that break this equivalence in the form of an inversion-symmetric potential (ISP). In previous work~\cite{herzog2024MFCI2,kwan2023MFCI3,PhysRevB.112.075110}, we used a simplified term of the form $ [H_{ISP}]_{l \sigma,l' \sigma'}= V_{ISP} \left|l - \frac{n-1}{2} \right| \delta_{ll'} \delta_{\sigma \sigma'}$ containing a single inversion-symmetric internal field $V_{ISP} = 16.65\,$meV. This value was obtained by fitting the piece-wise linear potential to the \emph{ab initio} band structure. In this work, we focus on the pentalayer system. Here, we extract $H_{ISP}$ from direct Wannierization of the \emph{ab initio} band structure~\cite{yiprivate} rather than the linear approximation. We obtain
\bea
\label{eq:ISP}
\null [H_{ISP}] = \text{diag}(0, 11.2, -29.2, -28.3, -32.2, -32.2, -28.3, -29.2, 11.2, 0)\,\text{meV} \ .
\eea
An overall chemical potential has been subtracted so that the outermost orbitals (the $A$ sublattice on the lowest layer $l=0$ and the $B$ sublattice on the highest $l=L-1$ layer) are at zero potential. The neighboring sublattices on these outer layers have potential $11.2\,$meV, and all other internal layers are lower in energy by roughly $30\,$meV. This shows that the internal layers are energetically favorable, giving a particle-in-box profile in the $z$-direction.    We remark that since the low-energy bands near neutrality are mostly supported on the $A$ sublattice on the bottom layer $l=0$, and the  $B$ sublattice on the top layer $l=L-1$, the effect of the $H_{ISP}$ is suppressed on the low-energy bands. In particular its effect vanishes at $\mbf{k}=0$. However, $H_{ISP}$ causes negative curvature (see Ref.~\cite{herzog2024MFCI2}) in the conduction bands since the wavefunction is supported more on the inner layers at larger $\mbf{k}$. Sample band structures are shown in \Fig{fig_app_h0_bands}.  The form of $H_{ISP}$ is in good agreement with \emph{ab initio} results on tetralayers~\cite{Park2023RMGhBNChernFlatBands}, where the outer layers have potentials $(0,\Delta_1)$ and all inner layers have potential $(\Delta_2,\Delta_2)$, where $\Delta_1 \approx 10$\,meV and $\Delta_2 \approx -30$\,meV. 

\begin{figure*}
\centering
\includegraphics[width=0.8\linewidth]{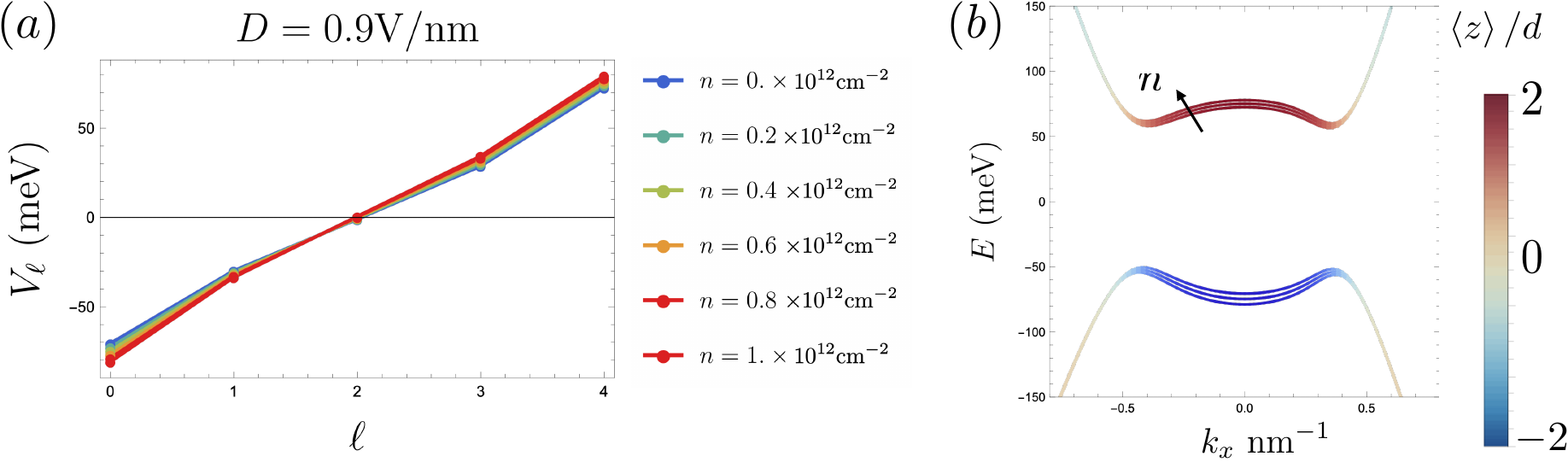}
\caption{(a) Self-consistent Hartree electrostatics at $D = 0.9\,\text{V nm}^{-1}$. (b) Band structures at $n=0,0.4,0.8 \times 10^{12}/\text{cm}^2$ (increasing $n$ is denoted by the arrow, and red/blue color denotes polarization to the top/bottom). We use an internal graphene capacitance per area of $C_g = 3 \eps_0/d$ and assume spin-valley polarization of the conduction band electrons (the valence bands are spin-valley degenerate).} 
\label{fig_app_D_field}
\end{figure*}

Next, we can apply an external electric field $D$. Since the applied field is constant, it gives a linear interlayer potential
\bea
\label{eq:H_D}
\null [H_{D}]_{l \sigma,l' \sigma'}= V \left(l - \frac{n-1}{2} \right) \delta_{ll'} \delta_{\sigma \sigma'}\ .
\eea
where, nominally, $V = e D d$ is the potential energy difference between the layers, with spacing $d = 0.33$nm. However, the high polarizability of rhombohedral graphene will strongly (and non-linearly) screen the $D$ field. Since our focus is on the large $D$ regime where the active electrons are highly polarized, it suffices to keep the linear potential but approximate $V$ by its renormalized value. To justify this, we can model the effects of screening by a Hartree approximation of the 3D Coulomb interaction across a range of electron densities and displacement fields. In this approximation, self-consistent values of the layer potential $V_\ell$ are obtained at a given electron density $n$ and displacement field $D$. They solve the Gauss law equations \cite{PhysRevB.108.195148,2026PhRvB.113g5131K,2025arXiv250507981T,2025arXiv251117423G,2025arXiv250718598K}
\bea
\label{eq:gauss}
V_{\ell+1}-V_{\ell} = -\frac{e^2}{C_g} (n_b + \sum_{\ell'=0}^\ell n_{\ell'}), \qquad \sum_\ell n_\ell = -(n_b + n_t)
\eea
where $n_\ell$ is the electron number density on the $\ell$th layer measured relative to charge neutrality in the self-consistent ground state, $n = -(n_b+n_t)$, and $n_b$ and $n_t$ are the charge densities on the bottom and top gates which create the field $D = (n_t-n_b)/{2\eps_0}$. Finally, $C_g = \eps_\perp/d$ is the graphene capacitance per area. We perform these self-consistent Hartree  calculations assuming spin-valley-polarized conduction bands (in a metallic state). The results in \Fig{fig_app_D_field} show weak $n$ dependence and fairly linear $V_\ell$ behavior. At $D = 0.9\,\text{Vnm}^{-1}$, typical values for the out-of-plane graphene dielectric constant ($\eps_\perp = 3$) give effective values of $V \in (30,40)$meV, compared to a bare value of $(n_t-n_b)/(2\eps_0/d) = 300$meV or $(n_t-n_b)/(2\eps_\perp/d) = 100$meV. For comparison, our ED calculations in Fig. 3 of the Main Text  identify FCIs for effective values of $V \in (20,30)$meV. This is in rough agreement with estimates of $V \in (30,40)$meV in the Hartree-only calculations, where the ground state is metallic.  

We now discuss the values of the graphene hopping parameters. It will be important to disentangle the bare kinetic terms and the effects of interactions, which renormalize the bands \cite{Gonzalez1994HoneycombRG,Gonzalez1999MarginalFermiLiquid,DasSarma2007ManyBodyGraphene,Barlas2007ChiralityCorrelations,Hwang2008QuasiparticleSpectralFunction,Polini2007PseudochiralFermiLiquid}. We will delay a careful derivation of these effects to \App{app:MBhamiltonian} and \App{app:fockcommentary} and simply summarize the parameters here. One method of extracting the hoppings is to fit them to experiments. Refs.~\cite{2021Natur.598..429Z,PhysRevB.111.075103} have extracted the hopping parameters from measurements of the Landau level crossings in bilayer and trilayer graphene. They obtain
\bea
\label{eq:parameters_exp}
v_F =660\text{meV nm}, \quad v_3 = 62\text{meV nm},\quad v_4 = 30\text{meV nm},\quad t_1 =  380 \text{meV}, \quad t_2 = -12\text{meV} \ .
\eea  
These values reflect the dispersion of the low-energy quasi-particles and hence include interaction effects. These values  (\Eq{eq:parameters_exp}) differ from those obtained from ab initio studies (see e.g.~Refs.~\cite{PhysRevB.89.035405,herzog2024MFCI2}) which obtain 
\bea
\label{eq:parameters_bare}
v_F =542\text{meV nm}, \quad v_3 = 34\text{meV nm}, \quad v_4 = 34\text{meV nm},\quad t_1 =  355\text{meV}, \quad t_2 = -7\text{meV} \ .
\eea
The differences between these parameter sets occur already in monolayer graphene, where DFT underestimates $v_F$ by $\sim20\%$. This disagreement can be resolved at the HF level. It has been found~\cite{Borghi2009FermiVelocityEnhancement} that the Fock term receives large contributions from the occupied valence bands at high $\mbf{k}$ up to the cutoff $\Lambda$, and induces long-range hoppings that significantly renormalize the Hartree-Fock dispersion relation. These ``non-local exchange" terms are poorly captured by ab initio studies in the local density approximation~\cite{PhysRevLett.101.226405,PhysRevLett.101.226405,2024PhRvB.109g5120G}.  Hence, within the scope of this work, the parameters in \Eq{eq:parameters_bare} will be viewed as ``bare" parameters.

In \App{app:MBhamiltonian} we will discuss the interacting Hamiltonian and show the effect of the high $\mbf{k}$ valence bands is to renormalize the hoppings by generating HF corrections. Starting with \emph{ab initio} bare values in \Eq{eq:parameters_bare} and performing HF calculations to renormalize the bands achieves good agreement with the bands computed directly from \Eq{eq:parameters_exp}. We will show examples of this agreement in \App{app:MBhamiltonian}. A detailed comparison is reserved for forthcoming work.  

Having discussed the Hamiltonian of pristine rhombohedral graphene, we now discuss the addition of hBN. Treating virtual hopping processes between the lowest graphene layer and hBN at second-order results in the hybridization potential written in terms of the couplings $V_0$ and $V_1 e^{i \psi}$~\cite{PhysRevB.90.155406,2015NatCo...6.6308J,PhysRevB.96.085442}
\bea
[V_{hbn}(\mbf{r})]_{l \sigma,l' \sigma'} &= V_0 \delta_{ll'} \delta_{l,0} \delta_{\sigma \sigma'} + V_1  \delta_{ll'} \delta_{l,0} e^{i\psi}\sum_{j=1}^3 e^{i \mbf{g}_j\cdot\mbf{r}}\bpm 1& \omega^{-j} \\ \omega^{j+1} &\omega \epm_{\sigma \sigma'}  + h.c.\ ,
\eea
 which only acts on the bottom layer $l=0$. We have introduced $\omega = e^{\frac{2\pi i}{3}}$, and the moir\'e reciprocal lattice vectors $\mbf{g}_j = R(\frac{2\pi(j-1)}{3}) (\mbf{q}_2-\mbf{q}_3)$, where $R(\phi)$ represents counter-clockwise rotation by $\phi$, for $j = 1,2,3$. They are defined in terms of the moir\'e scattering vectors $\mbf{q}_j$ via $\mbf{q}_{j+1} = R(\frac{2\pi}{3}) \mbf{q}_j$ and 
\bea
\label{eq:qvecmain}
\mbf{q}_1 = \mbf{K}_G - \mbf{K}_{hBN} = \frac{4\pi}{3 a_G}\left(1 - \frac{R(-\th)}{1+\delta_{hBN}} \right)\hat{x}, \quad \delta_{hBN} = 0.01673
\eea
\noindent  
where $\th$ is the twist angle (we focus on $\theta=0.77^\circ$ to model the experiment in Ref.~\cite{Lu2024fractional}), $\mbf{K}_G$ and $\mbf{K}_{hBN}$ are the valley $\eta=K$ Dirac momenta of graphene and hBN, $a_G = 2.46\AA$ is the graphene lattice constant, and $(1+\delta_{hBN})a_G$ is the hBN lattice constant. See \Fig{fig_app_geometry}b for an illustration of the geometry. We also define `standard' basis moir\'e reciprocal lattice vectors $\mbf{b}_{M,i}=\mbf{q}_3-\mbf{q}_i$ for $i=1,2$. 

There are two effects of the proximitized hBN. First, the $V_0$ term is a constant potential which acts on the bottom layer only but does not break the continuous translation invariance present in the continuum model of pristine R$L$G. In the two-center approximation, this term is independent of twist angle and is always present (even for large angles or ``unaligned" cases). The $V_1 e^{i \psi}$ term is periodic on the moir\'e scale and is the only microscopic term where moir\'e periodicity enters into our model. In realistic transport devices where the R$L$G is encased in hBN on both sides but aligned only on one side, we should include the $V_0$ term on both outer layers, leading to
\bea
\label{eq:Vxifinal}
H_{hBN} &= V_{tb} + V_M(\mbf{r}) \ , \\
[V_{tb}]_{l \sigma,l' \sigma'} &= V_{0} \delta_{ll'} (\delta_{l,0}+\delta_{l,L-1}) \delta_{\sigma \sigma'} \ ,\\
[V_M(\mbf{r})]_{l \sigma,l' \sigma'} &=  V_1  e^{i\psi} \delta_{ll'} \delta_{l,0}\sum_{j=1}^3 e^{i \mbf{g}_j\cdot\mbf{r}}\bpm 1& \omega^{-j} \\ \omega^{j+1} &\omega \epm_{\sigma \sigma'}  + h.c.\ ,
\eea
where $tb$ stands for `top-bottom'. In the regime of interest (large positive $D$ and small electron doping), the active conduction electrons are localized near the top of the device. Therefore, the $V_0$ term acting on the top layer plays a role in modifying the effective electrostatic potential coming from $H_D$ and $H_{ISP}$. It \emph{does not} induce moir\'e modulation directly on the top layer. The moir\'e potential $H_{hBN}^{-K}$ in the opposite valley is obtained by time-reversal taking $i \to -i$ in \Eq{eq:Vxifinal}.

Ideally, the values of $V_0, V_1, \psi$ would be determined from \emph{ab initio} calculations in the relaxed moir\'e structure. However, these calculations are beyond current capabilities   owing to the large number of atoms within a moir\'e unit cell. The simplest approach is to use the estimate from Refs.~\cite{PhysRevB.90.155406,herzog2024MFCI2}
\bea
\label{eq:hbnparamxi1}
V_0 = -3t_0^2(V_N^{-1} + V_B^{-1}) = 29 \text{meV}, \qquad V_1e^{i \psi} = -t_0^2(V_N^{-1} + \omega^* V_B^{-1}) = 21e^{i 16.55^\circ} \text{meV},
\eea
which is obtained from applying the Bistrizter-Macdonald two-center approximation to the graphene/hBN hopping assuming the structure is not relaxed. Here $t_0 = 150$\,meV is the tunneling matrix element between the hBN and the graphene, and $V_N = -1388\,$meV, $V_B=3352\,$meV are the energies of the hBN valence and conduction extrema at the hBN Dirac momentum measured relative to the graphene $K$ point. There are two inequivalent stacking orientations labeled by $\xi$ (corresponding to whether the carbon A site is aligned with the nitrogen ($\xi =1$) or boron ($\xi = 0$) atom) and $V_B,V_N$ should be swapped in \Eq{eq:hbnparamxi1} for the $\xi = 0$ stacking. The parameters $V_0$ and $V_1$ do not depend on the stacking, but $\psi$ does depend on the hBN stacking orientation (as can be seen \Eq{eq:hbnparamxi1} from switching $V_B$ and $V_N$). Throughout this work, we primarily consider $\xi=1$ where $\psi = 16.55^\circ$ which is believed to be the stacking configuration where Chern insulators are seen in experiment \cite{2025arXiv251015309H,2025arXiv250720647U}. The $\xi = 0$ stacking gives $\psi =-136.55^\circ$.  

In second quantization, in terms of the plane wave operators
\bea
\label{eq:fft}
c^\dag_{\mbf{k},\mbf{G},l \sigma,\eta s} = c^\dag_{\mbf{k}+\mbf{G},l \sigma,\eta s} = \frac{1}{\sqrt{\Omega_{tot}}}\int d^2r e^{i (\mbf{k}+\mbf{G}) \cdot \mbf{r}} c^\dag_{\mbf{r},l \sigma,\eta s} \ ,
\eea
the Hamiltonian reads
\bea\label{eq:H0_original}
H_0 &= \sum_{\mbf{k},\mbf{G},\mbf{G}', \eta s,l \sigma,l' \sigma'} c^\dag_{\mbf{k},\mbf{G},l \sigma,\eta s} [H_{\eta}(\mbf{k}+\mbf{G}) \delta_{\mbf{G},\mbf{G}'} + (H_{ISP} +H_D + V_{tb}) \delta_{\mbf{G},\mbf{G}'} + [V_M^{\eta}]_{\mbf{G},\mbf{G}'}]_{l \sigma, l' \sigma'} c_{\mbf{k},\mbf{G}',l' \sigma',\eta s} \\
&= \sum_{\mbf{k}, \eta s,n} E_{n,\eta}(\mbf{k}) \gamma^\dag_{\mbf{k},n,\eta s} \gamma_{\mbf{k},n,\eta s} \\
\eea  
where
\bea
\label{eq:moirematrixelement}
[V^{\eta=K}_M]_{\mbf{G}l \al,\mbf{G}' l' \be} = V_1 e^{i \psi}  \delta_{ll'}\delta_{l,0} \sum_{j=1,2,3} \delta_{\mbf{G},\mbf{G}'+\mbf{b}_j} \bpm 1 & \omega^{-j} \\ \omega^{j+1} & \omega \epm_{\sigma \sigma'} + h.c. \, \qquad [V^{K'}_M]_{\mbf{G},\mbf{G}'} = [V^{K}_M]^*_{-\mbf{G},-\mbf{G}'}
\eea
$\mbf{k}$ is taken in the moir\'e Brillouin zone, $\mbf{G},\mbf{G}'$ sum over the moir\'e reciprocal lattice vectors, $\eta = K,K'$ is the valley, and $s$ is the spin. The band basis creation operators are defined as
\bea
\label{eq:transform}
\gamma^\dag_{\mbf{k},n,\eta s} &= \sum_{\mbf{G},l \sigma} U^{\eta}_{\mbf{G} l \sigma, n}(\mbf{k}) c^\dag_{\mbf{k},\mbf{G},l \sigma,\eta s} 
\eea
with single-particle eigenvectors $[U^{\eta}_{n}(\mbf{k})]_{\mbf{G} l \sigma}$. We label the valence bands (with respect to charge neutrality) as $n < 0$ and the conduction bands with $n \geq 0$. The energies $E_{n,\eta}(\mbf{k})=E_{n,\eta}(\mbf{k}+\mbf{G})$ are periodic for $\mbf{G}$ a moir\'e reciprocal lattice vector. We can choose a periodic embedding $U^{\eta}_{\mbf{G}, l \sigma, n}(\mbf{k}+\mbf{G}')=U^{\eta}_{\mbf{G}+\mbf{G}', l \sigma, n}(\mbf{k})$ to define $U_n$ outside the first BZ and ensure $\gamma^\dag_{\mbf{k}+\mbf{G},n,\eta s}=\gamma^\dag_{\mbf{k},n,\eta s}$. 

One of the central messages of our work is that although the bare moir\'e hybridization $V_M(\mbf{r})$  has a quite weak effect, e.g. $\sim 2$ meV umklapp splitting at the mBZ corner,  on the conduction bands which are localized to the other side of the device in $D > 0$, the valence bands strongly respond to the hybridization potential $V_M(\mbf{r})$ and enhance the moir\'e on the top layer via interaction processes. We explain this below and derive the effective Hamiltonian acting on the conduction bands. 

\subsection{Interaction Hamiltonian and Low-energy Projection}
\label{app:MBhamiltonian}

We now define the interacting Hamiltonian. Since the family of multi-layer graphene continuum models is defined relative to charge neutrality of the half-filled $p_z$-bands of the microscopic graphene structure, the charge density must be measured relative to this point in order to define a long-range (electrostatic) Coulomb interaction with finite energy density (see Ref. \cite{giuliani2008quantum}) in the absence of gates. To do so, we define the Fourier components of the density operator in the $l$th layer for $\mbf{q} \in \mathbb{R}^2$ as
 \bea
\delta \rho_{\mbf{q},l} &=  \sum_{\mbf{k},\sigma \eta s} (c^\dag_{\mbf{k}+\mbf{q}, l \sigma \eta s}c_{\mbf{k}, l \sigma \eta s} - \frac{1}{2}\delta_{\mbf{q},0}) \ . \\
\eea
The $1/2$ term corresponds to a featureless neutrality reference at $D=0$. This can be viewed as a choice of normal ordering. The many-body Coulomb interaction is then
\bea
H_{int} &= \frac{1}{2 \Omega_{tot}} \sum_{\mbf{q} ll'} V_{ll'}(\mbf{q}) \delta \rho_{\mbf{q},l} \delta \rho_{-\mbf{q},l'}
\eea
where $V_{ll'}(\mbf{q})$ is the Fourier transform of the interaction potential between electrons on the $l$ and $l'$ layers. For concreteness, the interaction potential in the absence of gate screening, assuming an isotropic dielectric environment, is
\bea
V_{ll'}(\mbf{q}) = \frac{e^2}{2\eps |\mbf{q}|}e^{-d |l - l'| |\mbf{q}|}
\eea
where $\eps$ is the dielectric constant and $d = 0.33$\,nm is the interlayer distance. We will discuss the inclusion of gates and dielectric anisotropy later and keep $V_{ll'}(\mbf{q})$ general in the following.  

We now derive the correct form of the projected interaction. By projection, we mean that we restrict the low-energy Hilbert space to states in the form $\mathcal{O}_{\text{act.}} \ket{\text{valence}}$ where $\mathcal{O}_{\text{act.}}$ is an operator basis containing only $\gamma^\dag_{\mbf{k},n\geq 0}$ creation operators and
\bea
\ket{\text{valence}} &= \prod_{\mbf{k},n\in \text{val.}} \gamma^\dag_{\mbf{k},n} \ket{0} \ ,
\eea
corresponds to occupying all the valence states. The projected Hamiltonian $\overline{H_{int}}$ is defined by 
\bea
\bra{\text{valence}}  \mathcal{O}'^\dag_{\text{act.}} H_{int} \mathcal{O}_{\text{act.}} \ket{\text{valence}} &= \bra{0}  \mathcal{O'}^\dag_{\text{act.}} \overline{H_{int}} \mathcal{O}_{\text{act.}} \ket{0} + \text{const.},
\eea
where $\ket{0}$ is the fermion vacuum. In other words, we freeze the fully-occupied valence band Fermi sea. This approximation is controlled when the single-particle gap between the conduction and valence bands is larger than the interaction matrix element between them. Hence this is accurate in the large displacement field regime. In some other works, the effect of the valence band Fermi sea has been discarded \cite{dong2023anomalous,zhou2023fractional,dong2023theory,guo2023theory,huang2024selfconsistent}. However, our treatment below shows that, even though it is (approximately) inert, the valence bands still generate one-body terms that alter the conduction bands. 

Since the projection is accomplished in the band basis, we first transform the Hamiltonian into this basis using \Eq{eq:transform}. We write (now $\mbf{q}$ is in the moir\'e Brilloin zone)
\bea
\delta \rho_{\mbf{q}+\mbf{G},l} &=  \sum_{\mbf{k}, \mbf{G}',\sigma \eta s} (c^\dag_{\mbf{k}+\mbf{q},\mbf{G}'+\mbf{G}, l \sigma \eta s}c_{\mbf{k},\mbf{G}', l \sigma \eta s} - \frac{1}{2}\delta_{\mbf{q},0}\delta_{\mbf{G},0}) \\
&=  \sum_{\mbf{k}, \mbf{G}' \sigma,mn, \eta s} \gamma^\dag_{\mbf{k}+\mbf{q},m, \eta s} U^{\eta s}_{\mbf{G}'+\mbf{G}, l \sigma,m}(\mbf{k}+\mbf{q})^* U^{\eta}_{\mbf{G}' l \sigma,n}(\mbf{k})  \gamma_{\mbf{k}, n, \eta s} -  \sum_{\mbf{k},\mbf{G}',\sigma \eta s} \frac{1}{2}\delta_{\mbf{q},0} \delta_{\mbf{G},0} \\
&=  \sum_{\mbf{k},mn, \eta s} M^{l,\eta s}_{mn}(\mbf{k},\mbf{q}+\mbf{G})\gamma^\dag_{\mbf{k}+\mbf{q},m, \eta s} \gamma_{\mbf{k}, n, \eta s} -  \sum_{\mbf{k},\mbf{G}',\sigma \eta s} \frac{1}{2}\delta_{\mbf{q},0} \delta_{\mbf{G},0} \\
\eea
where we defined the layer-resolved form factors
\bea
M^{l,\eta s}_{mn}(\mbf{k},\mbf{q}+\mbf{G}) &= \sum_{\mbf{G}',\sigma} U^{\eta}_{\mbf{G}', l \sigma,m}(\mbf{k}+\mbf{q}+\mbf{G})^* U^{\eta}_{\mbf{G}' l \sigma,n}(\mbf{k}) \ . 
\eea
We obtain
\bea
\delta \rho_{\mbf{q}+\mbf{G},l} &= \sum_{\mbf{k},mn, \eta s} M^{l,\eta s}_{mn}(\mbf{k},\mbf{q}+\mbf{G}) (\gamma^\dag_{\mbf{k}+\mbf{q},m, \eta s} \gamma_{\mbf{k}, n, \eta s} - \frac{1}{2} \delta_{\mbf{q},0}  \delta_{mn})
\eea
from the completeness relation of the eigenvectors $ \delta_{\mbf{G},0}  =  \sum_m U^{\eta}_{\mbf{G}'+\mbf{G}, l \sigma,m}(\mbf{k}) U^{\eta}_{\mbf{G}' l \sigma,m}(\mbf{k})^*$. 
The sum is still over all bands, and no projection has occurred yet.

We now separate the interaction Hamiltonian into a conventionally normal-ordered two-body interaction (i.e.~of the form $\gamma^\dag\gamma^\dag\gamma\gamma$) and a one-body term. We find (ignoring overall constants)
\bea
\label{eq:HintmasterEq}
H_{int} &= \frac{1}{2 \Omega_{tot}} \sum_{\mbf{q} \mbf{G} ll'} \sum_{\mbf{k} mn \eta s} \sum_{\mbf{k}' m'n' \eta' s'}V_{ll'}(\mbf{q}+\mbf{G}) M^{l,\eta s}_{mn}(\mbf{k},\mbf{q}+\mbf{G})M^{l',\eta' s'}_{m'n'}(\mbf{k}',-\mbf{q}-\mbf{G}) \gamma^\dag_{\mbf{k}+\mbf{q},m, \eta s} \gamma_{\mbf{k}, n, \eta s} \gamma^\dag_{\mbf{k}'-\mbf{q},m', \eta' s'} \gamma_{\mbf{k}', n', \eta' s'}  \\
&\qquad -  \frac{1}{2 \Omega_{tot}} \sum_{\mbf{q} \mbf{G}, ll'} \sum_{\mbf{k} mn \eta s} \sum_{\mbf{k}' m'n' \eta' s'}V_{ll'}(\mbf{q}+\mbf{G}) M^{l,\eta s}_{mn}(\mbf{k},\mbf{q}+\mbf{G}) M^{l',\eta' s'}_{m'n'}(\mbf{k}',-\mbf{q}-\mbf{G})\gamma^\dag_{\mbf{k}+\mbf{q},m, \eta s} \gamma_{\mbf{k}, n, \eta s}  \delta_{\mbf{q},0} \delta_{m'n'} \\
&= \frac{1}{2 \Omega_{tot}} \sum_{\mbf{q} \mbf{G} ll'} \sum_{\mbf{k} mn \eta s} \sum_{\mbf{k}' m'n' \eta' s'}V_{ll'}(\mbf{q}+\mbf{G}) M^{l,\eta s}_{mn}(\mbf{k},\mbf{q}+\mbf{G})M^{l',\eta' s'}_{m'n'}(\mbf{k}',-\mbf{q}-\mbf{G}) \gamma^\dag_{\mbf{k}+\mbf{q},m, \eta s} \gamma^\dag_{\mbf{k}'-\mbf{q},m', \eta' s'} \gamma_{\mbf{k}', n', \eta' s'} \gamma_{\mbf{k}, n, \eta s}  \\
&\qquad -  \frac{1}{2 \Omega_{tot}} \sum_{ \mbf{G}, ll'} \sum_{\mbf{k} mn \eta s} \sum_{\mbf{k}' m' \eta' s'}V_{ll'}(\mbf{G}) M^{l,\eta s}_{mn}(\mbf{k},\mbf{G}) M^{l',\eta' s'}_{m'm'}(\mbf{k}',-\mbf{G})\gamma^\dag_{\mbf{k},m, \eta s} \gamma_{\mbf{k}, n, \eta s}   \\
&\qquad + \frac{1}{2 \Omega_{tot}} \sum_{\mbf{q} \mbf{G} ll'} \sum_{\mbf{k} mn \eta s} V_{ll'}(\mbf{q}+\mbf{G}) [M^{l,\eta s}(\mbf{k},\mbf{q}+\mbf{G})M^{l',\eta s}(\mbf{k}+\mbf{q},-\mbf{q}-\mbf{G})]_{mn} \gamma^\dag_{\mbf{k}+\mbf{q},m, \eta s}  \gamma_{\mbf{k}+\mbf{q}, n, \eta s}  \\
&= \frac{1}{2 \Omega_{tot}} \sum_{\mbf{q} \mbf{G} ll'} \sum_{\mbf{k} mn \eta s} \sum_{\mbf{k}' m'n' \eta' s'}V_{ll'}(\mbf{q}+\mbf{G}) M^{l,\eta s}_{mn}(\mbf{k},\mbf{q}+\mbf{G})M^{l',\eta' s'}_{m'n'}(\mbf{k}',-\mbf{q}-\mbf{G}) \gamma^\dag_{\mbf{k}+\mbf{q},m, \eta s} \gamma^\dag_{\mbf{k}'-\mbf{q},m', \eta' s'} \gamma_{\mbf{k}', n', \eta' s'} \gamma_{\mbf{k}, n, \eta s}  \\
&\qquad -  \frac{1}{2 \Omega_{tot}} \sum_{ \mbf{G}, ll'} \sum_{\mbf{k} mn \eta s} \sum_{\mbf{k}' m' \eta' s'}V_{ll'}(\mbf{G}) M^{l,\eta s}_{mn}(\mbf{k},\mbf{G}) M^{l',\eta' s'}_{m'm'}(\mbf{k}',-\mbf{G})\gamma^\dag_{\mbf{k},m, \eta s} \gamma_{\mbf{k}, n, \eta s}   \\
&\qquad + \frac{1}{2 \Omega_{tot}} \sum_{\mbf{q} \mbf{G} ll'} \sum_{\mbf{k} mn \eta s} V_{ll'}(\mbf{q}+\mbf{G}) [M^{l,\eta s}(\mbf{k},\mbf{q}+\mbf{G})^\dag M^{l',\eta s}(\mbf{k},\mbf{q}+\mbf{G})]_{mn} \gamma^\dag_{\mbf{k},m, \eta s}  \gamma_{\mbf{k}, n, \eta s}  \\
\eea
where we used $V_{ll'}(\mbf{q}) = V_{ll'}(-\mbf{q})$ and $M^{l,\eta s}(\mbf{k}-\mbf{q},\mbf{q}) = M^{l,\eta s}(\mbf{k},-\mbf{q})^\dag$. We have separated out the one-body terms (on the last two lines) from the conventionally normal-ordered term $:H_{int}:$ where all annihilation operators are placed to the right. So far this is an exact rewriting. Note that the expansion of $\delta \rho$ has generated Hartree- and Fock-like one body terms. 

We now compute the projected Hamiltonian on the active subspace. First we study the one-body terms. Consider a generic one-body term $H_1 = \sum_{\mbf{k},mn,\eta s} h_{mn}^{\eta s}(\mbf{k}) \gamma^\dag_{\mbf{k},m,\eta s} \gamma_{\mbf{k},n,\eta s}$, which obeys
\bea
\label{eq:onebody}
\bra{\text{valence}} \mathcal{O}'^\dag_{\text{act.}}  H_1 \mathcal{O}_{\text{act.}} \ket{\text{valence}} &= \bra{0} \mathcal{O}'^\dag_{\text{act.}}  \bar{H}_1 \mathcal{O}_{\text{act.}} \ket{0} + \bra{\text{valence}} H_1 \ket{\text{valence}}   \bra{0} \mathcal{O}'^\dag_{\text{act.}} \mathcal{O}_{\text{act.}} \ket{0}
\eea
where  $\bar{H}_1 = \sum_{\mbf{k},mn \in\text{act.},\eta s} h_{mn}^{\eta s}(\mbf{k}) \gamma^\dag_{\mbf{k},m,\eta s} \gamma_{\mbf{k},n,\eta s} $ is the standard projection. The second term in \Eq{eq:onebody} is a constant on the many-body conduction band Hilbert space which we neglect moving forward. \Eq{eq:onebody} is easily seen term-by-term by examining the cases $m,n$ both in the active bands (first term) and both in the valence bands (second term). The case of one $m,n$ in the active bands vanishes. This result covers the one-body terms in \Eq{eq:HintmasterEq}, which we will write explicit expressions for after dealing with $:H_{int}:$. 

Next we consider the normal-ordered two-body interaction $:H_{int}:$. There are $2^4$ types of terms since each $\gamma, \gamma^\dag$ operator can either be in the frozen or active space. All terms that do not conserve the total number of valence electrons vanish within the $\bra{\text{valence}} \mathcal{O}'^\dag_{\text{act.}}  :H_{int}: \mathcal{O}_{\text{act.}} \ket{\text{valence}} $ expectation value. There are 6 terms that remain: all four operators belong to the active subspace (the standard projected interaction), all four operators belong to the frozen subspace (an overall constant), and then four terms corresponding to virtual processes hopping a valence electron to the active subspace and back. These terms are computed to be 
\bea
\label{eq:twobody}
\bra{\text{valence}} \mathcal{O}'^\dag_{\text{act.}} \! :\!H_{int}\!: \mathcal{O}_{\text{act.}}\! \ket{\text{valence}} &= \frac{1}{2 \Omega_{tot}} \sum_{\mbf{q} \mbf{G} ll'} \sum_{\mbf{k} mn \eta s} \sum_{\mbf{k}' m'n' \eta' s'}V_{ll'}(\mbf{q}+\mbf{G}) M^{l,\eta s}_{mn}(\mbf{k},\mbf{q}+\mbf{G})M^{l',\eta' s'}_{m'n'}(\mbf{k}',-\mbf{q}-\mbf{G}) \\ \Big(
&\qquad \bra{0} \mathcal{O}'^\dag_{\text{act.}}  \gamma^\dag_{\mbf{k}+\mbf{q},m, \eta s} \gamma^\dag_{\mbf{k}'-\mbf{q},m', \eta' s'} \gamma_{\mbf{k}', n', \eta' s'} \gamma_{\mbf{k}, n, \eta s}\mathcal{O}_{\text{act.}} \ket{0} \\
&\qquad + \bra{0}\mathcal{O}'^\dag_{\text{act.}} \gamma^\dag_{\mbf{k}+\mbf{q},m, \eta s} \gamma_{\mbf{k}, n, \eta s}\mathcal{O}_{\text{act.}} \ket{0}  \bra{\text{valence}} \gamma^\dag_{\mbf{k}'-\mbf{q},m', \eta' s'} \gamma_{\mbf{k}', n', \eta' s'}  \ket{\text{valence}} \\
&\qquad  + \bra{0} \mathcal{O}'^\dag_{\text{act.}} \gamma^\dag_{\mbf{k}'-\mbf{q},m', \eta' s'} \gamma_{\mbf{k}', n', \eta' s'} \mathcal{O}_{\text{act.}} \ket{0} \bra{\text{valence}} \gamma^\dag_{\mbf{k}+\mbf{q},m, \eta s} \gamma_{\mbf{k}, n, \eta s} \ket{\text{valence}} \\
&\qquad -  \bra{0} \mathcal{O}'^\dag_{\text{act.}} \gamma^\dag_{\mbf{k}'-\mbf{q},m', \eta' s'}\gamma_{\mbf{k}, n, \eta s}\mathcal{O}_{\text{act.}} \ket{0} \bra{\text{valence}}  \gamma^\dag_{\mbf{k}+\mbf{q},m, \eta s}  \gamma_{\mbf{k}', n', \eta' s'} \ket{\text{valence}} \\
&\qquad - \bra{0} \mathcal{O}'^\dag_{\text{act.}} \gamma^\dag_{\mbf{k}+\mbf{q},m, \eta s}  \gamma_{\mbf{k}', n', \eta' s'} \mathcal{O}_{\text{act.}} \ket{0}  \bra{\text{valence}} \gamma^\dag_{\mbf{k}'-\mbf{q},m', \eta' s'}\gamma_{\mbf{k}, n, \eta s}  \ket{\text{valence}} \Big) \\
&\qquad + \text{const.} \\
\eea
  The one-body terms generated in \Eq{eq:twobody} also take Hartree- and Fock-like forms. Using $\bra{\text{valence}}  \gamma^\dag_{\mbf{k},m, \eta s}  \gamma_{\mbf{k}', n', \eta' s'} \ket{\text{valence}} = \delta_{\mbf{k},\mbf{k}'} \delta_{\eta \eta'} \delta_{ss'} \delta_{mn} \delta_{m \in \text{val.}}$, we obtain
 \bea
 \label{eq:twobodyexpanded}
\bra{\text{valence}} \mathcal{O}'^\dag_{\text{act.}} \! :\!H_{int}\!: \mathcal{O}_{\text{act.}}\! \ket{\text{valence}}  &=  \, : \bar{H}_{int}: \bra{0} \mathcal{O}'^\dag_{\text{act.}} \mathcal{O}_{\text{act.}} \ket{0}  \\
&\!\!\!\!\!\!\!\!\!\!\!\! \!\!\!\!\!\!\!\!\!\!\!\!\!\!\!\!\!\!\!\!\!\!\!\!\! \!\!\!\!\!\!\!\!\!\!\!\!\!\!\!\!\!\!\!\! +  \frac{1}{\Omega_{tot}} \sum_{\mbf{G} ll'} \sum_{\mbf{k} mn \eta s} \sum_{\mbf{k}' \eta' s',m' \in \text{val}}V_{ll'}(\mbf{G}) M^{l,\eta s}_{mn}(\mbf{k},\mbf{G})M^{l',\eta' s'}_{m'm'}(\mbf{k}',-\mbf{G}) \bra{0}\mathcal{O}'^\dag_{\text{act.}} \gamma^\dag_{\mbf{k},m, \eta s} \gamma_{\mbf{k}, n, \eta s}\mathcal{O}_{\text{act.}} \ket{0} \\
&\!\!\!\!\!\!\!\!\!\!\!\! \!\!\!\!\!\!\!\!\!\!\!\!\!\!\!\!\!\!\!\!\!\!\!\!\! \!\!\!\!\!\!\!\!\!\!\!\!\!\!\!\!\!\!\!\! - \frac{1}{\Omega_{tot}} \sum_{\mbf{G} ll'} \sum_{\mbf{k} mn \eta s} \sum_{\mbf{q},m' \in \text{val}}V_{ll'}(\mbf{q}+\mbf{G}) M^{l',\eta s}_{m m'}(\mbf{k}+\mbf{q},-\mbf{q}-\mbf{G}) M^{l,\eta s}_{m'n}(\mbf{k},\mbf{q}+\mbf{G})  \bra{0} \mathcal{O}'^\dag_{\text{act.}} \gamma^\dag_{\mbf{k},m, \eta s}\gamma_{\mbf{k}, n, \eta s}\mathcal{O}_{\text{act.}} \ket{0}
 \eea
which again are Hartree-like and Fock-like respectively (note the factor of 2).   

We have now dealt with the full projection of $H_{int}$ and separated out the conduction band interaction from the one-body terms. In \Eq{eq:HintmasterEq}, one-body terms were generated by expanding out the $\frac{1}{2} \delta_{\mbf{q},0} \delta_{mn}$ terms, and in \Eq{eq:twobody} they were generated by virtual processes. We now collect all terms from \Eq{eq:onebody} and \Eq{eq:twobody} and obtain our final result
\bea
\overline{H_{int}} &= \, :\!\overline{H}_{int} \!: \, +  \bar{H}_H + \bar{H}_F \\
\eea
where the interaction is
\bea
\label{eq:intHexplicit}
:\!\overline{H}_{int} \!: &= \frac{1}{2 \Omega_{tot}} \sum_{\mbf{q} \mbf{G} ll'} \sum_{\substack{\mbf{k}\mbf{k}' \eta s \eta' s' \\mnm'n' \in \text{act.}}}V_{ll'}(\mbf{q}+\mbf{G}) M^{l,\eta s}_{mn}(\mbf{k},\mbf{q}+\mbf{G})M^{l',\eta' s'}_{m'n'}(\mbf{k}',-\mbf{q}-\mbf{G}) \gamma^\dag_{\mbf{k}+\mbf{q},m, \eta s} \gamma^\dag_{\mbf{k}'-\mbf{q},m', \eta' s'} \gamma_{\mbf{k}', n', \eta' s'} \gamma_{\mbf{k}, n, \eta s}
\eea
and $\bar{H}_H$ and  $\bar{H}_F$ collect the one-body terms from \Eq{eq:HintmasterEq} and \Eq{eq:twobodyexpanded} to yield
\bea
\label{eq:HFterms}
\bar{H}_H &= \sum_{\mbf{k}\eta s, mn \in \text{act.}} \gamma^\dag_{\mbf{k},m, \eta s} \gamma_{\mbf{k}, n, \eta s} \sum_{\mbf{G}, ll'}  V_{ll'}(\mbf{G}) M^{l,\eta s}_{mn}(\mbf{k},\mbf{G}) \lp \frac{1}{\Omega_{tot}} \sum_{\mbf{k}' m' \eta' s'} M^{l',\eta' s'}_{m'm'}(\mbf{k}',-\mbf{G}) (\delta_{m' \in \text{val.}}  - \frac{1}{2})\rp \ , \\
\bar{H}_F&= - \sum_{\mbf{k} \eta s,  mn \in \text{act.}} \gamma^\dag_{\mbf{k},m, \eta s}\gamma_{\mbf{k}, n, \eta s}  \frac{1}{\Omega_{tot}} \sum_{\mbf{q} \mbf{G} m' ll'}  V_{ll'}(\mbf{q}+\mbf{G})M^{l',\eta s}_{m'm}(\mbf{k},\mbf{q}+\mbf{G})^* M^{l,\eta s}_{m'n}(\mbf{k},\mbf{q}+\mbf{G}) (\delta_{m' \in \text{val.}} - \frac{1}{2}) \\
\eea
where we used $V_{ll'}(\mbf{q})=V_{l'l}(-\mbf{q})$ and $M^{l \eta s}_{nm}(\mbf{k},\mbf{q})^* = M^{l \eta s}_{mn}(\mbf{k}+\mbf{q},-\mbf{q})$. These equations complete our derivation of the correct projected Hamiltonian. The key physical effect is the generation of one-body terms $\bar{H}_H$ and $\bar{H}_F$ which arise from the occupied valence bands. To understand their physical meaning, which may not be apparent from \Eq{eq:HFterms}, we will now rewrite them in a (projected) plane-wave basis.

\subsection{Plane-Wave Form}

We will first rewrite the Hartree term to make its physical meaning apparent. To start, we compute the Fourier component of the valence charge density:
\bea
\label{eq:backgrounddensity}
 \rho^{\text{val.}}_{\mbf{G},l'} &= \frac{1}{\Omega_{tot}} \braket{\text{valence}|\delta \rho_{\mbf{G},l}|\text{valence}} \\
 &=  \frac{1}{\Omega_{tot}}\sum_{\mbf{k},n \in \text{val}, \eta s} M^{l,\eta s}_{nn}(\mbf{k},\mbf{G}) -  \sum_{\mbf{k},\mbf{G}',\sigma \eta s} \frac{1}{2}\delta_{\mbf{q},0} \delta_{\mbf{G},0} \\
&=  \frac{1}{\Omega_{tot}}\sum_{\mbf{k}, \eta s} \lp \sum_{n \in \text{val}} M^{l,\eta s}_{nn}(\mbf{k},\mbf{G}) -  \sum_{\mbf{G}',\sigma} \frac{1}{2} \delta_{\mbf{G},0} \rp \\
&=  \frac{1}{\Omega_{tot}} \sum_{\mbf{k} \eta s, n}  M^{l,\eta s}_{nn}(\mbf{k},\mbf{G})  \lp \delta_{n \in \text{val}}  - \frac{1}{2}   \rp \\
\eea
using the completeness relation $ \delta_{\mbf{G},0}  =  \sum_m U^{\eta}_{\mbf{G}'+\mbf{G}, l \sigma,m}(\mbf{k}) U^{\eta}_{\mbf{G}' l \sigma,m}(\mbf{k})^*$. To show that this expression is manifestly gauge-invariant, let us introduce the valence manifold projector 
\bea
P^{\eta s}_{\text{val.}}(\mbf{k}) = \sum_{m \in \text{val.}} U_m^{\eta}(\mbf{k})U_m^{\eta}(\mbf{k})^\dag, \qquad \delta P^{\eta s}_{\text{val.}}(\mbf{k}) = P^{\eta s}_{\text{val.}}(\mbf{k}) - \frac{1}{2} \mathbb{1} \ .
\eea 
We see that we can write
\bea
\rho^{\text{val.}}_{\mbf{G},l} &=  \frac{1}{\Omega_{tot}} \sum_{\mbf{k} \eta s, \mbf{G}' \sigma, n}  U^\eta_{\mbf{G}'+\mbf{G}, l \sigma,n}(\mbf{k})^* U^\eta_{\mbf{G}' l \sigma,n}(\mbf{k})  \lp \delta_{n \in \text{val}}  - \frac{1}{2}   \rp \\ 
&= \frac{1}{\Omega_{tot}} \sum_{\mbf{k} \eta s} \sum_{\mbf{G}'\sigma} ([P^{\eta s}_{\text{val.}}(\mbf{k})]_{\mbf{G}' l \sigma,\mbf{G}'+\mbf{G},l \sigma}  -  \frac{1}{2}\delta_{\mbf{G},0}) \\
&= \frac{1}{\Omega_{tot}} \sum_{\mbf{k} \eta s} \sum_{\mbf{G}'\sigma} [\delta P^{\eta s}_{\text{val.}}(\mbf{k})]_{\mbf{G}' l \sigma,\mbf{G}'+\mbf{G},l \sigma}  \ .  \\
\eea
From \Eq{eq:HFterms}, we can now write
\bea
\bar{H}_H &= \sum_{\mbf{G}, ll'}  V_{ll'}(\mbf{G}) \sum_{\mbf{k}\eta s, mn \in \text{act.}} \gamma^\dag_{\mbf{k},m, \eta s} \gamma_{\mbf{k}, n, \eta s}  M^{l,\eta s}_{mn}(\mbf{k},\mbf{G}) \rho^{\text{val.}}_{-\mbf{G},l'} \ . \\
\eea
To show that $\bar{H}_H$ can be written in a fully gauge-invariant form, we will introduce the projected plane-wave operators 
\bea
\sum_{n \in \text{act.}} \gamma^\dag_{\mbf{k},n,\eta s} U^{\eta}_{\mbf{G}' l' \sigma', n}(\mbf{k})^* &= \sum_{\mbf{G},l \sigma}  c^\dag_{\mbf{k},\mbf{G},l \sigma,\eta s} [P^{\eta s}_{\text{con.}}(\mbf{k})]_{\mbf{G} l \sigma,\mbf{G}' l' \sigma'} \equiv \bar{c}^\dag_{\mbf{k},\mbf{G},l \sigma,\eta s}\\
\eea
using \Eq{eq:transform}, with $P^{\eta s}_{\text{con.}}(\mbf{k}) = \sum_{m \in \text{act.}} U_m^{\eta}(\mbf{k})U_m^{\eta}(\mbf{k})^\dag$. We can now write
\bea
\bar{H}_H &= \sum_{\mbf{G}, ll'}  V_{ll'}(\mbf{G}) \sum_{\mbf{k}\eta s, \mbf{G}' \sigma, mn \in \text{act.}} \gamma^\dag_{\mbf{k}, m, \eta s} \gamma_{\mbf{k}, n, \eta s}  
U^\eta_{\mbf{G}'+\mbf{G}, l \sigma,n}(\mbf{k})^* U_{\mbf{G}' l \sigma,n}(\mbf{k})
 \rho^{\text{val.}}_{-\mbf{G},l'} \\
 &= \sum_{\mbf{G}, ll'}  V_{ll'}(\mbf{G}) \sum_{\mbf{k}\eta s, \mbf{G}' \sigma, mn \in \text{act.}} \bar{c}^\dag_{\mbf{k},\mbf{G}+\mbf{G}', \sigma l, \eta s} \bar{c}_{\mbf{k},\mbf{G}', \sigma l, \eta s} 
 \rho^{\text{val.}}_{-\mbf{G},l'} \\
&=  \sum_{\mbf{k},\mbf{G}',\mbf{G},\sigma,ll', \eta s} \bar{c}^\dag_{\mbf{k},\mbf{G}',l \sigma, \eta s} \bar{c}_{\mbf{k},\mbf{G}'-\mbf{G},l \sigma, \eta s} V_{ll'}(\mbf{G}) \rho^{\text{val.}}_{-\mbf{G},l'} \ .
\eea
Transforming back to position space with \Eq{eq:fft}, we have the standard expression
\bea
\label{eq:hartreelayerdep}
\bar{H}_H &= \sum_{l \sigma, \eta s}  \int d^2r \, \mathcal{V}_{\text{val}., l}(\mbf{r}) \bar{c}^\dag_{\mbf{r},l \sigma,\eta s} \bar{c}_{\mbf{r}, l \sigma, \eta s} , \qquad  \mathcal{V}_{\text{val}., l}(\mbf{r}) = \sum_{\mbf{G} l'} e^{i \mbf{G} \cdot \mbf{r}} V_{ll'}(\mbf{G}) \rho^{\text{val.}}_{-\mbf{G},l'}  \ .
\eea
This form makes it clear that $H_H[\delta P_{\text{val.}}]$ acts as a scalar potential on the $l$th layer with Fourier harmonics $\sum_{l'} V_{ll'}(\mbf{G}) \rho^{\text{val.}}_{-\mbf{G},l'}$ where $\rho^{\text{val.}}_{\mbf{G},l'}$ are the layer-resolved Fourier coefficients of the valence charge density. 

The $\mbf{G}=0$ harmonic preserves continuous translation symmetry and does not create a moir\'e potential. Its only effect is to contribute a uniform layer-resolved potential that renormalizes the displacement field as discussed above in \Eq{eq:gauss}. One would expect $\rho^{\text{val.}}_{\mbf{G}=0,l}$ to give a good perturbative approximation to the self-consistent equations. However, this is not the case since $V_{ll'}(\mbf{G}) \sim 1/|\mbf{G}|$ is singular at $\mbf{G}=0$ for a 3D Coulomb interaction. For a gate-screened interaction, this term is finite but large. (In HF, this is leads to the phenomenon of ``charge-sloshing" that requires the optimal damping algorithm to converge \cite{HF_AIPreview}.)   Hence, we drop the $\mbf{G}=0$ term moving forward and take $V$ to be the effective interlayer potential. 

The novel terms are the $\mbf{G} \neq 0$ harmonics which break the continuous translation symmetry down to moir\'e translations as seen in \Eq{eq:hartreelayerdep}. We will develop a perturbation theory to calculate $\rho^{\text{val.}}_{\mbf{G},l}$ from a weak substrate potential in App.~\ref{app:moirePT} and \App{app:hartreecommentary}.

Next we study the Fock term. It is convenient to take $\mbf{q} + \mbf{G} \to \mbf{q} \in \mathds{R}^2$ and write
\bea\label{eqapp:HFock}
\bar{H}_F &= - \sum_{\mbf{k}n m \eta s} \gamma^\dag_{\mbf{k},m, \eta s}\gamma_{\mbf{k}, n, \eta s}  \frac{1}{\Omega_{tot}} \sum_{\mbf{q} \in \mathds{R}^2, m' ll'}  V_{ll'}(\mbf{q})M^{l',\eta s}_{m'm}(\mbf{k},\mbf{q})^* M^{l,\eta s}_{m'n}(\mbf{k},\mbf{q}) (\delta_{m' \in \text{val.}} - \frac{1}{2}) \\
&= - \sum_{\mbf{k},\mbf{G} \mbf{G}',\sigma \sigma',ll', \eta s} \bar{c}^\dag_{\mbf{k},\mbf{G},l \sigma, \eta s}\bar{c}_{\mbf{k}, \mbf{G}',l' \sigma', \eta s}  \frac{1}{\Omega_{tot}} \sum_{\mbf{q} \in \mathds{R}^2, m'}  V_{ll'}(\mbf{q})U_{m',\mbf{G} l \sigma}^{\eta s}(\mbf{k}+\mbf{q})(\delta_{m' \in \text{val.}} - \frac{1}{2})  U_{m',\mbf{G}' l' \sigma'}^{\eta s}(\mbf{k}+\mbf{q})^*\\
&= - \sum_{\mbf{k},\mbf{G} \mbf{G}',\sigma \sigma',ll', \eta s}  \bar{c}^\dag_{\mbf{k},\mbf{G},l \sigma, \eta s}\bar{c}_{\mbf{k}, \mbf{G}',l' \sigma', \eta s}  \frac{1}{\Omega_{tot}} \sum_{\mbf{q} \in \mathds{R}^2}  V_{ll'}(\mbf{q})[P^{\eta s}_{\text{val.}}(\mbf{k}+\mbf{q}) - \frac{1}{2} \mathbb{1}]_{\mbf{G} l \sigma, \mbf{G}' l' \sigma'} \ . \\
\eea
The Fock term is more subtle because this integral in fact diverges logarithmically at large $\mbf{q}$ and hence its value is dependent on the cutoff $\Lambda$.  Note that this divergence comes from the $1/|\mbf{q}$ tail of the interaction and is a short-range phenomenon, which is not cut-off by long-range gate screening. This is an artifact of the continuum theory that is well understood from the simplest case of monolayer graphene and is responsible for the renormalization of the hopping parameters, e.g. $v_F$ (see Refs. \cite{Borghi2009FermiVelocityEnhancement,PhysRevLett.118.266801,2025arXiv250812825S} for a Hartree-Fock-focused review). We will address it in more detail in \App{app:fockcommentary}. 

\subsection{Moir\'e Perturbation Theory}
\label{app:moirePT}

The main object we need to compute for the background Hartree and Fock terms is $P_{\text{val.}}(\mbf{k})$, which is the one-body density matrix of the occupied valence bands. (We work in the $K$ valley in this section. The results for valley $K'$ can be obtained by time-reversal.) Since the bare hBN potential $V_1 = 21$\,meV is weak  compared to the gap $V(L-1) \sim 100$\,meV, we can obtain analytical expressions using perturbation theory. A direct way to compute $P_{\text{val.}}$ is via scattering theory. First, recall the Riesz projector formula for a Hermitian matrix $h = \sum_n E_n P_n$:
\bea
\oint_{z<0} \frac{dz}{2\pi i} \frac{1}{z - h} &= \sum_n P_n \oint_{z<0} \frac{dz}{2\pi i} \frac{1}{z - E_n} = \sum_{E_n<0} P_n = P
\eea
from which we obtain the matrix geometric series 
\bea
P_{\text{val.}}(\mbf{k}) &= \oint_{z<0} \frac{dz}{2\pi i} \frac{1}{z - h(\mbf{k}) -V_M } =  \oint_{z<0} \frac{dz}{2\pi i} \lp \frac{1}{z - h(\mbf{k})} + \frac{1}{z - h(\mbf{k})} V_M \frac{1}{z - h(\mbf{k})} + \dots \rp \\
&= P^0_{\text{val.}}(\mbf{k}) + P^1_{\text{val.}}(\mbf{k}) + \dots
\eea
where $h(\mbf{k}) = H_K + H_{ISP} + H_D + V_{tb}$ is the kinetic energy including all terms obeying continuous translation symmetry, and hence is block-diagonal in the reciprocal basis vectors $\mbf{G},\mbf{G}'$. Finally, $V_M $ is the moir\'e periodic term (repeated from \Eq{eq:moirematrixelement})
\bea
[V_M]_{\mbf{G}l \sigma,\mbf{G}' l' \sigma'} = V_1 e^{i \psi}  \delta_{ll'}\delta_{l,1} \sum_{j=1,2,3} \delta_{\mbf{G},\mbf{G}'+\mbf{b}_j} \bpm 1 & \omega^{-j} \\ \omega^{j+1} & \omega \epm_{\sigma \sigma'} + h.c. 
\eea
We take the contour in $z$ to surround all occupied bands separated by the neutrality gap $\sim (L-1)V$ opened by the displacement field $V$. This is a perturbation theory in $\sim V_M/((L-1)V)$  in terms of the bare $2L\times 2L$ rhombohedral graphene energies $\la_n(\mbf{k})$  and eigenstate projectors $P_n(\mbf{k})$, where $n$ is the band index. Note that we also assume that the interaction scale is also much less than the neutrality gap. 

The zeroth order term $P^0_{\text{val.}}(\mbf{k})$ is simply the bare rhombohedral valence projector which has continuous translation symmetry. The first order term is
\bea
\label{eq:P1folded}
P^1_{\text{val.}}(\mbf{k}) &= \sum_{mn} \oint_{z<0} \frac{dz}{2\pi i} \frac{P_m(\mbf{k})}{z - \lambda_m(\mbf{k})} V_M \frac{P_n(\mbf{k})}{z - \lambda_n(\mbf{k})} \\
&= \sum_{mn}  P_m(\mbf{k}) V_M P_n(\mbf{k}) \oint_{z<0} \frac{dz}{2\pi i} \frac{1}{(z - \lambda_m(\mbf{k})) (z - \lambda_n(\mbf{k}))} \ . \\
&=\sum_{m\in\text{val}, n\in\text{con}} \frac{P_m(\mbf{k}) V_M P_n(\mbf{k})+P_n(\mbf{k}) V_M P_m(\mbf{k})}{\lambda_m(\mbf{k}) - \lambda_n(\mbf{k})}  \ . \\
\eea
In the last line, we dropped the $m,n \in \text{val.}$ terms from the sum since the numerator is symmetric in $m\leftrightarrow n$ but the denominator is anti-symmetric.   This expression is manifestly Hermitian and shows clearly that moir\'e modulation occurs from hybridization between the valence (val.) and conduction (con.) manifold at charge neutrality (controlled by $V_M/((L-1)V)$). As $V \to \infty$, this term vanishes, and its effect is also reduced in thicker (larger $L$) devices.   However, for realistic $V\sim25$\,meV in pentalayer graphene, the gap is $\sim100$\,meV and the perturbation is $V_M \sim 20$\,meV, leading to a non-negligible induced moir\'e potential. 

We now develop more explicit expressions to show the convergence of this term since the valence and conduction bands are unbounded in the absence of a finite cutoff $\Lambda$. To do this, it is convenient to reindex the expression \Eq{eq:P1folded} in an un-folded scheme. To be precise, we index the eigenvalues of $h(\mbf{k})$ by $n \to n,\mbf{G}$ with eigenvalue $\lambda_{\mbf{G},n} = \lambda_n(\mbf{k}+\mbf{G})$ where $\mbf{k} \in \text{mBZ}$, $\mbf{G}$ are the reciprocal lattice vectors, and $n = 1, \dots, 2L$. The projectors are labeled by $P_{\mbf{G},n}(\mbf{k}) = P_n(\mbf{k}+\mbf{G})$, and their matrix elements are $[P_n(\mbf{k}+\mbf{G})]_{\mbf{b},\mbf{b}'} = \delta_{\mbf{b},\mbf{G}}\delta_{\mbf{b}',\mbf{G}}P_n(\mbf{k}+\mbf{G})$. Thus we obtain
\bea
\label{eq:perturbexpression}
[P^1_{\text{val.}}(\mbf{k})]_{\mbf{b},\mbf{b}'} &=\sum_{\mbf{G},m\in\text{val}, \mbf{G}', n\in\text{con}} \frac{[P_m(\mbf{k}+\mbf{G}) V_M P_n(\mbf{k}+\mbf{G}')]_{\mbf{b},\mbf{b}'}+[P_n(\mbf{k}+\mbf{G}') V_M P_m(\mbf{k}+\mbf{G})]_{\mbf{b},\mbf{b}'}}{\lambda_m(\mbf{k}+\mbf{G}) - \lambda_n(\mbf{k}+\mbf{G}')} \\
&=\sum_{\mbf{G},m\in\text{val}, \mbf{G}', n\in\text{con}} \frac{\delta_{\mbf{b},\mbf{G}} P_m(\mbf{k}+\mbf{G}) [V_M]_{\mbf{b},\mbf{b}'} P_n(\mbf{k}+\mbf{G}') \delta_{\mbf{b}',\mbf{G}'}  +\delta_{\mbf{b},\mbf{G}'} P_n(\mbf{k}+\mbf{G}') [V_M]_{\mbf{b},\mbf{b}'} P_m(\mbf{k}+\mbf{G}) \delta_{\mbf{b}',\mbf{G}}}{\lambda_m(\mbf{k}+\mbf{G}) - \lambda_n(\mbf{k}+\mbf{G}')} \\
&=\sum_{m\in\text{val}, n\in\text{con}} \frac{P_m(\mbf{k}+\mbf{b}) [V_M]_{\mbf{b},\mbf{b}'} P_n(\mbf{k}+\mbf{b}') }{\lambda_m(\mbf{k}+\mbf{b}) - \lambda_n(\mbf{k}+\mbf{b}')} + \frac{P_n(\mbf{k}+\mbf{b}) [V_M]_{\mbf{b},\mbf{b}'} P_m(\mbf{k}+\mbf{b}') }{\lambda_m(\mbf{k}+\mbf{b}') - \lambda_n(\mbf{k}+\mbf{b})}\\
\eea
 from which we see that at first order, $[P_{\text{val.}}^1(\mbf{k})]_{\mbf{G},\mbf{G}'}$ is only nonzero for $\mbf{G} - \mbf{G}' = \pm \mbf{g}_j$, the six reciprocal lattice vectors in the first shell, since $[V_M]_{\mbf{G},\mbf{G}'}$ only has nearest-neighbor terms in momentum space (first harmonic). 
\subsection{Treatment of the Fock term}
\label{app:fockcommentary}

We first consider the Fock term (Eq.~\ref{eqapp:HFock}), which is nontrivial even in the moir\'e-less limit. We write it in general as an integral
\bea
[h_{\text{val.},F}(\mbf{k})]_{\mbf{G} l \sigma,\mbf{G}' l' \sigma'} &= -\int \frac{d^2q}{(2\pi)^2} V_{ll'}(\mbf{q})[P^{\eta s}_{\text{val.}}(\mbf{k}+\mbf{q}) - \frac{1}{2} \mathbb{1}]_{\mbf{G} l \sigma, \mbf{G}' l' \sigma'} \ .
\eea
At zeroth order in perturbation theory (where no moir\'e effects are included), this integral diverges since the continuum Dirac model has unbounded dispersion. The Fock integral divergence is known from  monolayer graphene (see Refs. \cite{Borghi2009FermiVelocityEnhancement,PhysRevLett.118.266801,2025arXiv250812825S} for a Hartree-Fock-focused review), where the logarithmic divergence requires a cutoff on the scale of $1/a$, the inverse graphene lattice spacing. This is responsible for the renormalization of the Fermi velocity in monolayer graphene. To review this example briefly, we take $h_{0,\text{bare}} =v_0 \mbf{k} \cdot \pmb{\sigma}$ and recall the integral \cite{Borghi2009FermiVelocityEnhancement}
\bea
\int^\Lambda \frac{d^2q}{(2\pi)^2} V_{00}(\mbf{q}) \delta P_{\text{val.}}(\mbf{k}+\mbf{q}) = v(\mbf{k}) \mbf{k} \cdot \pmb{\sigma}, \qquad v(\mbf{k}) = v_0 \frac{e^2}{16 \pi \eps_\parallel} \log \frac{4\Lambda/\sqrt{e}}{|\mbf{k}|} + O(\frac{|\mbf{k}|}{\Lambda})
\eea
where $\delta P_{\text{val.}}(\mbf{k}) = P_{\text{val.}}(\mbf{k}) - \frac{1}{2} \mathbb{1}$. For energies below the cutoff, even though $v(\mbf{k})$ is not analytic at $\mbf{k}=0$, it can be approximated as a constant but renormalized velocity to good accuracy, as confirmed by quantum twisting microscopy at small bias voltages \cite{2026NanoL..26.4046L}. (Note that non-linearities due to the Fock integral are observed  experimentally at larger voltages in Ref. \cite{2026NanoL..26.4046L}.)  One way to choose $v_F$ is to minimize the least-squares deviation $(v(\mbf{k})|\mbf{k}| - v_F |\mbf{k}|)^2$ over $k \in (0,\Lambda')$ where $\Lambda'$ is the relevant low-energy scale. The result is $v_F/v_0 -1 = \frac{e^2}{16 \pi \eps_{\parallel}}(\frac{1}{4}+\log \frac{4\Lambda/\sqrt{e}}{\Lambda'})$. This gives $v(\mbf{k}) \sim v_F = 1.2 v_0$ using $v_0 \Lambda' = 150$meV and $\eps_\parallel = 8.2$ as the effective dielectric constant in the hBN environment. In other words, the approximation
\bea
\label{eq:approxmono}
h_{0,\text{bare}}(\mbf{k}) + h_{\text{val.},F}(\mbf{k}) \approx h_{0,\text{eff.}}(\mbf{k})
\eea
is known to be reliable over a low energy range in monolayer graphene (as benchmarked in recent QTM experiments \cite{2026NanoL..26.4046L}), where the bare velocity is $542$meV nm in \Eq{eq:parameters_bare} and the effective velocity is $660$meV nm in \Eq{eq:parameters_exp}. 

At large $\mbf{k}$ in rhombohedral graphene where $v k \gg t_1$, the same divergence occurs since the Hamiltonian is approximately given by decoupled monolayer blocks. Thus we must prescribe a method for treating this divergence. Ideally, we would return to the full tight-binding description and compute the Fock integral in the atomic Brillouin zone. In monolayer graphene, detailed calculations~\cite{PhysRevLett.118.266801} have shown that this reproduces the continuum result with a fixed value of the cutoff $v_0 \Lambda \sim 3\,$eV or equivalently $\Lambda = 6.0$nm${}^{-1}$. It is beyond the scope of this work to perform similar calculations in rhombohedral graphene. Instead, we will make use of estimates of the effective parameters extracted from experiments. 

\begin{figure*}
\centering
\includegraphics[width=0.95\linewidth]{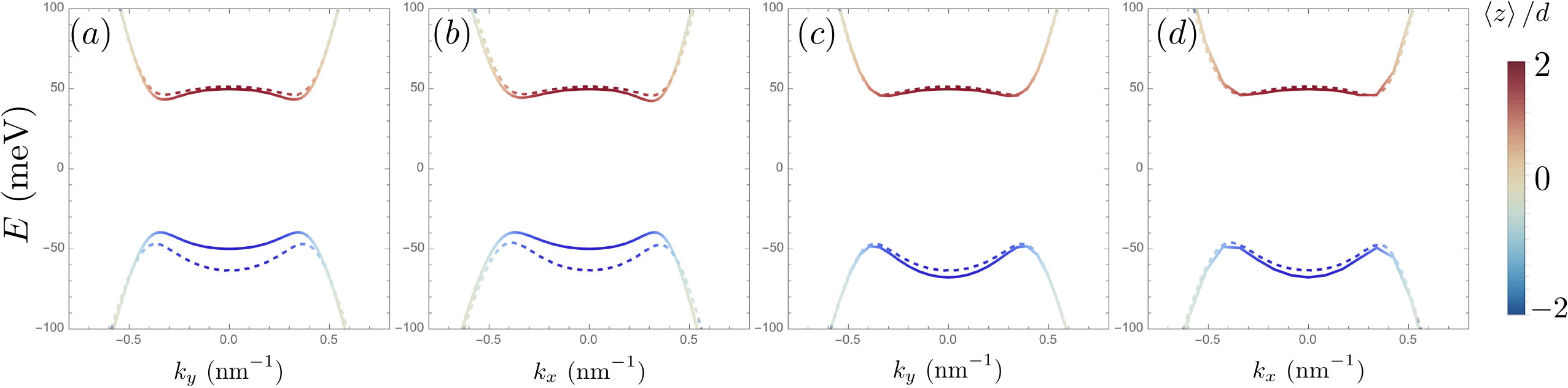}
\caption{Comparison of effective parameters and interacting HF calculations at charge neutrality. In $(a)$ and $(b)$, we show the single-particle band structure using the effective parameters (\Eq{eq:parameters_exp}) as solid lines with $V=25\,$meV. In $(c)$ and $(d)$, we show the fully self-consistent HF bands (computed using the bare single-particle parameters and 3D layer-dependent interactions) computed at neutrality as solid lines. These calculations use $\Lambda = 6.0\text{nm}^{-1}$, $\eps_\perp = 3$, and $\eps_\parallel = 8$ as inferred from recent QTM experiments~\cite{2026NanoL..26.4046L}. To match the gap at $\mbf{k}=0$, we use $D=0.72\text{V nm}^{-1}$. The dashed bands in $(a)-(d)$ correspond to bare non-interacting parameters (\Eq{eq:parameters_bare}) with the zeroth order Fock term in \Eq{eq:backgroundfock} but no Hartree term. We use $V=12$meV to match the gap at $\mbf{k}=0$. Thus, up to renormalization of the values of $V$, we see that the different methods of computing the conduction band are in agreement with each other.} 
\label{fig_app_schf}
\end{figure*}

We show that imposing a finite cutoff on the Fock term can reproduce the effective models. We need the expression
\bea
\label{eq:backgroundfock}
[h^0_{\text{val.},F}(\mbf{k})]_{\mbf{G} l \sigma,\mbf{G}' l' \sigma'} &= \delta_{\mbf{G},\mbf{G}'}\int \frac{d^2q}{(2\pi)^2} V_{ll'}(\mbf{q})[P^{\eta s}_{\text{val.}}(\mbf{k}+\mbf{G}+\mbf{q}) - \frac{1}{2} \mathbb{1}]_{l \sigma,l' \sigma'} \\
&= \delta_{\mbf{G},\mbf{G}'}\int \frac{d^2q}{(2\pi)^2} V_{ll'}(\mbf{k}+\mbf{G}-\mbf{q})[P^{\eta s}_{\text{val.}}(\mbf{q}) - \frac{1}{2} \mathbb{1}]_{l \sigma,l' \sigma'} \ . \\
\eea
This term renormalizes the bare dispersion only, and does not create any moir\'e Umklapp effects at $0$th order. Since it is not the focus of this paper to rigorously determine the bare parameters, we will simply approximate $h_0 + h^0_{\text{val.},F} = h_{\text{eff}}$, where $h_{\text{eff}}$ is the non-interaction Hamiltonian using the parameters in \Eq{eq:parameters_exp} which are taken from experimental fitting. To test this approximation, we perform self-consistent HF calculations on $h_0$ with the bare parameters in \Eq{eq:parameters_bare} and a large momentum cutoff $\Lambda = 6.0\text{nm}^{-1}$ and $\eps_\parallel = 8.2$ \cite{2026NanoL..26.4046L} to compute the Fock integral $h^0_{\text{val.},F}$ via \Eq{eq:backgroundfock}. We then compare the band structure obtained in this way to the effective single-particle model $h_{\text{eff}}$. We show in \Fig{fig_app_schf} that, up to a renormalized value of the displacement field, the dispersions are in reasonable agreement with each other, validating this approximation.  

Next we show that the first order moir\'e Fock correction is small and can be neglected. The expression is
\bea
[h^1_{\text{val.},F}(\mbf{k})]_{\mbf{G} l \sigma,\mbf{G}' l' \sigma'} &= \int \frac{d^2q}{(2\pi)^2} V_{ll'}(\mbf{q}-\mbf{k})[P^1_{\text{val.}}(\mbf{q}) ]_{\mbf{G} l \sigma, \mbf{G}' l' \sigma'} \ .
\eea
The maximum eigenvalue of $h^1_{\text{val.},F}(\mbf{k})$ when projected to the top two layers is less than $0.1\,$meV. In contrast, we will find that the moir\'e Hartree term has an effect of order 10\,meV on the conduction bands. One can understand this difference from \Eq{eq:HFterms}, where we see that the Hartree term contains forms factors where the conduction and valences bands are contracted between themselves, whereas the Fock term is contracted between each other. At zeroth order in $V_M$, the $l=l
'$ part of the Fock term is dominated by the large $\mbf{q}$ part of the integral, where the valence wavefunctions are not layer polarized, and hence this term has a significant effect on the valence bands. In contrast, at first order in $V_M$, $P^1_{\text{val.}}(\mbf{q})$ is mostly supported on the lowest layer $l=0$ (due to $V_M$ being nonzero only on the bottom layer, and the top and bottom layers \emph{not} being strongly hybridized in $P^0$), so the integrand of the Fock term is suppressed by the exponential decay of $V_{0,L-1}(\mbf{q})$. 

To summarize, the dominant effect of the Fock background is to renormalize the effective dispersion of the conduction band. This effect derives completely from the moir\'e-less limit, and hence we use the experimentally fitted hoppings to capture this effect and simplify the modeling. We showed that the moir\'e part of the Fock contribution at first order is small, and henceforth we completely neglect such terms. We will now show that the moir\'e part of the Hartree contribution is strong and cannot be neglected. 

\subsection{Hartree Contribution: Analytical form}
\label{app:hartreecommentary}

We now derive an analytical expression for the background Hartree term using perturbation theory in a weak bare moir\'e potential acting on the bottom layer. We will use this expression to show that the Hartree term is convergent and gives an unambiguous expression for an electrostatic moir\'e potential on the top layer. 

From \Eq{eq:perturbexpression} for the first order moir\'e correction to the density matrix of the occupied valence bands, and \Eq{eq:backgrounddensity} for the expression of the valence charge density for the lowest harmonics $\mbf{G} = \pm \mbf{g}_j$, we obtain (accounting for the spin-valley degeneracy of the valence bands)
\bea
\label{eq:MCEappintegral}
\rho^{\text{val.}}_{-\mbf{G},l} &= \frac{4}{\Omega_{tot}} \sum_{\mbf{k},\mbf{G}'\sigma}\sum_{\substack{m\in\text{val} \\ n\in\text{con}}}  \left[ \frac{P_{m}(\mbf{k}+\mbf{G}') [V_M]_{\mbf{G}',\mbf{G}'-\mbf{G}} P_n(\mbf{k}+\mbf{G}'-\mbf{G})}{\lambda_{m}(\mbf{k}+\mbf{G}') - \lambda_{n}(\mbf{k}+\mbf{G}'-\mbf{G})} + \frac{P_{n}(\mbf{k}+\mbf{G}') [V_M]_{\mbf{G}',\mbf{G}'-\mbf{G}} P_m(\mbf{k}+\mbf{G}'-\mbf{G})}{\lambda_{m}(\mbf{k}+\mbf{G}'-\mbf{G}) - \lambda_{n}(\mbf{k}+\mbf{G}')} \right]_{l \sigma,l\sigma}  \\
&= \sum_{\substack{m\in\text{val} \\ n\in\text{con}}}  \frac{4}{\Omega_{tot}} \sum_{\mbf{q}\in\mathbb{R}^2,\sigma} \left[ \frac{P_{m}(\mbf{q}) [V_M]_{\mbf{G}} P_n(\mbf{q}-\mbf{G})}{\lambda_{m}(\mbf{q}) - \lambda_{n}(\mbf{q}-\mbf{G})} + \frac{P_{n}(\mbf{q}) [V_M]_{\mbf{G}} P_m(\mbf{q}-\mbf{G})}{\lambda_{m}(\mbf{q}-\mbf{G}) - \lambda_{n}(\mbf{q})} \right]_{l \sigma,l\sigma}  \\
 &= \sum_{\substack{m\in\text{val} \\ n\in\text{con}}}  \frac{4}{\Omega_{tot}} \sum_{\mbf{q}\in\mathbb{R}^2} \frac{\Tr [P_{m}(\mbf{q})]_{l,0} [V_M]_{\mbf{G}} [P_n(\mbf{q}-\mbf{G})]_{0,l}}{\lambda_{m}(\mbf{q}) - \lambda_{n}(\mbf{q}-\mbf{G})} + \frac{\Tr [P_{n}(\mbf{q})]_{l,0} [V_M]_{\mbf{G}} [P_m(\mbf{q}-\mbf{G})]_{0,l}}{\lambda_{m}(\mbf{q}-\mbf{G}) - \lambda_{n}(\mbf{q})}  \\
  &= 4 \sum_{\substack{m\in\text{val} \\ n\in\text{con}}} \int \frac{d^2q}{(2\pi)^2} \frac{\Tr [P_{m}(\mbf{q})]_{l,0} [V_M]_{\mbf{G}} [P_n(\mbf{q}-\mbf{G})]_{0,l}}{\lambda_{m}(\mbf{q}) - \lambda_{n}(\mbf{q}-\mbf{G})} + \frac{\Tr [P_{n}(\mbf{q})]_{l,0} [V_M]_{\mbf{G}} [P_m(\mbf{q}-\mbf{G})]_{0,l}}{\lambda_{m}(\mbf{q}-\mbf{G}) - \lambda_{n}(\mbf{q})}  \\
\eea
where we used that $V_M$ is nonzero only on the bottom layer, and the trace is over the sublattice indices. We have reduced our expression to integrals over the energies and eigenvectors of pristine rhombohedral graphene. We now show that these integrals converge as the cutoff $\Lambda$ is taken to infinity (unlike the Fock term), demonstrating that the Hartree term is unambiguously defined. 

At large $\mbf{q}$, the valence bands have energies $\la_m(\mbf{q}) \sim - v_F |\mbf{q}|$ whereas the conduction bands behave as $\la_n(\mbf{q}) \sim v_F |\mbf{q}|$. Thus the energy denominator is $O(\mbf{q})$. For the integral to converge as $\Lambda\rightarrow\infty$, we require the numerator to be $O(1/q^2)$, so it suffices to show that the $O(1/q)$ term vanishes. To do this, first note that at large $\mbf{q}$, the wavefunctions are those of monolayer-graphene, weakly hybridized by the interlayer coupling at $O(t_\perp^2/(v_F |\mbf{q})$. Next, the numerator contains the overlap $\sum_\sigma U^\dag_{m,l \sigma}(\mbf{q}+\mbf{G}) U_{l\sigma, n}(\mbf{q})$  in which only the intra-layer component appears. This means we can evaluate $\sum_\sigma U^\dag_{m,l \sigma}(\mbf{q}+\mbf{G}) U_{l\sigma, n}(\mbf{q})$ using monolayer graphene wavefunctions to evaluate the $O(1/q)$ term. We find
\bea
\label{eq:converge}
U^{\text{mono.}\dag}_m(\mbf{q}+\mbf{G}) U^{\text{mono}}_n(\mbf{q}) &= 0 + \frac{i}{2} \frac{\mbf{G} \times \mbf{q}}{|\mbf{q}|^2} + O(1/q^2) .
\eea
The $O(1/q)$ term vanishes upon integration over $\mathbf{q}$ since it is odd. This completes our proof. 

\begin{figure*}
\centering
\includegraphics[width=0.95\linewidth]{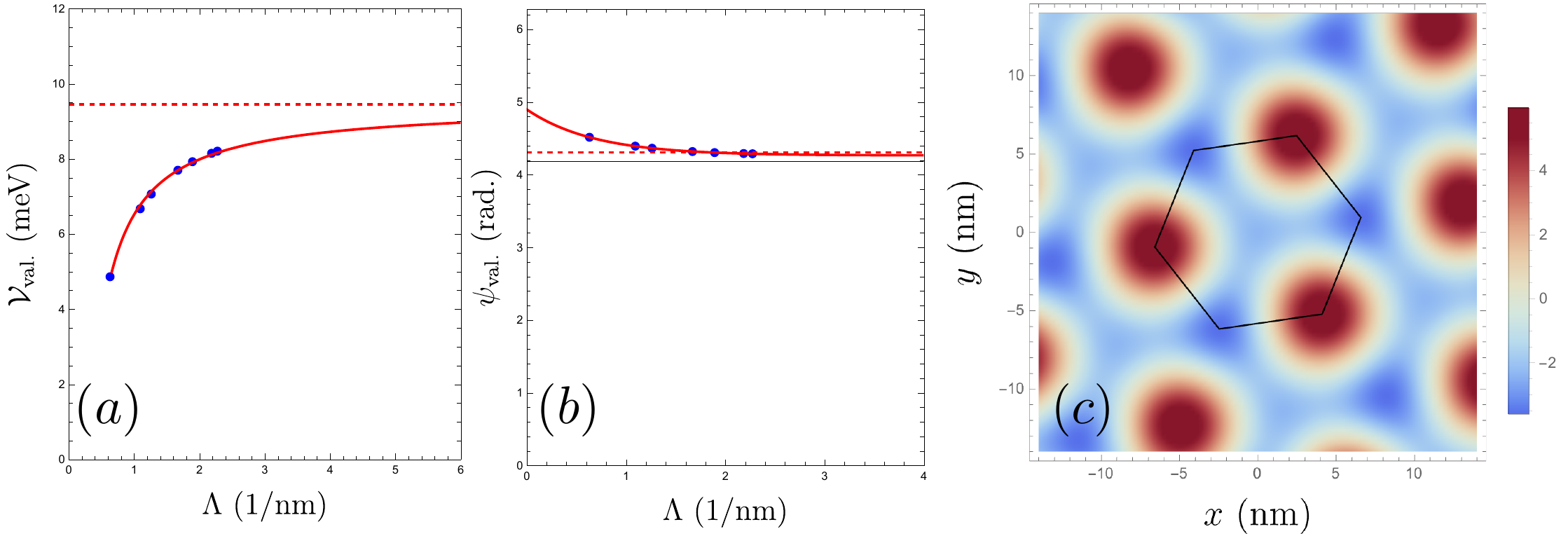}
\caption{Convergence of $\rho^{\text{val}.}_{\mbf{G},l=0}$. In $(a)$ and $(b)$ we plot (blue dots) the amplitude and phase of $\mathcal{V}_{\text{val.}} e^{i  \psi_{\text{val.}}} $ in \Eq{eq:mathcalV} as a function of the cutoff $\Lambda$. We perform a fit to $c_1 + c_2/\Lambda$ in $(a)$, shown in red, and observe clear $1/\Lambda$ convergence as explained in \Eq{eq:converge}. The dashed red line is its asymptotic value which is close to $10\,$meV. In $(b)$, we fit the phase to an exponential function $c_1 + c_2 e^{-c_3 \Lambda}$ (red line), which results in $\psi_{\text{val.}} =4.268$ which we round to $4.3$ (dashed line).  This value is close to the $C_6$-symmetric limit $4\pi/3 = 4.18$ (black line). Slight deviations from $C_6$ symmetry are visible in $(c)$ where we plot $\mathcal{V}_{\text{val.}}(\mbf{r})/\mathcal{V}_{\text{val.}}$ (black hexagon denotes a moir\'e unit cell). 
In this plot, take $V=30\,$meV, $V_1 = 21\,$meV, $\theta = 0.77^\circ$ and $\eps_\perp = 3$. 
}  
\label{fig:rhoGintegralnumerics}
\end{figure*}

One can check explicitly in Bernal graphene that there is a non-vanishing $|\mbf{G}|/|\mbf{q}|^2$ term in the numerator, and thus the Hartree integral (\Eq{eq:MCEappintegral}) in general will converge like $\int^\Lambda q dq\, \frac{1/q^2}{q} = 1/\Lambda$. This is confirmed numerically in \Fig{fig:rhoGintegralnumerics}. \Fig{fig:rhoGintegralnumerics}a shows the $1/\Lambda$ convergence of the amplitude whereas \Fig{fig:rhoGintegralnumerics}b shows exponential convergence of the phase.  For convenience, the real-space plot of the density is shown in \Fig{fig:rhoGintegralnumerics}c. 

\begin{figure*}
\centering
\includegraphics[width=0.95\linewidth]{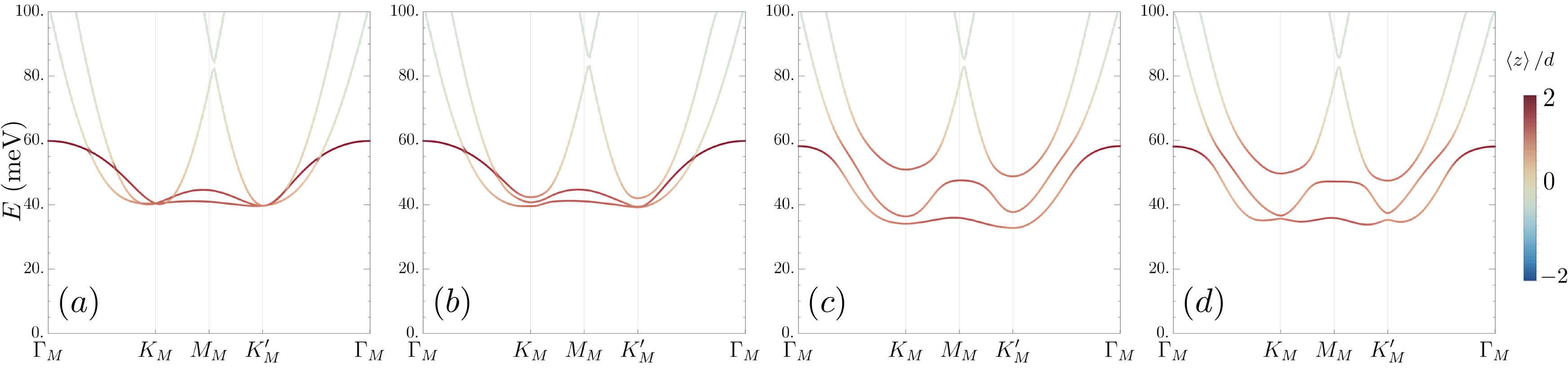}
\caption{Dispersion of the effective single-particle conduction bands at $V=30\,$meV. Color coding denotes layer polarization with red/blue being fully polarized to the top/bottom layer.  We show $(a)$ the moir\'e-less band structure, $(b)$ the band structure with the bare hybridization on the bottom layer, and $(c)$ the band structure with the layer-uniform induced potential in \Eq{eq:mathcalV} with $\mathcal{V}_{\text{val.}} = 12\,$meV, $\psi_{\text{val.}} = \frac{4\pi}{3}+0.12$. In $(d)$, we show the bands computed using the $l$-dependent $\mathcal{V}_{\text{val}.,l} = \frac{e^2}{2 \eps_\perp |\mbf{g}_1|} e^{- d |l-1| |\mbf{g}_1|} \rho^{\text{val.}}_{-\mbf{g}_1,0}$ (see \Eq{eq:hartreelayerdep}).}  
\label{fig:moirebands}
\end{figure*}

\begin{figure*}
\centering
\includegraphics[width=0.9\linewidth]{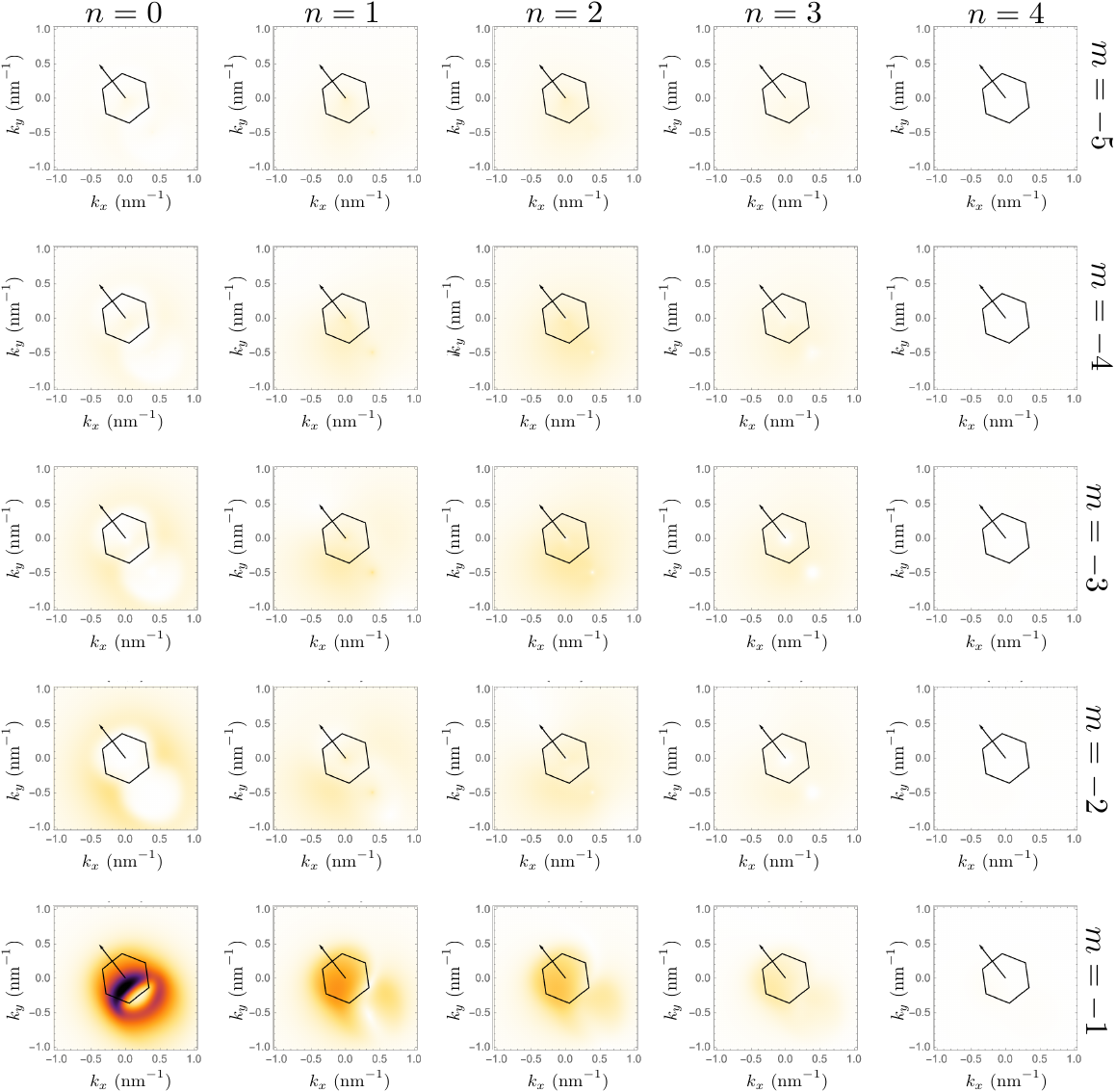}
\caption{Heat-map of the contributions to $\rho_{\mbf{G},0}$ plotted as a function of $\mbf{k}$ for each inter-band transition from a valence band $m=-5,\dots,-1$ to a conduction band $n=0,\dots,4$ at $V=25\,$meV (darkest color is normalized to 1). The dominant contribution comes from the low energy transition $m =-1$ to $n=0$, i.e.~the surface-polarized bands nearest charge neutrality. The mBZ is marked by a hexagon and $\mbf{G}$ is indicated with an arrow. We observe that the largest contributions appear where $\mbf{G}$ connects bands the bottom-polarized low energy valence band to the dispersive layer-hybridized bands, which both have non-negligible weight on the bottom layer.
 }  
\label{fig:rhoGheatmap}
\end{figure*}

Having established that the integral converges, we study the behavior of $\rho^{\text{val.}}_{\mbf{G},l}$ for different $l$  . For a typical value of the displacement field $V = 30$meV, we find that the layer-resolved densities $\rho^{\text{val.}}_{\mbf{G},l}$ are dominated by the bottom layers as expected: $90\%$ of $\sum_l \rho^{\text{val.}}_{\mbf{G},l}$ is contributed by  the bottom layer $l=0$. Hence, for simplicity, we can keep only the $l=0$ (bottom) contribution. This density creates a layer-dependent potential that decays with $l$  according to $\mathcal{V}_{\text{val}., l}(\mbf{r})$ in \Eq{eq:hartreelayerdep}, which is $\sim e^{-|\mbf{G}_M| d l}$ for 3D Coulomb. Since the conduction bands are highly polarized to the top layer, we can approximate $\mathcal{V}_{\text{val}., l}(\mbf{r}) \simeq \mathcal{V}_{\text{val}., L-1}(\mbf{r})$, taking the potential to be uniform with its strength given by the  value $\mathcal{V}_{\text{val}., L-1}(\mbf{r})$ on the top layer. These approximations are benchmarked in \Fig{fig:moirebands}, where we find that they are accurate within $2-4\,$meV compared to the overall $20\,$meV splitting of the conduction bands. 
Therefore, from \Eq{eq:hartreelayerdep}, we arrive at a simple potential that captures the key effect of the hBN:
\bea
\mathcal{V}_{\text{val}.}(\mbf{r}) &= \sum_{\mbf{G} } e^{i \mbf{G} \cdot \mbf{r}} V_{0,L-1}(\mbf{G}) \rho^{\text{val.}}_{-\mbf{G},0}  \ .
\eea
Hermiticity and $C_3$ symmetry enforce $\rho^{\text{val.}}_{-\mbf{G},0} = \rho^{\text{val.}*}_{\mbf{G},0}$ and $\rho^{\text{val.}}_{C_3\mbf{G},0} = \rho^{\text{val.}}_{\mbf{G},0}$. From these constraints, we derive
\bea
\label{eq:mathcalV}
\mathcal{V}_{\text{val.}}(\mbf{r}) &= 2 \mathcal{V}_{\text{val.}}\sum_{j=1}^3 \cos (\mbf{g}_j \cdot \mbf{r} + \psi_{\text{val.}}), \qquad \mathcal{V}_{\text{val.}} e^{i  \psi_{\text{val.}}} = V_{0,L-1}(\mbf{g}_1) \rho^{\text{val.}}_{-\mbf{g}_1,0} = \frac{e^2}{2 \eps_\perp |\mbf{g}_1|} e^{- d |L-1| |\mbf{g}_1|} \rho^{\text{val.}}_{-\mbf{g}_1,0} \\
\eea
where we used the 3D Coulomb interaction $V_{ll'}(\mathbf{q})$ with perpendicular dielectric constant $\eps_\perp$. We observe that the bare moir\'e parameters factor out, giving the final expression
\bea
\label{eq:rhoexplicit}
\rho^{\text{val.}}_{-\mbf{g}_1,0} &= 4 V_{1} e^{i \psi} \sum_{\substack{m\in\text{val} \\ n\in\text{con}}}  \frac{1}{\Omega_{tot}} \sum_{\mbf{q}\in\mathbb{R}^2} \frac{\Tr [P_{m}(\mbf{q})]_{0,0} T_1 [P_n(\mbf{q}-\mbf{g}_1)]_{0,0}}{\lambda_{m}(\mbf{q}) - \lambda_{n}(\mbf{q}-\mbf{g}_1)} - (m \leftrightarrow n), \qquad T_1 = \bpm 1 & \omega^* \\ \omega^* & \omega \epm  \ . \\
\eea
The contributions for each $m\to n$ transition are shown in \Fig{fig:rhoGheatmap}. 

\begin{figure*}
\centering
\includegraphics[width=0.99\linewidth]{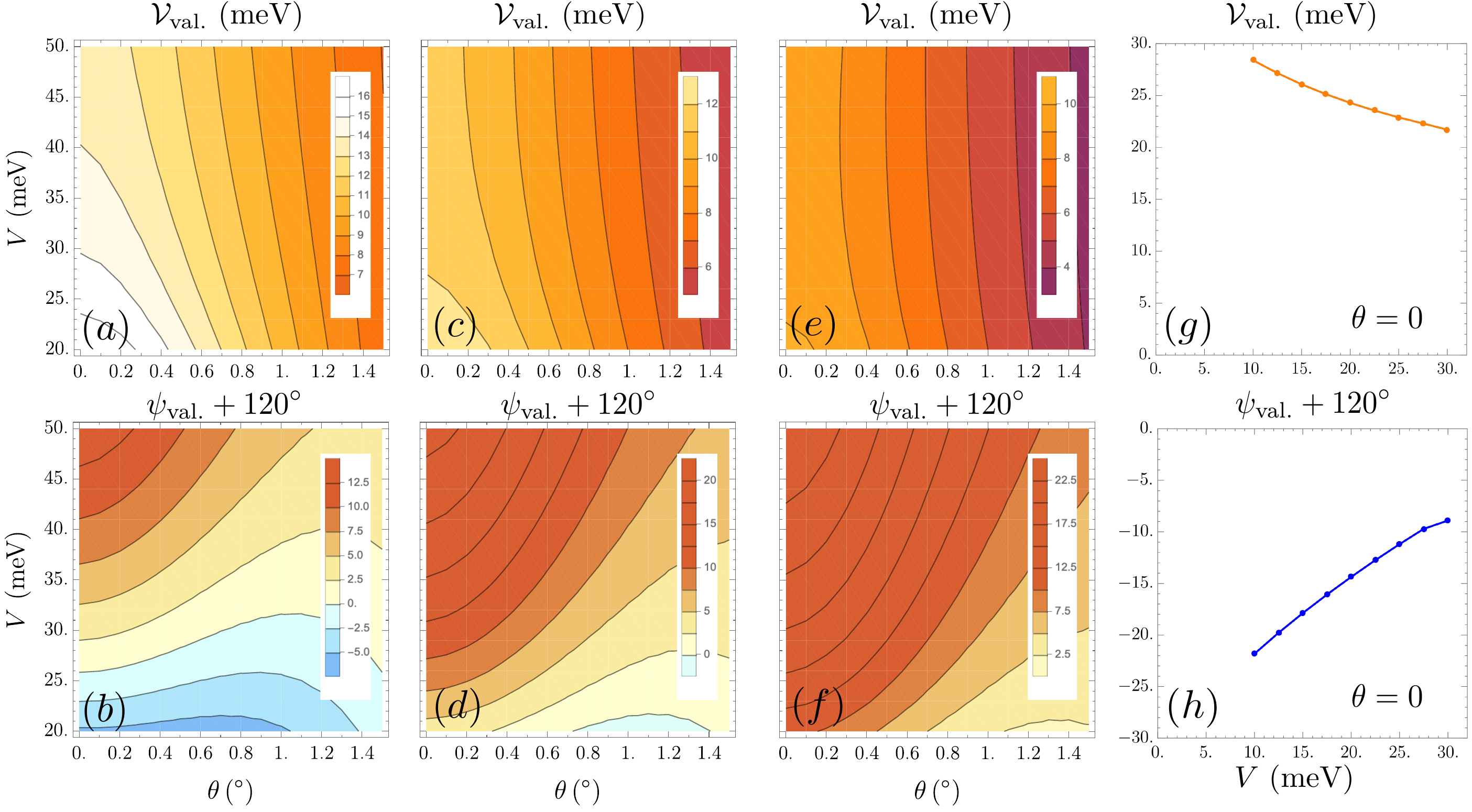}
\caption{Behavior of induced potential $\mathcal{V}_{\text{val.}}$ as a function of twist angle $\th$ and interlayer potential $V$ in R$L$G. $(a),(c),(e)$ show $\mathcal{V}_{\text{val.}}$ in $L=4,5,6$ respectively,
and $(b),(d),(f)$ show $\psi_{\text{val.}}+120^\circ$ in $L=4,5,6$ respectively. In $(g),(h)$, we show line cuts for $L=3$ at a fixed $\theta = 0$ (aligned hBN). Over a wide range of the parameter space, we see that $\mathcal{V}_{\text{val.}}\simeq 10$meV and $\psi_{\text{val.}}\simeq -2\pi/3$. We see that $\mathcal{V}_{\text{val.}}$ is decreasing with $\th$ and $V$.  }  
\label{fig:MCE1}
\end{figure*}

As an order of magnitude estimate for $\mathcal{V}_{\text{val.}}$ in $\theta=0.77^\circ$ R5G/hBN, we approximate the energy gap denominator of Eq.~\ref{eq:rhoexplicit} by the displacement-field-induced gap and drop the wavefunction overlaps in the numerator, giving 
\bea
\mathcal{V}_{\text{val}.} &= 4 V_{1}  \frac{e^2}{2 \eps_\perp |\mbf{g}_1|} e^{- d |L-1| |\mbf{g}_1|} \left| \sum_{\substack{m\in\text{val} \\ n\in\text{con}}}  \frac{1}{\Omega_{tot}} \sum_{\mbf{q}\in\mathbb{R}^2} \frac{\Tr [P_{m}(\mbf{q})]_{0,0} T_1 [P_n(\mbf{q}-\mbf{g}_1)]_{0,0}}{\lambda_{m}(\mbf{q}) - \lambda_{n}(\mbf{q}-\mbf{g}_1)} - (m \leftrightarrow n) \right| \\
&\approx 4 V_{1} \frac{e^2}{2 \eps_\perp |\mbf{g}_1|} e^{- d |L-1| |\mbf{g}_1|}   \frac{1}{\Omega_{tot}} \sum_{\mbf{q}\in\text{mBZ}} \frac{1}{V(L-1)} \\
&= \frac{e^2}{2 \eps_\perp |\mbf{g}_1|} e^{- d |L-1| |\mbf{g}_1|}  \frac{4 V_1}{(L-1)V} \frac{\Omega_{mBZ}}{(2\pi)^2} = 12\,\text{meV}
\eea
where $\Omega_{mBZ}$ is the area of the moir\'e Brillouin zone, and we used $V=30\,$meV and $\eps_{\perp} = 3$. In \Fig{fig:rhoGintegralnumerics}, we show the numerical results and find $\mathcal{V}_{\text{val.}} = 9.5\,$meV for the same parameters, demonstrating that the order of magnitude estimate is accurate. The dependence on $V$ is weak within the range of interest (for $V=20\,$meV, we find $\mathcal{V}_{\text{val.}} = 10.2\,$meV) and will be neglected throughout this work.  Numerically, we find that $\psi_{\text{val.}} \simeq \frac{4\pi}{3}$ over $V \in (20,30)$meV for $\psi = 16.55^\circ$ (see \Fig{fig:MCE1}). A similar trend is observed for $L=4,6$. We also observe that $\mathcal{V}_{\text{val.}}$ is broadly decreasing with $L$. 

\subsubsection{Other Layer Numbers}
\label{app:othersystems}

We now discuss other materials where experiments have performed spatially-resolved measurements: monolayer graphene/hBN~\cite{2026Natur.650..875K} and R3G/hBN~\cite{2025arXiv251009548S}. In \Fig{fig:MCE1}(g,h), we show the evolution of $\mathcal{V}_{\text{val}.}e^{i \psi_{\text{val}.}}$ computed numerically in R3G/hBN with $\theta = 0$. We again see that $\psi_{\text{val}.} \sim 4\pi/3$. We can gain some understanding for this value from the simplest possible example: monolayer graphene. In this simple case, we can compute the integral in \Eq{eq:MCEappintegral} analytically. Note that we previously argued that the neutrality gap $(L-1)V$ justified perturbation theory in multilayer rhombohedral graphene aligned with hBN, while in monolayer there is no gap. In the latter case, perturbation theory is justified when the first order correction is controllably small. We will show post-hoc that this corresponds to small $V_1 |\mbf{G}|/v_F$, and hence is justifiable in the weakly interacting limit $v_F \to \infty$, as can be expected. 

Since there is only a single occupied band and one layer, \Eq{eq:MCEappintegral} reduces to (all integrals are over $\mathbb{R}^2$)
\bea
\rho^{\text{val.}}_{-\mbf{G}} 
 &= \frac{4}{\Omega_{tot}} \sum_{\mbf{q}\in\mathbb{R}^2} \frac{\Tr P(\mbf{q}) [V_M]_{\mbf{G}} Q(\mbf{q}-\mbf{G})}{-v_F|\mbf{q}| - v_F|\mbf{q}-\mbf{G}|} + \frac{\Tr Q(\mbf{q}) [V_M]_{\mbf{G}} P(\mbf{q}-\mbf{G})}{-v_F |\mbf{q}-\mbf{G}| - v_F |\mbf{q}|}  \\
  &= -4 \int \frac{d^2q}{(2\pi)^2} \Tr \frac{ P(\mbf{q}) [V_M]_{\mbf{G}} Q(\mbf{q}-\mbf{G}) + Q(\mbf{q}) [V_M]_{\mbf{G}} P(\mbf{q}-\mbf{G})}{v_F|\mbf{q}| + v_F|\mbf{q}-\mbf{G}|}   \\
\eea 
where $P(\mbf{q}) = \frac{1}{2}(\mathbb{1}_2 - \frac{\mbf{k} \cdot \pmb{\sigma}}{|\mbf{k}|})$ is the occupied band projector and $Q(\mbf{q}) = \mathbb{1}_2 - P(\mbf{q})$. Expanding out the Pauli matrices and performing the trace, we find  
\bea
\rho^{\text{val.}}_{-\mbf{G}}  &= -4 \int \frac{d^2q}{(2\pi)^2} \frac{1}{2} \Tr \frac{[V_M]_\mbf{G} - \hat{\mbf{q}} \cdot \pmb{\sigma} [V_M]_\mbf{G} \frac{\mbf{q}-\mbf{G}}{|\mbf{q}-\mbf{G}|} \cdot \pmb{\sigma} }{v_F|\mbf{q}| + v_F|\mbf{q}-\mbf{G}|} = -\frac{2 \Tr [V_M]_\mbf{G}}{v_F} \int \frac{d^2q}{(2\pi)^2}  \frac{1- \frac{\mbf{q} \cdot (\mbf{q}-\mbf{G})}{|\mbf{q}||\mbf{q}-\mbf{G}|} }{|\mbf{q}| + |\mbf{q}-\mbf{G}|}   \\
\eea
where we used that only the identity component of $[V_M]_\mbf{G}$ leads to an even-$\mbf{q}$ term in the integral. From \Eq{eq:Vxifinal}, we now see that $-\Tr [V_M]_\mbf{G} = - V_1 e^{i \psi} (1+\omega) = V_1 e^{i \psi} \omega^*$. Continuing, we see that the integral is rotationally symmetric so we can choose $\mbf{G} = (G,0) = G \hat{x}$ and rescale variables $\mbf{q} \to \mbf{x} G$
\bea
\rho^{\text{val.}}_{-\mbf{G}} &= \frac{2 V_1 e^{i \psi} \omega^*}{v_F} \int \frac{d^2q}{(2\pi)^2}  \frac{1- \frac{\mbf{q} \cdot (\mbf{q}-\mbf{G})}{|\mbf{q}||\mbf{q}-\mbf{G}|} }{|\mbf{q}| + |\mbf{q}-\mbf{G}|} 
= \frac{2 V_1 |\mbf{G}| e^{i \psi} \omega^*}{v_F} \int \frac{d^2x}{(2\pi)^2}  \frac{1- \frac{\mbf{x} \cdot (\mbf{x}-\hat{x})}{ |\mbf{x}||\mbf{x}-\hat{x}|} }{|\mbf{x}| + |\mbf{x}-\hat{x}|} = \frac{2 V_1 |\mbf{G}| e^{i \psi} \omega^*}{v_F} \frac{1}{16} = \frac{V_1 |\mbf{G}| e^{i \psi} \omega^*}{8 v_F}
\eea
where the final dimensionless integral may be found in App. A of Ref. \cite{2024PhRvB.109g5120G} and is evaluated in terms of elliptic coordinates (see also Refs. \cite{PhysRevB.78.035119,PhysRevB.89.235431}).  Our key observation is that the $\omega^*$ appearing from the graphene integrals gives an intrinsic phase of $4\pi/3$ that adds to the phase $\psi$ coming from the hBN potential. 

\subsubsection{Connection to the Positive-Semi-definite Form}
\label{app:PSDform}

We briefly relate our derivation of the projected Hamiltonian to the positive semi-definite family of Hamiltonians including twisted bilayer graphene \cite{TBG3}. Define the projected density with respect to charge neutrality,
\bea
\bar{\rho}_{\mbf{q}+\mbf{G},l} &= \sum_{\substack{\mbf{k}\eta s\\mn\in\mathrm{act.}}}
M^{l,\eta s}_{mn}(\mbf{k},\mbf{q}+\mbf{G}) (
\gamma^\dagger_{\mbf{k}+\mbf{q},m,\eta s}
\gamma_{\mbf{k},n,\eta s}
-
\frac{1}{2}\delta_{\mbf{q},0} \delta_{mn}) \ .
\eea
The positive-semi-definite interaction is
\bea
\overline{H}_{\mathrm{PSD}} &=
\frac{1}{2\Omega_{\mathrm{tot}}}
\sum_{\mbf{q}\mbf{G}ll'}
V_{ll'}(\mbf{q}+\mbf{G}) \bar{\rho}^\dag_{\mbf{q}+\mbf{G},l} \bar{\rho}_{\mbf{q}+\mbf{G},l'}
 \eea
Expanding the density subtraction and normal ordering the product of densities gives
\bea
\overline{H}_{\mathrm{PSD}} &=:\!\overline{H}_{\mathrm{int}}\!: + \overline{H}'_H + \overline{H}'_F
+ \text{const.}.
\eea
where the one-body terms are
\bea
\overline{H}'_H &=\sum_{\substack{\mbf{k}\eta s\\mn\in\mathrm{act.}}}\gamma^\dagger_{\mbf{k},m,\eta s}\gamma_{\mbf{k},n,\eta s}\sum_{\mbf{G},ll'}
V_{ll'}(\mbf{G})M^{l,\eta s}_{mn}(\mbf{k},\mbf{G})
\lp
\frac{1}{\Omega_{\mathrm{tot}}}
\sum_{\mbf{k}'\eta's'm'}
M^{l',\eta's'}_{m'm'}(\mbf{k}',-\mbf{G}) (-\frac{1}{2} \delta_{m' \in \text{act.}})\rp \\
\overline{H}'_F &=
-\sum_{\substack{\mbf{k}\eta s\\mn\in\mathrm{act.}}}
\gamma^\dagger_{\mbf{k},m,\eta s}
\gamma_{\mbf{k},n,\eta s}
\frac{1}{\Omega_{\mathrm{tot}}} \sum_{\mbf{q}\mbf{G}ll' m'}
V_{ll'}(\mbf{q}+\mbf{G}) M^{l',\eta s}_{m'm}(\mbf{k},\mbf{q}+\mbf{G})^* M^{l,\eta s}_{m'n}
(\mbf{k},\mbf{q}+\mbf{G}) (- \frac{1}{2} \delta_{m' \in \text{act.}}) \ .
\eea
We will now compare these expressions to \Eq{eq:HFterms}. We first show that the Hartree terms are identical in the case of particle-hole symmetry. We consider the general circumstance where the valence manifold contains bands $< -N$, the active manifold is $(-N,N)$, and bands $\geq N$ are empty. Then we have
\bea
\bar{H}_H - \bar{H}_H' &= \sum_{\mbf{k}\eta s, mn \in \text{act.}} \gamma^\dag_{\mbf{k},m, \eta s} \gamma_{\mbf{k}, n, \eta s} \sum_{\mbf{G}, ll'}  V_{ll'}(\mbf{G}) M^{l,\eta s}_{mn}(\mbf{k},\mbf{G}) \lp \frac{1}{\Omega_{tot}} \sum_{\mbf{k}' m' \eta' s'} M^{l',\eta' s'}_{m'm'}(\mbf{k}',-\mbf{G}) (\delta_{m' \in \text{val.}}  - \frac{1}{2} + \frac{1}{2} \delta_{m'\in \text{act.}})\rp \ . \\
\eea
Next, observe that $(\delta_{m' \in \text{val.}}  - \frac{1}{2} + \frac{1}{2} \delta_{act}) = \pm \frac{1}{2}$ for $m'$ a valence/empty band and  $(\delta_{m' \in \text{val.}}  - \frac{1}{2} + \frac{1}{2} \delta_{act}) = 0$ if $m'$ is an active band. If there is a particle-hole symmetry that ensures $\sum_{\mbf{k}' \eta' s'} M^{l,\eta s}_{m'm'}(\mbf{k}',\mbf{G}) = \sum_{\mbf{k}' \eta' s'} M^{l,\eta s}_{-m',-m'}(\mbf{k}',\mbf{G})$, then $\bar{H}_H - \bar{H}_H' = 0$. 

However, particle-hole does not equate $\overline{H}'_F$ and $\overline{H}_F$ because
\bea
 &\frac{1}{\Omega_{tot}} \sum_{\mbf{q} \mbf{G} m' ll'}  V_{ll'}(\mbf{q}+\mbf{G})M^{l',\eta s}_{m'm}(\mbf{k},\mbf{q}+\mbf{G})^* (\delta_{m' \in \text{val.}}  - \frac{1}{2} + \frac{1}{2} \delta_{m'\in \text{act.}}) M^{l,\eta s}_{m'n}(\mbf{k},\mbf{q}+\mbf{G})  \\
 &=\frac{1}{\Omega_{tot}} \sum_{\mbf{q} \mbf{G}ll'}  V_{ll'}(\mbf{q}+\mbf{G}) \lp \sum_{m'<-N} M^{l',\eta s}_{m'm}(\mbf{k},\mbf{q}+\mbf{G})^*  M^{l,\eta s}_{m'n}(\mbf{k},\mbf{q}+\mbf{G}) - \sum_{m'>N} M^{l',\eta s}_{m'm}(\mbf{k},\mbf{q}+\mbf{G})^*  M^{l,\eta s}_{m'n}(\mbf{k},\mbf{q}+\mbf{G}) \rp  \\
 &=\frac{1}{\Omega_{tot}} \sum_{\mbf{q} \mbf{G}ll'}  V_{ll'}(\mbf{q}+\mbf{G}) \sum_{m'<-N}  \lp M^{l',\eta s}_{m'm}(\mbf{k},\mbf{q}+\mbf{G})^*  M^{l,\eta s}_{m'n}(\mbf{k},\mbf{q}+\mbf{G}) - M^{l',\eta s}_{-m',m}(\mbf{k},\mbf{q}+\mbf{G})^*  M^{l,\eta s}_{-m',n}(\mbf{k},\mbf{q}+\mbf{G}) \rp  \\
\eea
is generically nonzero. This can be seen simply in the case of monolayer graphene, where $\overline{H}_{\mathrm{PSD}}$ does not include the Fock renormalization of the Fermi velocity (which preserves particle-hole symmetry). 

\subsection{Analytical Wavefunctions}
\label{app:exacteigenstates}

To develop analytical understanding, we show that the continuum model for rhombohedral graphene can be solved analytically for any layer number if we only retain the $v_Fk$ and $t_1$ terms. This is through a mapping to the SSH model, which was solved in Ref. \cite{2011PhRvB..84s5452D} (see also \cite{2019CmPhy...2..164S,2009PhRvB..80p5409K,2011JETPL..93...59H,2024PNAS..12110714Z}). We will also solve the system with edge potentials that model the applied displacement field.  We neglect higher hoppings $v_3, v_4, t_2$ for solvability, although they could be incorporated in perturbation theory. For brevity, we drop the subscripts in $v_F,t_1$ in the remainder of this section. The Hamiltonian we study is
\bea
H(k) &= \bpm 
0 & v k & 0 & 0 &\dots \\
v \bar{k} & 0 & t & 0&  \\
0 & t & 0 & v k &  \\
\vdots &  & v \bar{k} & \ddots &  \\
\epm
\eea 
which is $2L\times 2L$ for an $L$ layer system, and we have defined $k=k_x+ik_y$ and $\bar{k}=k_x-ik_y$. It describes an SSH chain with open boundary conditions. Since the outer hopping is $vk$, this SSH chain is in the topological phase for $v|k| < t$, so that we understand the two flat bands as the edge states of the SSH chain, which merge into the bulk as $v|k| \sim t$. This model has $SO(2)$ symmetry generated by
\bea
\label{eq:SO2rotation}
L_z = \text{diag}(0,1,1,2,2,\dots,L-1)
\eea
such that $e^{i \th L_z} H(|k|e^{i \th}) e^{-i \th L_z} = H(|k|)$. Going forward, we assume $k$ is real without loss of generality. 

The eigenvalue equation $H(k) \psi = E \psi$, with $\psi = (A_1,B_1,A_2,B_2,\dots)$, reduces to
\bea
vk A_n + t A_{n+1} &= E B_n, \qquad t B_{n-1} + vk B_n = E A_n \\
vk B_1 &= E A_1, \qquad vk A_{L} = E B_L
\eea
for the `bulk' and `boundaries' respectively.  We now make the reflecting wave ansatz parameterized by $z$
\bea
\label{eq:refwave}
A_n &= A z^n + \tilde{A} z^{-n}, \qquad B_n &= B z^n + \tilde{B} z^{-n} \ .
\eea 
The bulk equation reduces to
\bea
vk (A z^n + \tilde{A} z^{-n}) + t (z A z^n + \frac{1}{z}\tilde{A} z^{-n}) &= E (B z^n + \tilde{B} z^{-n})\\
t (\frac{1}{z} B z^n + z \tilde{B} z^{-n}) + vk (B z^n + \tilde{B} z^{-n}) &= E (A z^n + \tilde{A} z^{-n}) \\
\eea
which, upon gathering powers of $z$, gives
\bea
v k A + t z A &= E B \\
v k \tilde{A} + \frac{t}{z} \tilde{A} &= E \tilde{B} \\
\frac{t}{z} B + v k B &= E A \\
z t \tilde{B} + v k \tilde{B} &= E \tilde{A}.  \\
\eea
These equations are consistent and lead to a solution
\bea
\pm E &= \pm (v k + t z) \sqrt{\frac{ \frac{t}{z} + v k}{v k + t z}}= \pm \sqrt{(\frac{t}{z} + v k) (v k + t z)}  = \pm \sqrt{(vk)^2 + t^2+ v k t (z+z^{-1}) }\\
A &= \frac{E}{v k+ t z} B \\
\tilde{A} &= \frac{E}{v k+ t/z} \tilde{B}. \\
\eea
The boundary equations for layer $0,L-1$ are
\bea
v k (B z + \tilde{B} z^{-1} ) = E (A z + \tilde{A} z^{-1}), \qquad  v k (A z^L + \tilde{A} z^{-L} ) = E (B z^L + \tilde{B} z^{-L}) \ .
\eea
In total we have 5 equations, plus normalization, for the 4 wavefunction amplitudes, $z$, and $E$. Note that each energy comes in $\pm$ pairs due to chiral symmetry. Plugging in the bulk equations into the boundary, we obtain
\bea
\bpm 
\label{eq:LAAbareqn}
z^{L+1} & z^{-L-1} \\
z(v k + t z) & t + v k z
\epm \bpm A \\ \tilde{A} \epm &= 0 .
\eea
For the equation to have a solution, the matrix determinant must be zero, and we thus obtain the quantization equation in rational form
\bea
\label{eq:quantizationeq}
\frac{z \left(1-z^{2 L}\right)}{z^{2 L+2}-1} &= \frac{v k}{t} \ .
\eea
This equation simplifies under the substitution $z = e^{i\kappa}$ to
\bea
\frac{\sin \kappa L}{ \sin \kappa (L+1)} &=- \frac{v k}{t} \ .
\eea
We will use the $\kappa$ variable for the high-energy dispersive bands (the dimer states) and the $z$ variable for the low-energy bands. When $|vk/t| > 1 + 1/L$, there are only $n-1$ real solutions of $\kappa$. The flat part of the R$L$G band is surface surface-localized and occurs when $|z| < 1$. Recall that each value of $\kappa,z$ corresponds to two bands with energy $\pm E$.

 We can obtain the eigenstates (defined by $A,\bar{A}$ in \Eq{eq:LAAbareqn}) by observing that the quantization equation \Eq{eq:quantizationeq}  (for any $z$) obeys
\bea
z^{2L} &= \frac{1}{z^2} \frac{v k + t z}{v k + t/z} \\
z^{L+1} &= \pm \frac{v k + t z}{v k + t/z} \\
\eea
from which we find that the matrix in \Eq{eq:LAAbareqn} has a null space given by
\bea
A &= -(v k + t /z), \quad \tilde{A} = v k + z t \\
\eea
and therefore
\bea
B &= -\frac{(v k + t /z) (v k + t z)}{E} = -E \\
\tilde{B} &= \frac{(v k + t /z) (v k + t z)}{E} = E \ . \\
\eea
To understand the solution for $z$ in the flat band regime, consider the case of $L\to \infty$. The boundary mode  then has the solution
\bea
-z&= \frac{v k}{t}, \qquad |z| < 1 \\
A &= -(v k - t^2/vk), \quad \tilde{A} = 0, \quad B = \tilde{B} = E = 0  \ .\\
\eea
Since only $A$ is nonzero, we obtain the eigenstate $A_n = A (- \frac{vk}{t})^n$. This state, which we refer to as the chiral state, has large overlap ($\approx 95\%$ for $|\mbf{k}| < 0.4\text{nm}^{-1}$ \cite{herzog2024moire}. 

To study the effect of displacement field, we identify a solvable limit where $V$ is modeled as an inversion-breaking potential only on the inner and outer orbitals. The Hamiltonian is
\bea
H(k) &= \bpm 
V & v k & 0 & 0 &\dots \\
v \bar{k} & 0 & t & 0&  \\
0 & t & 0 &  &  \\
\vdots &  & & \ddots & v k \\
& & &  v \bar{k}  & -V \\
\epm \ .
\eea 
It can be thought of as an extreme case of electrostatic screening with $\pm V$ on a single sublattice on either end. It describes an SSH chain with open boundary conditions and an inversion-breaking edge impurity. This model preserves $SO(2)$ symmetry in \Eq{eq:SO2rotation}. 

In this case, the eigenvalue equation $H(k) \psi = E \psi$, with $\psi = (A_1,B_1,A_2,B_2,\dots)$, reduces to
\bea
vk A_n + t A_{n+1} &= E B_n, \quad (n = 1, \dots L-1), \qquad t B_{n-1} + vk B_n = E A_n, \qquad  (n = 2, \dots L)\\
VA_1 + vk B_1 &= E A_1, \qquad vk A_{L} - V B_L = E B_L
\eea
for the bulk and boundaries respectively.  The reflecting wave ansatz from \Eq{eq:refwave}  (but with $z$ for now undetermined) still holds in the bulk since those equations are unaltered, so we have
\bea
\pm E &= \pm \sqrt{(vk)^2 + t^2+ v k t (z+z^{-1}) }\\
A &= \frac{E}{v k+ t z} B, \qquad \tilde{A} = \frac{E}{v k+ t/z} \tilde{B}. \\
\eea
Importantly, chiral symmetry is preserved so each solution for $z$ corresponds to two states with $\pm E$. The boundary equations are
\bea
v k (B z + \tilde{B} z^{-1} ) = (E-V) (A z + \tilde{A} z^{-1}), \qquad  v k (A z^L + \tilde{A} z^{-L} ) = (E+V) (B z^L + \tilde{B} z^{-L}) \ .
\eea
We now plug in the bulk expressions for $E,B,\tilde{B}$ into the boundary equations to obtain an eigenvalue equation for $A,\tilde{A},z$. We obtain, analogously to \Eq{eq:LAAbareqn},
\bea
\left(
\begin{array}{cc}
 z^L \left(\frac{V \sqrt{z} \sqrt{t z+v k}}{\sqrt{t+v k z}}+t z\right) & z^{-L-1} \left(\frac{V \sqrt{z} \sqrt{t+v k z}}{\sqrt{t z+v k}}+t\right) \\
 V z-\frac{t \sqrt{z} \sqrt{t z+v k}}{\sqrt{t+v k z}} & V/z-\frac{t \sqrt{z} \sqrt{t+v k z}}{z\sqrt{t z+v k}} \\
\end{array}
\right) \bpm A \\ \tilde{A} \epm = 0
\eea
which has a solution only if the determinant is zero. After some algebraic manipulations, the determinant equation simplifies to
\bea
\frac{v k}{t} = \frac{z(1-z^{2L})(d^2-1)}{-1 + z^{2(L+1)} + d^2z^2(1-z^{2(L-1)})}, \qquad y = \frac{V}{t} \ .
\eea
This is an algebraic equation for $z$, which implicitly defines $E$. There are two classes of solutions as before, each corresponding to two states with energy $\pm E$. One is the bulk state class where $z = e^{i \kappa}$ has $L-1$ solutions for real $\kappa$, which simplifies to
\bea
- \frac{t}{vk} &= \cos \kappa  + \frac{1+y^2}{1-y^2} \cot \kappa L \sin \kappa \ .
\eea
The other solutions are the surface states where  $z = -e^{-\kappa}$ has solutions for real $\kappa$:
\bea
\frac{t}{vk} &= \cosh \kappa  + \frac{1+y^2}{1-y^2} \coth \kappa L \sinh \kappa. \\
\eea
Note that the right-hand side approaches $1 + \frac{1}{L} \frac{1+y^2}{1-y^2}$ as $\kappa \to 0$, so the surface state is only defined when 
\bea
\frac{t}{vk} \geq 1 + \frac{1}{L} \frac{1+y^2}{1-y^2} \ .
\eea
This gives us a rigorous way to determine when the low-energy bands are surface polarized. Since $y = V/t \ll 1$ for realistic parameters, we can conclude that $\frac{t}{vk} \geq 1 + \frac{1}{L}$ is a good approximation to the surface-polarized regime even when the surface bands are gapped. For $v=660$meV nm and $t=380$meV, we have $k_c = t/v = 0.57\text{nm}^{-1}$ and $k_c/(1+1/L) = 0.48\text{nm}^{-1}$ for $L=5$. These are both larger than the moir\'e vector $|\mbf{q}_1| = 0.36\text{nm}^{-1}$.

\subsection{Effective Hamiltonian and Single-Particle Phase Diagram}
\label{app:phasediagram}

We now summarize the results of our discussion in \App{app:MBhamiltonian}.  We have shown that the valence bands of R$L$G/hBN can be integrated out when the single-particle neutrality gap is larger than the interaction and moir\'e potential strength, yielding two effects: (1) the moir\'e-less Fock contribution is accounted for by renormalizing the bare hoppings in \App{app:fockcommentary}, and (2) the first-order moir\'e Hartree term generates a moir\'e-modulated electrostatic potential acting on the top layers in \App{app:hartreecommentary}. Hence we have arrived at a sequence of approximations 
\bea
\bar{H}_0 + \bar{H}_{int} = \bar{H}_0 + :\!\overline{H}_{int} \!: \, +  \bar{H}_H + \bar{H}_F \approx \bar{H}^{\text{eff}}_0 + :\!\overline{H}_{int} \!: \equiv \bar{H}_{\text{con}}
\eea
such that all one-body terms are grouped into the following (approximate) effective  single-particle Hamiltonian
\bea\label{appeq:H0eff}
H^{\text{eff.}}_0 &= \sum_{\mbf{k},\mbf{G},\mbf{G}', \eta s,l \sigma,l' \sigma'} c^\dag_{\mbf{k},\mbf{G},l \sigma,\eta s} [H_{\eta}(\mbf{k}+\mbf{G}) \delta_{\mbf{G},\mbf{G}'} + (H_{ISP} +H_D + V_{tb}) \delta_{\mbf{G},\mbf{G}'} + \mathcal{V}^{\text{val}.,\eta}_{\mbf{G},\mbf{G}'}]_{l \sigma, l' \sigma'} c_{\mbf{k},\mbf{G}',l' \sigma',\eta s} 
\eea
where $H_{\eta}$ uses the effective hoppings in \Eq{eq:parameters_exp} and 
\bea
[\mathcal{V}^{\text{val}.,\eta=K}]_{\mbf{G}l \sigma,\mbf{G}' l' \sigma} = \mathcal{V}_{\text{val.}} e^{i \psi_{\text{val.}}}  \delta_{ll'} \sum_{j=1,2,3} \delta_{\mbf{G},\mbf{G}'+\mbf{b}_j} \delta_{\sigma \sigma'} + h.c., \qquad [\mathcal{V}^{\text{val}.,\eta=-K}]_{\mbf{G}l \sigma,\mbf{G}' l' \sigma} = [\mathcal{V}^{\text{val}.,\eta=K}]^*_{-\mbf{G}l \sigma,-\mbf{G}' l' \sigma}
\eea
is the Hartree-generated electrostatic potential. We have completely dropped the bare moir\'e hybridization potential $V_M$ in $H^{\text{eff.}}_0$ since it is suppressed on the conduction bands. We take the conduction bands of $H^{\text{eff.}}_0$ as the active subspace in the interaction term $: \bar{H}_{int}:$ defined in \Eq{eq:intHexplicit}. Thus we obtain a simple model that captures the effects of the valence band background.

We make one final simplification and use a 2D gate screened interaction in $\overline{H}_{int}^{\text{con}.}$ instead of the 3D Coulomb interaction which was used to compute the one-body terms in \App{app:fockcommentary} and \App{app:hartreecommentary}. This simplifies the computation of the numerical HF and ED matrix elements, and is justified since the moir\'e conduction bands kept in our many-body calculations are highly polarized to the top layer. It is \emph{not} a good approximation to use a 2D interaction when computing the background terms involving matrix elements between the conduction and valence bands, which are localized to opposite surfaces.

We now analyze the single-particle band structure of $\bar{H}^{\text{eff.}}_0$ before proceeding to a HF analysis of $\bar{H}^{\text{con}.}$ at $\nu=1$ in App.~\ref{app:HF}. Note that the interaction $:\bar{H}_{int}:$ is normal ordered with respect to the charge neutrality gap, so the band structure of $H^{\text{eff.}}_0$ is the dispersion of a single electron on top of charge neutrality. It does not reflect the dispersion at $\nu=1$, which will be significantly altered by further interaction effects.

\begin{figure*}
\centering
\includegraphics[width=0.95\linewidth]{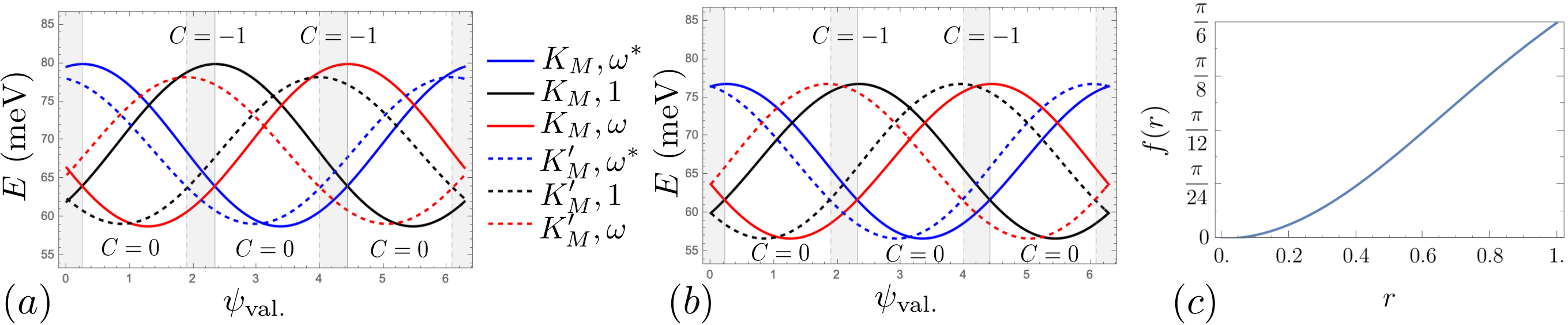}
\caption{Topological phase diagram of the lowest effective single-particle conduction band. We show the numerical energies of the different $C_3$ irreps at $K_M$ and $K'_M$ with $V=30\,$meV and $\mathcal{V}_{\text{val}.} = 12\,$meV for $(a)$ the full model with all hoppings and $(b)$ the simplified rotationally symmetric model with only $v_F$ and $t_1$ terms. The phase boundaries are almost unchanged. In $(c)$ we show the function $f(\mbf{r})$  which determines the phase shift that generates the $C=-1$ phase.} 
\label{fig:backgroundpd}
\end{figure*}

The bands shown in \Fig{fig:moirebands}(c) show two clear features. First, the $\Gamma_M$ point is essentially unchanged by $\mathcal{V}_{\text{val}.}$ since it couples $\Gamma_M$ with the nearest-neighbor $\mbf{G}$ shells, but these are very high energy.  At the $K_M$ and $K'_M$ points, we observe large splittings due to the effective moir\'e potential. These splittings induce topological phase transitions as a function of $\psi_{\text{val.}}$, as demonstrated in \Fig{fig:backgroundpd}, which shows the energies of the three different $C_3$ irreps at $K_M$ and $K'_M$. These determine the Chern number mod 3, which we find is either $0$ or $-1$ mod 3. (We confirm numerically that the true Chern number without modding by 3 is indeed $0$ or $-1$.) The splitting of the $K_M$ and $K'_M$ bands  is well described by a tripod model \cite{2011PNAS..10812233B,2021PhRvB.103t5411B}, which we now solve analytically.

At the $K_M$ point, the tripod Hamiltonian consisting of momenta $K_M, C_3 K_M, C_3^2 K_M$ coupled by $\mathcal{V}_{\text{val}.}$ is a $6L \times 6L$ matrix. As we show in \Fig{fig:backgroundpd}b, a good approximation to the energies is obtained by only keeping the $v_F,t_1$ hoppings and neglecting all others. Then we can project to the chiral state~\cite{herzog2024moire} (recall \App{app:exacteigenstates} with $\bar{k} = k_x - i k_y$) $\ket{\mbf{k}}$ with $\braket{L-\ell, B|\mbf{k}} = (-v_F \bar{k}/t_1)^\ell/\sqrt{N(\mbf{k})}$. This gives a $3\times3$ matrix 
\bea
h_K &= [\braket{C_3^i K_M|H_0^{\text{eff}}|C_3^j K_M}]_{i,j=1,\dots,3}\\
&= \bpm E_{K} & \mathcal{V}_{\text{val.}} e^{-i \psi_{\text{val.}}} M^* & \mathcal{V}_{\text{val.}} e^{i \psi_{\text{val.}}} M \\
\mathcal{V}_{\text{val.}} e^{i \psi_{\text{val.}}} M & E_K & \mathcal{V}_{\text{val.}} e^{-i \psi_{\text{val.}}} M^* \\ \mathcal{V}_{\text{val.}} e^{-i \psi_{\text{val.}}} M^* & \mathcal{V}_{\text{val.}} e^{i \psi_{\text{val.}}} M & E_K \epm, \qquad M = \braket{C_3\mbf{K}_M|\mbf{K}_M} = \frac{1}{N(\mbf{K}_M)} \sum_\ell \lp \omega |v_F K_M/t_1|^2 \rp^\ell
\eea
where $E_K$ is the energy at the $K_M$ point, $E_K \approx (L-1)V/2+ V_{tb} + |v_F \mbf{K}_M/t_1|^{2L}$.  Since $K_M$ is a $C_3$ symmetric point, this matrix can be diagonalized into $C_3$ irreps $\omega^j$. Its energies $E_i$ and eigenvectors $u_j$ are  are
\bea
\label{eq:Kpointens}
E_j &= E_K + 2 \mathcal{V}_{\text{val.}} |M| \cos \lp \psi_{\text{val.}} - \arg M + \frac{2\pi}{3}j \rp, \qquad u_j = (1, \omega^{-j},\omega^j)/\sqrt{3} \ .
\eea
This form explains the sinusoidal oscillation with $\psi_{\text{val.}}$ as observed in \Fig{fig:backgroundpd}. The amplitude is proportional to $\mathcal{V}_{\text{val.}}$, but suppressed by $|M| < 1$, and the offset in the phase is determined by $\arg M$. To simplify the geometric series defining the form factor $M$, we take $L \to \infty$ and obtain
\bea
M &= (1-\omega r)^{-1} / (1-r)^{-1} = \frac{1-r}{\sqrt{1+r+r^2}} e^{i f(r)}, \qquad  f(r)=  \arctan \frac{\sqrt{3} r}{2+r}, \qquad r = |v_F K_M/t_1|^2 \approx 0.4 .
\eea
Note that $f(r)$  goes from $0$ to $\pi/6$ as $r$ goes from 0 to 1 as shown in \Fig{fig:backgroundpd}c. At $\theta = 0.77^\circ$, this result gives $\arg M = 0.28\,$rad, in good agreement with the numerical result $0.22\,$rad. Finally, the $K'_M$ point is related by conjugation, taking $\arg M \to - \arg M$ and $j \to -j$. In \Fig{fig:backgroundpd}, we label the energies according to their $C_3$ eigenvalues in the convention
\bea
D[C_3]_{\mbf{G} l \sigma,\mbf{G}' l' \sigma'} &= \delta_{\mbf{G},C_3 \mbf{G}'} \delta_{ll'} \delta_{\al \be} e^{i \frac{2\pi}{3}(l-1+\al)} \ .
\eea
The overall phase of $D[C_3]$ is a convention, which the Chern number does not depend on. For concreteness, the irreps at $\psi_{\text{val.}} = 4\pi/3 + 0.12$ are $\omega^*,\omega^*,\omega$ at $\Gamma_M, K_M, K'_M$ in the lowest band. The second band has irreps $1, \omega^*$ at $K_M,K'_M$ respectively, and is nearly gapless at $\Gamma_M$. From \Fig{fig:backgroundpd}, we see that the full phase diagram is
\bea
C &= \begin{cases}
-1, & \psi_{\text{val.}} \in (\frac{2\pi}{3} j - \arg M, \frac{2\pi}{3} j + \arg M) \\
0, &\text{otherwise} \ .
\end{cases}
\eea
Note that this phase diagram does not depend on $E_K$ or the magnitude of $\mathcal{V}_{\text{val.}}$. Since $\arg M$ is increasing (see \Fig{fig:backgroundpd}c) with the size of the moir\'e Brillouin zone, larger angles increase the size of the $C=-1$ region. Moreover at $\arg M = 0$, the $C=-1$ regions shrink to 0, and at $\arg M = \pi/6$, the $C=-1$ regions have maximal width $\pi/3$ in $\psi_\text{val.}$ with boundaries $\frac{2\pi}{3}j \pm \pi/6$. Note that the gap in the $C=-1$ region is always smaller than the $C=0$ region. Taylor expanding \Eq{eq:Kpointens} in small $\arg M$, we find
\bea
\Delta_{C=-1} \approx 2 \mathcal{V}_{\text{val.}} |M| \cdot 2 \sqrt{3} \arg M, \qquad \Delta_{C=0} \approx 2 \mathcal{V}_{\text{val.}} |M| \cdot 3
\eea
such that $\frac{\Delta_{C=-1}}{\Delta_{C=0}} = \frac{2  \arg M}{\sqrt{3}} \in (0,\frac{\pi/3}{\sqrt{3}}) \approx (0,0.6)$. Since our calculation of $\psi_{\text{val.}}$ is close to $4\pi/3$, the bands have $C=-1$ with a small gap of $\sim 4$meV in \Fig{fig:backgroundpd}a . This means that band mixing between the lowest two bands is expected to be large at the mBZ corners. Indeed, in App.~\ref{app:HF} we show that at $\nu=1$, HF obtains a $C=+1$ phase --- absent in the single-particle phase diagram --- that occurs from double band inversion at $K_M$ and $K_M'$. As found in Ref. \cite{herzog2024moire}, increasing the bare moir\'e strength results in a $C=5$ Chern number, and we have shown here that $\mathcal{V}_{\text{val.}}$ creates $C=0,-1$. In no case do we obtain the $C=+1$ band found in $\nu=1$ Hartree-Fock. 

\clearpage
\newpage

\section{Hartree-Fock and Time-Dependent Hartree-Fock}
\label{app:HF}

\subsection{Methods}

We first recall the interacting conduction band Hamiltonian derived in App.~\ref{app:AppA} that will be studied using HF and TDHF. The effective single-particle model $H_0^\text{eff.}$ is given by Eq.~\ref{appeq:H0eff}, which is diagonalized to obtain moir\'e band energies $\tilde{E}_{n,\eta}(\mbf{k})$ and moir\'e Bloch functions $\tilde{U}^{\eta}_{\mbf{G} l \sigma, n}(\mbf{k})$. In this appendix, we use tildes to distinguish quantities computed using the effective single-particle conduction band model in Eq.~\ref{appeq:H0eff}, from the starting single-particle Hamiltonian in Eq.~\ref{eq:H0_original}. We remind the reader that the term $\mathcal{V}^{\text{val}.,\eta}$ in Eq.~\ref{appeq:H0eff} corresponds to the moir\'e-modulated electrostatic potential generated via the moir\'e capacitor effect, and acts on all layers. $\mathcal{V}^{\text{val}.,\eta}$ does not include the bare moir\'e potential $V_M^{\eta}$ induced by hybridization between the hBN and the bottom layer graphene, which has a negligible direct effect on the low-energy conduction states for a large interlayer potential $V>0$. On the other hand, Eq.~\ref{eq:H0_original} only includes $V_M^{\eta}$ but not $\mathcal{V}^{\text{val}.,\eta}$. 

We project onto the conduction states of Eq.~\ref{appeq:H0eff} so that the single-particle Hamiltonian becomes
\begin{equation}\label{eq:H_con_sp}
    H^\text{con. sp.}=\sum_{\mbf{k},n,\eta,s}\tilde{E}_{n,\eta}(\mbf{k})\tilde\gamma^\dagger_{\mbf{k},n,\eta, s}\tilde\gamma_{\mbf{k},n,\eta, s},
\end{equation}
where $\tilde\gamma^\dagger_{\mbf{k},n,\eta ,s}$ is band creation operator, and $n$ is a conduction band index. Note that Eq.~\ref{eq:H_con_sp} does not include the valence band states (their impact on the conduction states has already been incorporated via the moir\'e capacitor effect).

The two-body term consists of density-density interactions normal-ordered with respect to the Fock vacuum of the projected Hilbert space of conduction bands. We emphasize that this choice of normal-ordering is employed because the effects of the occupied valence bands, such as the Fermi velocity renormalization and the moir\'e capacitor effect, have already been folded into the effective parameters (see Eq.~\ref{eq:parameters_exp}) of $H^\text{eff.}_0$. The interaction takes the form
\begin{align}
    H^\text{con. int.}=&\frac{1}{2\Omega_\text{tot}}\sum_{\mbf{q},\mbf{G}}\sum_{\mbf{k},\mbf{k}',m,m',n,n',\eta,\eta',s,s'}V(\mbf{q}+\mbf{G}) \tilde M^\eta_{mn}(\mbf{k},\mbf{q}+\mbf{G})\tilde M^{\eta'^*}_{n'm'}(\mbf{k}'-\mbf{q},\mbf{q}+\mbf{G})\\
    &\quad\quad\quad\times\tilde\gamma^\dagger_{\mbf{k}+\mbf{q},m,\eta,s}\tilde\gamma^\dagger_{\mbf{k}'-\mbf{q},m',\eta',s'}\tilde\gamma_{\mbf{k}',n',\eta',s'}\tilde\gamma_{\mbf{k},n,\eta,s}
\end{align}
where the form factor is
\begin{equation}
    \tilde M^{\eta}_{mn}(\mbf{k},\mbf{q}+\mbf{G}) = \sum_{\mbf{G}',l\sigma} \tilde U^{\eta}_{\mbf{G}', l \sigma,m}(\mbf{k}+\mbf{q}+\mbf{G})^* \tilde U^{\eta}_{\mbf{G}' l \sigma,n}(\mbf{k}).
\end{equation}
For simplicity, we have used a layer-independent 2D interaction, which is expected to be a good approximation in the large displacement field regime (of interest here) where the low-energy electrons are strongly polarized towards the top layer. The dual-gate-screened interaction potential is 
\begin{equation}\label{eq:dual_gate_Vq}
    V(\mbf{q})=\frac{e^2}{2\epsilon_0\epsilon_r q}\tanh(qd_\text{sc})
\end{equation}
where $\epsilon_r$ is the relative dielectric and $d_\text{sc}$ is the gate-to-sample distance. Note that the gate screening has a very weak effect on the amplitude $\mathcal{V}_{\text{val.}}$ of the moir\'e capacitor effect because the gate distance $d_\text{sc}\sim 10$\,nm is much larger than the device thickness $\sim 1$\,nm. To see this, we recall that the 3D Coulomb interaction in the presence of gate-screening corrections is~\cite{kwan2023MFCI3} 
\begin{equation}
    V_{ll'}(\mbf{q}) =\frac{e^2}{2\epsilon_0\epsilon_r q}\lp e^{-q |z_l-z_{l'}|} - 2 e^{-2 q d_{\text{sc}}} \cosh(q (z_l + z_{l'})) + O(e^{-4 q d_{\text{sc}}})\rp.
\end{equation}
The gate-screening correction to $\mathcal{V}_{\text{val.}}$ is negligible since $2 |\mbf{G}| d_{\text{sc}} = 12.6$ for $d_{\text{sc}} = 10$nm. 

Unless otherwise stated, the HF and TDHF calculations will use $\epsilon_r=5$ and $d_\text{sc}=10\,$nm. (Note that strictly speaking, if the R$n$G is placed symmetrically between the two gates, the low-energy conduction electrons, which are mainly polarized on the top layer, will have a different distance to the two gates. We assume that $d_\text{sc}$ is significantly larger than the graphene interlayer distance such that Eq.~\ref{eq:dual_gate_Vq} remains accurate.)

The total effective Hamiltonian is then
\begin{equation}
    H^\text{con.}=H^\text{con. sp.}+H^\text{con. int.}.
\end{equation}
In numerical calculations, we project onto the lowest $n_c$ moir\'e conduction bands per spin and valley, and consider a system with $N_1\times N_2$ moir\'e unit cells.

\subsubsection{Hartree-Fock}\label{secapp:HF_formalism}

We refer the reader to Refs.~\cite{kwan2023MFCI3,HF_AIPreview} for a detailed description of the self-consistent HF procedure. The HF calculation iteratively minimizes the total energy over the space of single Slater determinants parameterized by the one-body HF density matrix.
In this work, we restrict the HF density matrix to preserve moir\'e translation invariance and spin-valley $U(1)$ symmetries, i.e.
\begin{equation}
    \langle \tilde\gamma^\dagger_{\mbf{k},n,\eta, s}\tilde\gamma_{\mbf{k}',n',\eta', s'} \rangle=\delta_{\mbf{k}\mbf{k}'}\delta_{\eta\eta'}\delta_{ss'}P_{nn'}(\mbf{k},\eta,s).
\end{equation}
For later use, we write the mean-field decoupled interaction term 
\begin{align}\label{eqapp:HHF_int}
    H_{\text{HF,int}}[P]&=\sum_{\mbf{G},\mbf{k},m,n,\eta,s}\frac{V(\mbf{G})}{\Omega_\text{tot}}\tilde M^\eta_{mn}(\mbf{k},\mbf{G})\left(\sum_{\mbf{k}'m'\eta's'}\tilde M^{\eta'^*}_{n'm'}(\mbf{k}',\mbf{G})P_{m',n'}(\mbf{k},\eta',s')\right)\tilde \gamma^\dagger_{\mbf{k},m,\eta,s}\tilde \gamma_{\mbf{k},n,\eta,s}\\
    &-\sum_{\mbf{G},\mbf{q},\mbf{k},m,m',n,n',\eta,s}\frac{V(\mbf{q}+\mbf{G})}{\Omega_\text{tot}}\tilde M^{\eta^*}_{n'm}(\mbf{k},\mbf{q}+\mbf{G})\tilde M^\eta_{m'n}(\mbf{k},\mbf{q}+\mbf{G})P_{m',n'}(\mbf{k}+\mbf{q},\eta,s)\tilde \gamma^\dagger_{\mbf{k},m,\eta,s}\tilde \gamma_{\mbf{k},n,\eta,s}.
\end{align}

For each parameter, we perform $\nu=1$ HF calculations with at least 10 initial seeds. Included within these seeds are `vortex ansatze' that specifically target $C=0$ and $C=1$ solutions (see App.~\ref{appsec:vortex_ansatz} for details). At $\nu=1$, any insulating HF solution must be fully spin-valley polarized given the constraints of conduction-band projection, moir\'e translation invariance, and preservation of spin-valley $U(1)$ symmetries. 
Without loss of generality, we apply symmetry transformations to such solutions so that they are polarized in valley $K$ and spin $\uparrow$. The HF ground state corresponds to the solution with the lowest total energy.

\begin{figure*}
\centering
\includegraphics[width=0.95\linewidth]{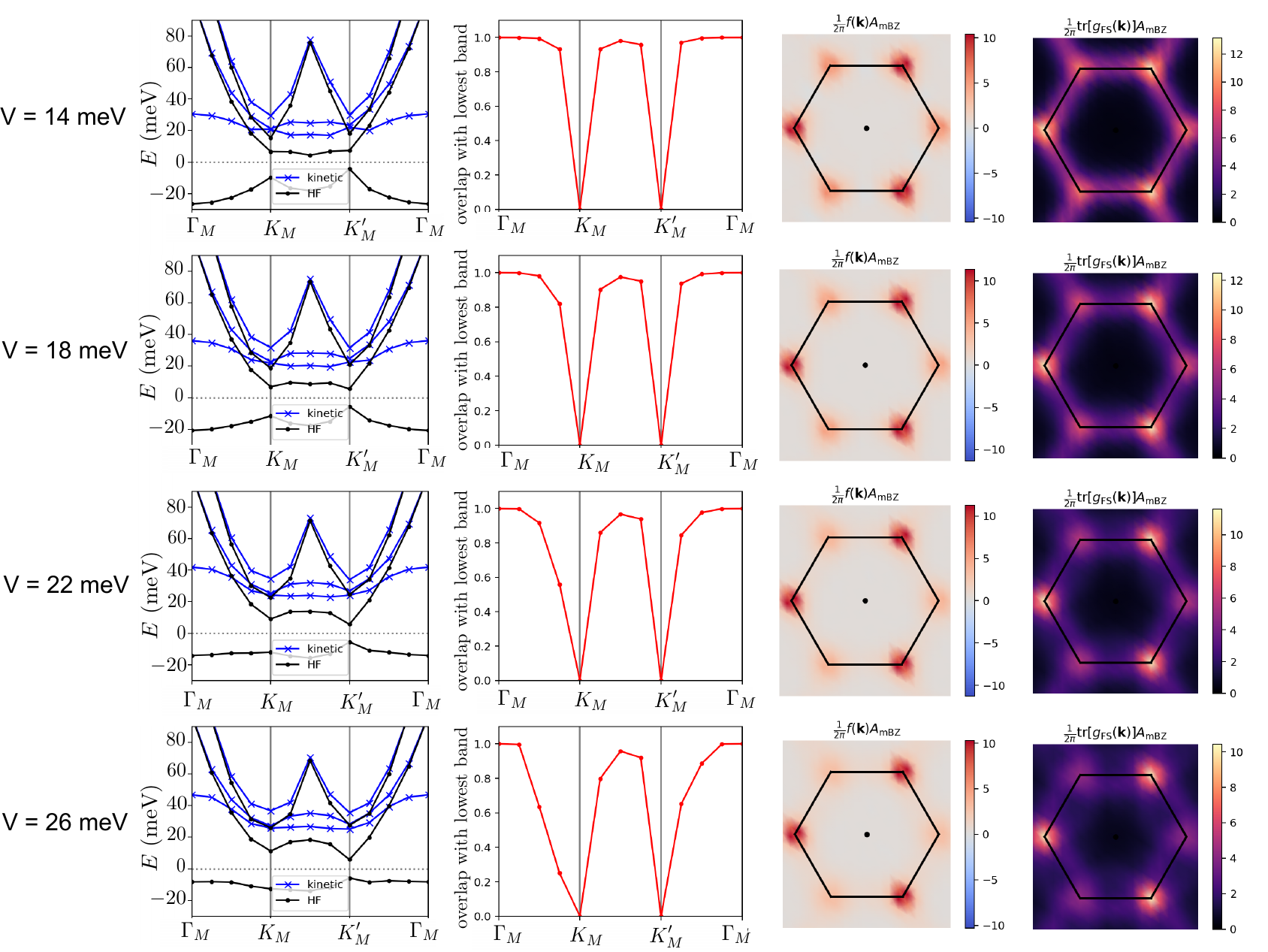}
\caption{HF calculations of the $C=1$ state at $\nu=1$, for $V=14,16,22,26\,$meV. First column shows the HF (black) and non-interacting (blue) dispersion within the $K\uparrow$ flavor which accommodates all the doped electrons at $\nu=1$. Grey dotted line at $E=0\,$meV denotes the Fermi level in the HF calculation. The non-interacting dispersion has been shifted so that the highest non-interacting energy coincides with the highest HF eigenvalue. Second column shows the wavefunction overlap of the lowest HF band and the lowest non-interacting band along the path in the mBZ as the first column. Third column shows the Berry curvature of the lowest HF band within the mBZ. Fourth column shows the trace of the Fubini-Study metric of the lowest HF band within the mBZ. Parameters are $V_{tb}=32\,$meV, $\psi=(\frac{4\pi}{3}+0.12)\,$rad, $\mathcal{V}_\text{val.}=8\,$meV, and $\epsilon_r=5$. System size is $12\times12$ and $n_c=3$ conduction bands are kept.} 
\label{fig:app_bandstruct_olap_QG}
\end{figure*}

As we are interested in understanding FCIs in this work, the bandwidth of the system plays an important role. In particular, we expect FCIs to require a narrow bandwidth. However, in a strongly-interacting multi-band system such as R$L$G/hBN, there is no unique definition for the bandwidth. In fact, the system is essentially gapless without the Hartree potential generated by the valence bands, and there is no obvious way to define the bandwidth. With goal of understanding the stability of the $\nu=2/3$ FCI, we consider the following three candidate definitions for the bandwidth:
\begin{enumerate}
    \item \underline{Non-interacting bandwidth:} The bandwidth of the lowest non-interacting conduction band of $H^\text{con. sp.}$ is the simplest definition, but suffers from several severe drawbacks. First, the lowest non-interacting conduction band usually has a Chern number $C\neq 1$ (see App.~\ref{app:phasediagram}) that disagrees with the interacting Chern number $C=1$ at $\nu=1$, disqualifying it from serving as a natural parent band for the $\nu=2/3$ FCI. Second, interaction effects at finite-doping substantially reshape the dispersion, as shown for example in the $\nu=1$ HF calculations of Fig.~\ref{fig:app_bandstruct_olap_QG}. The non-interacting bandwidth does not account for such renormalization effects.
    \item \underline{$\nu=1$ HF bandwidth:} The most commonly-adopted theoretical approach to $\nu=2/3$ FCIs in R$L$G/hBN begins by performing an HF calculation at $\nu=1$ in order to obtain a parent $C=1$ band. [Note we are not claiming that the $\nu=2/3$ FCI is built entirely out of orbitals in the parent $C=1$ band. Indeed, as shown later in the ED calculations of App.~\ref{app:ED_23} (see in particular Fig.~\ref{fig_PES_3b7_oneband}b), the stable FCI within multiband calculations has finite occupation on the higher HF bands. However, the FCI is still predominantly ($\sim 90\%$) within the lowest HF band, so we anticipate that the dispersion of this HF band plays a significant role in the stability of the FCI.]
    
    While the approach of considering the $C=1$ HF band yields a parent band with the correct topology, the $\nu=1$ HF dispersion likely overestimates 
    the interaction-induced corrections to the bandwidth as relevant for a putative FCI at $\nu=2/3$. The reasoning is that the FCI of interest occurs at a more dilute filling of $\nu=2/3$, along with the fact that interaction-induced renormalization of the dispersion grows with the electronic density (for example, the Hartree and Fock corrections to the interacting dispersion in mean-field theory are proportional to the density, as shown in Eq.~\ref{eqapp:HHF_int}). 
    \item \underline{Effective bandwidth $W_{2/3}$:} We introduce the effective bandwidth $W_{2/3}$, which is designed to correct for the overestimate of interaction corrections in the $\nu=1$ HF bandstructure. We first motivate the construction of $W_{2/3}$. [Again, we re-iterate that the stable FCI in multiband calculations involves finite occupation of higher HF bands, so the following discussion should viewed as a heuristic argument motivated from the limit where the FCI lies predominantly within the lowest HF band. Phrased differently, the stable multiband FCI involves mixing with higher HF bands, but the dispersion of the lowest $C=1$ HF band still contributes to the FCI stability.]
    \end{enumerate}
    
    Robust and well-behaved FCIs in isolated Chern bands typically possess an electron occupation factor that is nearly uniform over the Brillouin zone. If the creation operator for the Chern band is $d^\dagger_{\mbf{k}}$, this would imply that $P^\text{FCI}(\mbf{k})=\langle d^\dagger_{\mbf{k}}d_{\mbf{k}}\rangle\simeq \nu$, where $\nu$ is the filling factor corresponding to the FCI of interest. Note that this is simply the one-body density matrix of the fully-occupied Chern band, but scaled by a factor of $\nu$. We can construct an `effective' band structure $E^{\text{FCI}}(\mbf{k})$ by mean-field decoupling the Hamiltonian using $P^\text{FCI}(\mbf{k})$. While this neglects the higher-body correlations in the FCI, this construction correctly captures the Hartree and Fock effects. In particular, if there is to be a stable FCI, we expect that $E^{\text{FCI}}(\mbf{k})$ should be relatively narrow compared to the interaction scale. Otherwise, the system may be unstable to a Fermi liquid by fully occupying (emptying) the regions where $E^{\text{FCI}}(\mbf{k})$ is low (high). 

    For the present calculation of R$L$G/hBN, we anticipate that the HF $C=1$ band generated at $\nu=1$ serves as a candidate parent Chern band for the $\nu=2/3$ FCI. Let $P^{\nu=1}(\mbf{k})$ denote the (spin-valley polarized) density matrix of the $C=1$ state, and consider the one-body term
    \begin{equation}\label{eqapp:Hsp+2/3HHF}
        H_{\text{HF}}[\frac{2}{3}P^{\nu=1}(\mbf{k})]=H^\text{con. sp.}+H_{\text{HF,int}}[\frac{2}{3}P^{\nu=1}(\mbf{k})],
    \end{equation} 
    where $H^\text{con. sp.}$ is the single-particle Hamiltonian for the conduction bands (see Eq.~\ref{eq:H_con_sp})
    and $H_\text{HF,int}$ is defined in Eq.~\ref{eqapp:HHF_int} (see also Ref.~\cite{pi2026engineeringtopologicalflatbands}). Compared to the HF Hamiltonian of the $\nu=1$ state, the above expression rescales the interaction part by a factor of $2/3$. To construct $W_{2/3}$, we first express Eq.~\ref{eqapp:Hsp+2/3HHF} in the $\nu=1$ HF basis. Let the unitary transformation between the moir\'e band basis creation operator $\tilde{\gamma}^\dagger_{\mbf{k},n}$ and $\nu=1$ HF basis creation operator $\gamma^\dagger_{\text{HF},\mbf{k},\alpha}$ be 
    \begin{equation}
        \gamma^\dagger_{\text{HF},\mbf{k},x}=\sum_{n}v_{\text{HF},x,n}(\mbf{k})\tilde{\gamma}^\dagger_{\mbf{k},n},
    \end{equation}
    where $x$ indexes HF bands, and we have dropped the spin and valley indices, since we only need to consider the sole spin-valley flavor that is occupied at $\nu=1$. Then we have
    \begin{align}
        H_{\text{HF}}[\frac{2}{3}P^{\nu=1}(\mbf{k})]&=\sum_{\mbf{k},x,y}\tilde{E}_{n}(\mbf{k})v^*_{\text{HF},x,n}(\mbf{k})v_{\text{HF},y,n}(\mbf{k})\gamma^\dagger_{\text{HF},\mbf{k},x}\gamma_{\text{HF},\mbf{k},y}\\
        +\frac{2}{3}\sum_{\mbf{G},\mbf{k},m,n,x,y}\frac{V(\mbf{G})}{\Omega_\text{tot}}&\tilde M_{mn}(\mbf{k},\mbf{G})\left(\sum_{\mbf{k}'m'}\tilde M^*_{n'm'}(\mbf{k}',\mbf{G})P^{\nu=1}_{m',n'}(\mbf{k})\right)v^*_{\text{HF},x,m}(\mbf{k})v_{\text{HF},y,n}(\mbf{k})\gamma^\dagger_{\text{HF},\mbf{k},x}\gamma_{\text{HF},\mbf{k},y}\\
    -\frac{2}{3}\sum_{\mbf{G},\mbf{q},\mbf{k},m,m',n,n',x,y}\frac{V(\mbf{q}+\mbf{G})}{\Omega_\text{tot}} &\tilde M^*_{n'm}(\mbf{k},\mbf{q}+\mbf{G})\tilde M_{m'n}(\mbf{k},\mbf{q}+\mbf{G})P^{\nu=1}_{m',n'}(\mbf{k}+\mbf{q})v^*_{\text{HF},x,m}(\mbf{k})v_{\text{HF},y,n}(\mbf{k})\gamma^\dagger_{\text{HF},\mbf{k},x}\gamma_{\text{HF},\mbf{k},y}
    \end{align}
    Generally, this will involve hybridization terms between the different HF bands $x\neq y$. We then define $W_{2/3}$ as the bandwidth of Eq.~\ref{eqapp:Hsp+2/3HHF} when projected to the lowest HF band, i.e.~evaluating the above equation when $x=y=0$. Note that this procedure does not require rediagonalizing any matrices.

\subsubsection{Vortex Ansatz}\label{appsec:vortex_ansatz}

The vortex ansatz is a specially designed one-body density matrix that is useful for targeting insulating states at $\nu=+1$ with specific Chern numbers in the self-consistent HF calculations.  We assume fully flavor polarization into valley $\eta=+$ and spin $s=\uparrow$, so we drop these indices for notational brevity.

To motivate the vortex ansatz, we first consider the conduction subspace of $H^\text{con. sp.}$ in the limit  where we turn off the moir\'e capacitor effect (i.e.~$\mathcal{V}_\text{val.}=0$) so that continuous translation symmetry is recovered. Consider the lowest continuum conduction band of $H^\text{con. sp.}$ with creation operator $d^\dagger_{\mbf{k}}$
\begin{equation}
    d^\dagger_{\mbf{k}}=\sum_{l\sigma}u_{l\sigma}(\mbf{k})c^\dagger_{\mbf{k},l\sigma}
\end{equation}
where $c^\dagger_{\mbf{k},l\sigma}$ is the creation operator for a plane wave with momentum $\mbf{k}$ (relative to the Dirac momentum in valley $\eta=+$) on layer $l$ and sublattice $\sigma$, and $u_{l\sigma}(\mbf{k})$ are the Bloch coefficients. 
We fix the gauge such that the $(n-1,B)$ component of $u_{l\sigma}(\mbf{k})$ is real and positive. Out of this conduction band, we can generally construct a single `moir\'e band' of creation operators $\tilde{\Gamma}^\dagger_{\mbf{k}}$ by specifying the envelope function $\mathcal{F}(\mbf{k}+\mbf{G})$
\begin{gather}
\tilde{\Gamma}^\dagger_{\mbf{k}} = \sum_{\mbf{G}} \tilde{\mathcal{F}}(\mbf{k}+{\mbf{G}})d^\dagger_{\mbf{k}+\mbf{G}}\\
d^\dagger_{\mbf{k}+\mbf{G}}=\tilde{\mathcal{F}}^*(\mbf{k}+\mbf{G})\tilde{\Gamma}^\dagger_{\mbf{k}}+\ldots
\end{gather}
where the $\ldots$ indicate higher `moir\'e bands' that are not needed for the vortex ansatz, and $\tilde{\mathcal{F}}(\mbf{k}+\mbf{G})$ is a `normalized' version of $\mathcal{F}(\mbf{k}+\mbf{G})$
\begin{equation}
    \tilde{\mathcal{F}}(\mbf{k}+\mbf{G})=\frac{1}{\sqrt{\sum_{\mbf{G}'}|{\mathcal{F}}(\mbf{k}+\mbf{G}')|^2}}{\mathcal{F}}(\mbf{k}+\mbf{G})=\frac{1}{\sqrt{\mathcal{N}(\mbf{k})}}{\mathcal{F}}(\mbf{k}+\mbf{G}).
\end{equation}
Note that $\tilde{\Gamma}^\dagger_{\mbf{k}}$ and the normalization factor $\mathcal{N}(\mbf{k})$ are periodic under $\mbf{k}\rightarrow \mbf{k}+\mbf{G}$.  

The vortex ansatz corresponds to the following parameterization of the envelope function
\begin{equation}
    \mathcal{F}(\mbf{k})=w(|\mbf{k}|)\left(\frac{k}{|\mbf{k}|}\right)^C,
\end{equation}
where $k=k_x+ik_y$. The phase winding $\sim k^C$ 
controls the Chern number $C$ of the band $\tilde\Gamma^\dagger_{\mbf{k}}$. We also discuss the `domain wall function' $w(|\mbf{k}|)$. One choice is 
\begin{equation}\label{eq:domain_wall_wk}
    w(|\mbf{k}|)=\frac{1}{2}\left[1+\tanh\left(-\frac{|\mbf{k}|-k_0}{d}\right)\right]
\end{equation}
where $k_0$ sets the domain wall position, and $d$ the domain wall width. A typical choice is $k_0=q_1$ and $d=0.2q_1$. This choice of $w(|\mbf{k}|)$ is chosen so that the wave function of $\tilde\Gamma^\dagger_{\mbf{k}}$ is primarily localized within the lowest mBZ shells in order to avoid a large kinetic energy penalty. Note that the precise choice of $w(|\mbf{k}|)$ is not important, since the role of the vortex ansatz is to aid the convergence of the numerical HF calculations, rather than provide actual trial wavefunctions for many-body calculations.

The density matrix of the vortex ansatz is characterized by $\langle \tilde\Gamma^\dagger_{\mbf{k}}\tilde\Gamma_{\mbf{k}}\rangle=1 $. Equivalently, this corresponds to
\begin{equation}
    \langle c^\dagger_{\mbf{k+\mbf{G}},l\sigma}c_{\mbf{k}+\mbf{G}',l'\sigma'} \rangle=u^*_{l\sigma}(\mbf{k}+\mbf{G})u_{l'\sigma'}(\mbf{k}+\mbf{G}')\tilde{\mathcal{F}}^*(\mbf{k}+\mbf{G})\tilde{\mathcal{F}}(\mbf{k}+\mbf{G}').
\end{equation}

We now need to project the vortex ansatz onto the active conduction moir\'e bands of the effective Hamiltonian $H^{\text{con s.p.}}$ (with the actual desired value of $\mathcal{V}_\text{val.}$). We recall that the latter are described by creation operators $\tilde\gamma^\dagger_{\mbf{k},n}$
\begin{eqnarray}
    &\tilde\gamma^\dagger_{\mbf{k},n}=\sum_{\mbf{G},l\sigma }
    \tilde{U}_{\mbf{G},l\sigma,n}(\mbf{k})
    c^\dagger_{\mbf{k}+\mbf{G},l\sigma},
\end{eqnarray}
where $n$ is a band label. The projected vortex ansatz is described by its one-body density matrix in the moir\'e band basis
\begin{align}
    P_{mn}(\mbf{k})&=\langle \tilde\gamma^\dagger_{\mbf{k},m}\tilde\gamma_{\mbf{k},n}\rangle =\sum_{\mbf{G},l\sigma}\sum_{\mbf{G}',l'\sigma'}\tilde{U}_{\mbf{G} l\sigma m}(\mbf{k}) \tilde{U}^*_{\mbf{G}'l'\sigma' n}(\mbf{k})\langle c^\dagger_{\mbf{k}+\mbf{G},l\sigma} c_{\mbf{k}+\mbf{G}',l'\sigma'} \rangle \\
    &=\sum_{\mbf{G},l\sigma}\sum_{\mbf{G}',l'\sigma'}\tilde{U}_{\mbf{G}l\sigma m}(\mbf{k}) \tilde{U}^*_{\mbf{G}'l'\sigma' n}(\mbf{k})u^*_{l\sigma}(\mbf{k}+\mbf{G})u_{l'\sigma'}(\mbf{k}+\mbf{G}')\tilde{\mathcal{F}}^*(\mbf{k}+\mbf{G})\tilde{\mathcal{F}}(\mbf{k}+\mbf{G}').
\end{align}
For our choices of domain wall function (e.g.~Eq.~\ref{eq:domain_wall_wk} with $k_0=q_1$ and $d=0.2q_1$), $P_{mn}(\mbf{k})$ is close to a projector, reflecting that the vortex ansatz is mainly constructed out of states lying within the active Hilbert space. 

Finally, we need to discuss various possibilities for the real-space origin of the vortex ansatz in order to capture the different combinations of $C_3$ irreps at high symmetry momenta. For our conventions where $\mbf{b}_1$ and $\mbf{b}_2$ are related by a $2\pi/3$ rotation, the corresponding real space lattice vectors $\mbf{a}_1$ and $\mbf{a}_2$ are separated by a rotation $\pi/3$. Hence, we should consider the vortex ansatz shifted by three possible real-space shifts: $\mbf{r}_1=\mbf{0},\mbf{r}_2=\frac{1}{3}(\mbf{a}_1+\mbf{a}_2),\mbf{r}_3=\frac{2}{3}(\mbf{a}_1+\mbf{a}_2)$.  For a shift of $\mbf{R}$, we adjust the vortex ansatz by taking
\begin{equation}
    \tilde{\mathcal{F}}(\mbf{k}+\mbf{G})\rightarrow
     \tilde{\mathcal{F}}(\mbf{k}+\mbf{G})e^{-i(\mbf{k}+\mbf{G})\cdot\mbf{R}}.
\end{equation} 
In the absence of a moir\'e potential, the choice of $\mbf{R}$ in the vortex ansatz is unimportant due to continuous translation symmetry of the underlying Hamiltonian. However, when there exists a moir\'e potential (e.g.~via the moir\'e capacitor effect), vortex ans\"atze with distinct $\mbf{R}$ have generally distinct energies. Hence, when specifying the initial density matrix in numerical HF calculations, it is useful to consider the $C_3$-symmetric configurations $\mbf{R}=\mbf{r}_1,\mbf{r}_2,\mbf{r}_3$.

\subsubsection{Time-dependent Hartree-Fock}\label{secapp:TDHF_formalism}

The time-dependent HF (TDHF) framework enables calculation of the charge-neutral collective mode energies $\Omega^\alpha$ ($\alpha$ indexes the collective modes) of a self-consistent HF state. We refer the reader to Refs.~\cite{kwan2023MFCI3,HF_AIPreview,ring2004nuclear} for a detailed description of the TDHF procedure. Here, we first briefly summarize the TDHF formalism for a general Hamiltonian, before specializing to the case of R$L$G/hBN at $\nu=+1$.

Consider a general interacting Hamiltonian
\begin{equation}\label{appeq:TDHF_startH}
    \hat{H}=\sum_{ij}T_{ij}c^\dagger_ic_j+\frac{1}{2}\sum_{ijkl}V_{ij,kl}c^\dagger_jc^\dagger_ic_kc_l,
\end{equation}
where $i,j,k,l$ index the single-particle orbitals. Consider a self-consistent HF solution $\ket{\text{HF}}$ with total mean-field energy $E_{\text{HF}}$. The associated HF basis of creation operators is
\begin{equation}
d^\dagger_{a}=\sum_{i}v_{a,i}c^\dagger_i,
\end{equation}
where $a$ indexes the HF orbitals with HF energies $\mathscr{E}_a$. We divide the HF orbitals into the occupied subspace $\mathscr{H}_\text{occ.}$ with dimension $N_\text{occ.}$, and the unoccupied subspace $\mathscr{H}_\text{unocc.}$ with dimension $N_\text{unocc.}$. We also introduce the complete set of $N_\text{occ.}\times N_{\text{unocc.}}$ particle-hole (ph) labels $\phi=(\phi_\text{p},\phi_\text{h})$, where $\phi_\text{p}$ ($\phi_\text{h}$) indexes an unoccupied (occupied) HF orbital.

Note that if we had access to the exact many-body ground state $\ket{\text{GS}}$ with energy $E_0$, and exact excited states $\ket{\alpha}$ (indexed by $\alpha$) with energy $E_\alpha$, we could create (generally non-local) neutral excitations with the operator $Q_\alpha^\dagger=\ket{\alpha}\bra{\text{GS}}$ and excitation energy $\Omega^a=E_\alpha-E_0$. In the equation-of-motion language~\cite{ring2004nuclear}, these mode creation operators $Q_\alpha^\dagger$ can be straightforwardly shown to satisfy the commutator relation
\begin{equation}\label{appeq:doublecom}
    \bra{\text{GS}}[\delta Q,[\hat{H},Q^\dagger_\alpha]]\ket{\text{GS}}=({E}_\alpha-{E}_0)\bra{\text{GS}}[\delta Q,Q^\dagger_\alpha]\ket{\text{GS}}.
\end{equation}
where $\delta Q$ is an arbitrary many-body operator.

However in practice, we do not have access to the exact many-body ground state and excitations. In the random phase approximation (RPA)~\cite{ring2004nuclear}, we make an approximation by restricting to an excitation operator of the form 
\begin{equation}\label{appeq:Q_gen}
    Q^\dagger_\alpha=\sum_{\phi}\left(X^\alpha_{\phi}d^\dagger_{\phi_\text{p}}d_{\phi_\text{h}}-Y^\alpha_{\phi}d^\dagger_{\phi_\text{h}}d_{\phi_\text{p}}\right).
\end{equation}
The RPA ground state $\ket{\text{RPA}}$ is implicitly defined by the annihilation $Q_\alpha\ket{\text{RPA}}=0$ (the RPA state is not explicitly constructed in practice). Note that $\ket{\text{RPA}}$ is in general different from $\ket{\text{HF}}$ since $\sum_\phi Y_\phi^{\alpha*} d^\dagger_{\phi_\text{p}}d_{\phi_\text{h}}\ket{\text{HF}}\neq 0$. In the equation-of-motion approach in Ref.~\cite{ring2004nuclear}, we obtain the following conditions by considering Eq.~\ref{appeq:doublecom} with $\delta Q$ a general linear combination of $d^\dagger_{\phi_\text{p}}d_{\phi_\text{h}}$ and $d^\dagger_{\phi_\text{h}}d_{\phi_\text{p}}$
\begin{gather}
    \bra{\text{RPA}}\left[d^\dagger_{\phi_\text{h}}d_{\phi_\text{p}},\left[\hat{H},Q^\dagger_\alpha\right]\right]\ket{\text{RPA}}=\Omega^\alpha \bra{\text{RPA}}\left[d^\dagger_{\phi_\text{h}}d_{\phi_\text{p}},Q^\dagger_\alpha\right]\ket{\text{RPA}}\\
        \bra{\text{RPA}}\left[d^\dagger_{\phi_\text{p}}d_{\phi_\text{h}},\left[\hat{H},Q^\dagger_\alpha\right]\right]\ket{\text{RPA}}=\Omega^\alpha \bra{\text{RPA}}\left[d^\dagger_{\phi_\text{p}}d_{\phi_\text{h}},Q^\dagger_\alpha\right]\ket{\text{RPA}}.
\end{gather}
There is one pair of equations for every possible ph label $\phi$.
In the quasi-boson approximation~\cite{ring2004nuclear}, commutators of fermion bilinears above are evaluated according using the HF state
\begin{equation}\label{appeq:QBA}
    \bra{\text{RPA}}[d^\dagger_{\phi_\text{h}} d_{\phi_\text{p}},d^\dagger_{\phi'_\text{p}} d_{\phi'_\text{h}}]\ket{\text{RPA}}\simeq \bra{\text{HF}}[d^\dagger_{\phi_\text{h}} d_{\phi_\text{p}},d^\dagger_{\phi'_\text{p}} d_{\phi'_\text{h}}]\ket{\text{HF}}=\delta_{{\phi_\text{h}}{\phi'_\text{h}}}\delta_{{\phi_\text{p}}{\phi'_\text{p}}},
\end{equation}
leading to the block-matrix equation
\begin{gather}\label{appeq:L_general}
\mathcal{L}\begin{pmatrix}X^\alpha\\Y^\alpha\end{pmatrix}=\begin{pmatrix}
    A & B \\ -B^* & -A^*
    \end{pmatrix}\begin{pmatrix}X^\alpha\\Y^\alpha\end{pmatrix}=\Omega^a\begin{pmatrix}X^\alpha\\Y^\alpha\end{pmatrix},
\end{gather}
where $\mathcal{L}$ is the TDHF kernel. Each block acts on the set particle-hole labels, and the $A$ and $B$ matrices are
\begin{align}
\begin{split}\label{appeq:Amatrix_general}
    A_{\phi,\phi'}=&(\mathscr{E}_{\phi_\text{p}}-\mathscr{E}_{\phi_\text{h}})\delta_{{\phi_\text{p}}{\phi'_\text{p}}}\delta_{{\phi_\text{h}}{\phi'_\text{h}}}\\
    &+V_{{\phi_\text{p}}{\phi'_\text{h}},{\phi_\text{h}}{\phi'_\text{p}}}-V_{{\phi_\text{p}}{\phi'_\text{h}},{\phi'_\text{p}}{\phi_\text{h}}}\\
    B_{\phi,\phi'}=&V_{{\phi_\text{p}}{\phi'_\text{p}},{\phi_\text{h}}{\phi'_\text{h}}}-V_{{\phi_\text{p}}{\phi'_\text{p}},{\phi'_\text{h}}{\phi_\text{h}}}.
\end{split}
\end{align}
In the above expressions, the interaction tensor $V_{ij,kl}$ has been unitarily transformed to the HF basis. The matrix $A$ ($B$) is Hermitian (symmetric). 

Let the eigenvectors and eigenvalues of $\mathcal{L}$ be $w^\alpha=(X^\alpha,Y^\alpha)$ and $\Omega^a$ respectively.  Since $\mathcal{L}$ satisfies $
    \mathcal{L}=-\sigma_x \mathcal{L}^* \sigma_x$ (where $\sigma_x$ acts on the block structure), an eigenvector $(X^\alpha,Y^\alpha)$ with eigenvalue $\Omega^\alpha$ implies the existence of an eigenvector $-(Y^{\alpha*},-X^{\alpha*})$ with eigenvalue $-\Omega^{\alpha*}$. Therefore, finite real eigenvalues come in plus/minus pairs. In the quasiboson approximation, the overlap between two excited states $\ket{\alpha},\ket{\alpha'}$ is computed as
\begin{equation}\label{appeq:excited_TDHF_olap}
    \langle \alpha | \alpha'\rangle=\bra{\text{RPA}}Q_\alpha Q_{\alpha'}^\dagger\ket{\text{RPA}}=\bra{\text{RPA}}[Q_\alpha,Q_{\alpha'}^\dagger]\ket{\text{RPA}}=(X^\alpha)^\dagger X^{\alpha'}-(Y^\alpha)^\dagger Y^{\alpha'}.
\end{equation}
Since we should demand $\langle \alpha | \alpha'\rangle=\delta_{\alpha\alpha'}$ for the excited eigenstates, the physical excitation operators have coefficients $X,Y$ that satisfy $(X^\alpha)^\dagger X^{\alpha'}-(Y^\alpha)^\dagger Y^{\alpha'}=\delta_{\alpha\alpha'}$.

Related to the TDHF kernel $\mathcal{L}$ is the Hermitian Thouless stability matrix~\cite{THOULESS1960225}
\begin{equation}
    \mathcal{H}=\begin{pmatrix}
    A & B \\ B^* & A^*
    \end{pmatrix}.
\end{equation}
Negative eigenvalues of $\mathcal{H}$ signal a local instability of the HF state. The collective mode eigenvalues $\Omega^\alpha$ are only guaranteed to be real and non-negative if $\mathcal{H}$ does not have negative eigenvalues.

We now specialize the above formalism to R$L$G/hBN at $\nu=+1$, where the goal is the calculation of the charge-neutral collective mode energies $\Omega^\alpha_{\mathbf{q}}$, with $\alpha$ indexing the collective modes for a given momentum transfer $\mbf{q}$.
Here, the starting point for TDHF will always be a self-consistent HF solution for a $C=1$ state at $\nu=1$. Since this is fully spin-valley polarized, and we are considering spin-valley polarized FCIs in this work, we drop the spin-valley indices and restrict to the polarized flavor (we have checked numerically that the spin-flip and valley-flip neutral excitations are always positive semi-definite). 

Consider the HF basis where the occupied conduction band at $\nu=1$ has band index 0, and the $n_c-1$ unoccupied bands (recall that we project to $n_c$ bands per spin/valley in the HF calculation) in the same flavor are indexed by $r$. We use the indices $x,y$ to index HF bands generally, independent of if they are occupied or unoccupied. 
We express the mode creation operator as
\begin{equation}
    Q^\dagger_\alpha(\mbf{q})=\sum_{\mbf{k}}\sum_{r}\left(X^\alpha_{r}(\mbf{k},\mbf{q})\gamma^\dagger_{\text{HF},\mbf{k}+\mbf{q},r}\gamma_{\text{HF},\mbf{k},0}-Y^\alpha_{r}(\mbf{k},-\mbf{q})\gamma^\dagger_{\text{HF},\mbf{k},0}\gamma_{\text{HF},\mbf{k}-\mbf{q},r}\right),
\end{equation}
where we recall that $\gamma^\dagger_{\text{HF},\mbf{k},x}$ is the HF basis creation operator. We define $A$ and $B$ matrices according to
\begin{align}
    A_{\mbf{k},r;\mbf{k}',r'}(\mbf{q})=&\delta_{\mbf{k},\mbf{k}'}\delta_{r,r'}\big(E^\text{HF}_{r}(\mbf{k}+\mbf{q})-E^\text{HF}_{0}(\mbf{k})\big)\\
    &+\frac{1}{\Omega_\text{tot}}\sum_{\mbf{G}}\bigg[
    V(\mbf{q}+\mbf{G})\tilde{M}_{r0}(\mbf{k},\mbf{q}+\mbf{G})\tilde{M}_{0r'}(\mbf{k}'+\mbf{q},-\mbf{q}-\mbf{G})\\
    &\quad\quad\quad \quad \quad \quad -V(\mbf{k}-\mbf{k}'+\mbf{G})\tilde{M}_{rr'}(\mbf{k}'+\mbf{q},\mbf{k}-\mbf{k}'+\mbf{G})\tilde{M}_{00}(\mbf{k},-\mbf{k}+\mbf{k}'-\mbf{G})
    \bigg]\\
    =[&A_{\mbf{k}',r';\mbf{k},r}(\mbf{q})]^*
\end{align}
\begin{align}
    B_{\mbf{k},r,\mbf{k}',r'}(\mbf{q})
    &=\frac{1}{\Omega_\text{tot}}\sum_{\mbf{G}}\bigg[
    V(\mbf{q}+\mbf{G})\tilde{M}_{r0}(\mbf{k},\mbf{q}+\mbf{G})\tilde{M}_{r',0}(\mbf{k}',-\mbf{q}-\mbf{G})\\
    &\quad\quad\quad \quad \quad \quad -V(\mbf{k}-\mbf{k}'+\mbf{q}+\mbf{G})\tilde{M}_{r0}(\mbf{k}',\mbf{k}-\mbf{k}'+\mbf{q}+\mbf{G})\tilde{M}_{r'0}(\mbf{k},-\mbf{k}+\mbf{k}'-\mbf{q}-\mbf{G})
    \bigg]\\
    &=B_{\mbf{k}',r',\mbf{k},r}(-\mbf{q}).
\end{align}
where $E^\text{HF}_x(\mbf{k})$ denotes the HF eigenvalues, and the form factor in the HF basis is
\begin{equation}
    \tilde M_{xy}(\mbf{k},\mbf{q}+\mbf{G}) = \sum_{\mbf{G}',l\sigma} \tilde U^\text{HF}_{\mbf{G}', l \sigma,x}(\mbf{k}+\mbf{q}+\mbf{G})^* \tilde U^\text{HF}_{\mbf{G}' l \sigma,y}(\mbf{k}),
\end{equation}
with $\tilde U_{\mbf{G}, l \sigma,x}(\mbf{k})$ denoting the HF Bloch coefficients.

For every momentum transfer $\mbf{q}$, the Thouless stability matrix $\mathcal{H}(\mathbf{q})$ is~\cite{THOULESS1960225}
\begin{equation}
    \mathcal{H}(\mathbf{q})=\begin{bmatrix}
        A(\mbf{q}) & B(\mbf{q}) \\ B(\mbf{q})^\dagger & A^*(-\mbf{q})
    \end{bmatrix},
\end{equation}
where both $A(\mbf{q})$ and $B(\mbf{q})$ are viewed as square matrices of linear dimension $N_1N_2(n_c-1)$  (recall that the system consists of $N_1\times N_2$ moir\'e unit cells).
If $\mathcal{H}(\mathbf{q})$ has negative eigenvalues, this implies that the HF state has a local instability at wavevector $\mathbf{q}$ within the space of single Slater determinants. 

For a time-reversal invariant momentum $\mbf{q}$, we define the TDHF kernel $\mathcal{L}(\mbf{q})$ as
\begin{equation}
    \mathcal{L}_{\mbf{q}\in\text{TRIM}}=\begin{bmatrix}
        A(\mbf{q}) & B(\mbf{q}) \\ -B^*(\mbf{q}) & -A^*(\mbf{q})
    \end{bmatrix}.
\end{equation}
Let the right eigenvectors of $\mathcal{L}_{\mbf{q}\in\text{TRIM}}$
be of the form $(X,Y)^T$, where $X,Y$ have length $N_1\times N_2\times(n_c-1)$, with corresponding eigenvalue $\omega$. Note that $(Y^*,X^*)^T$ is also an eigenvector with eigenvalue $-\omega^*$. To this pair of eigensolutions, we can assign a collective mode energy $\Omega^\alpha_{\mathbf{q}}=\omega\,\text{sgn}(X^\dagger X-Y^\dagger Y)$ (see discussion around Eq.~\ref{appeq:excited_TDHF_olap}).
If $\mathcal{H}(\mathbf{q})$ is positive semi-definite, then all the $\Omega^\alpha_{\mathbf{q}}$ will be real and non-negative.

For momenta $\mathbf{q}$ which are not time-reversal invariant, the $\pm\mbf{q}$ sectors are grouped together according to
\begin{equation}
    \mathcal{L}_{\mbf{q}\notin\text{TRIM}}=\begin{bmatrix}
        A(\mbf{q}) & 0 & 0 &  B(\mbf{q}) \\ 0 & A(-\mbf{q}) & B^T(\mbf{q}) & 0 \\ 
        0 & -B^*(\mbf{q}) & -A^*(\mbf{q}) & 0
        \\ -B^\dagger(\mbf{q}) & 0 & 0 & -A^*(-\mbf{q})
    \end{bmatrix}.
\end{equation}
The block structure of the matrix above is arranged so that it acts on $(X(\mbf{q}),Y(\mbf{q}),X(-\mbf{q}),Y(-\mbf{q}))^T$. The coupling between the $+\mbf{q}$ and $-\mbf{q}$ can be seen by tracking the index structure of Eq.~\ref{appeq:Amatrix_general}. (For example, if $\phi$ corresponds to a ph pair with particle momentum $\mbf{k}+\mbf{q}$ and hole momentum $\mbf{k}$, and $\phi'$ corresponds to a ph pair with particle momentum $\mbf{k}'+\mbf{q}'$ and hole momentum $\mbf{k}'$, then $B_{\phi,\phi'}$ would be forced to vanish unless $\mbf{q}'=-\mbf{q}$ due to momentum conservation).

We can study just the outer sub-block
\begin{equation}
    \mathcal{L}_{\mbf{q},\text{outer}}=\begin{bmatrix}
        A(\mbf{q}) & B(\mbf{q}) \\ -B^\dagger(\mbf{q}) & -A^*(-\mbf{q})
    \end{bmatrix}.
\end{equation}
This is because the inner sub-block is related to the outer sub-block via
\begin{equation}
    \mathcal{L}_{\mbf{q},\text{inner}}=-\sigma_x\mathcal{L}_{\mbf{q},\text{outer}}^*\sigma_x.
\end{equation}

Let $\omega$ be an eigenvalue of $\mathcal{L}_{\mbf{q},\text{outer}}$, and $(X,Y)^T$ the corresponding eigenvector. We assign $\omega$ to a collective mode energy according to the following procedure (see discussion around Eq.~\ref{appeq:excited_TDHF_olap}):
\begin{itemize}
    \item If $\text{sgn}(X^\dagger X-Y^\dagger Y)>0$, we assign $\omega$ to $+\mbf{q}$
    \item If $\text{sgn}(X^\dagger X-Y^\dagger Y)<0$, we assign $-\omega$ to $-\mbf{q}$
\end{itemize}
If $\mathcal{H}(\pm\mathbf{q})$ is positive semi-definite, then all the $\Omega^\alpha_{\pm\mbf{q}}$ will be real and non-negative.

We define the TDHF gap $\Delta_\text{TDHF}$ as the minimum collective mode energy $\Omega^\alpha_{\mbf{q}}$ over all $\mbf{q}$. If the set of $\Omega^\alpha_{\mbf{q}}$ contains negative or complex values, then we say that $\Delta_{\text{TDHF}}<0$ and the HF state is locally unstable. In such cases, we do not assign a specific value to $\Delta_{\text{TDHF}}$, since we are interested in avoiding unstable regions anyway.

\subsection{Results}

\subsubsection{Discussion of criteria for candidate $\nu=2/3$ FCIs}\label{secapp:discussion_criteria}

In this section,
we discuss criteria for identifying promising parameter regions for $\nu=2/3$ FCIs in R$L$G/hBN, based on HF and TDHF calculations. We first consider the celebrated example of the $\nu=2/3$ fractional quantum Hall (FQH) state in spinless Landau levels. The `conventional wisdom' for realizing FCIs in narrow isolated Chern bands arises from three key aspects of the above FQH setting:
\begin{enumerate}
    \item \underline{Existence of parent Chern band:} The ground state at the nearest integer filling $\nu=1$ basically consists of fully occupying the lowest Landau level, which yields a Chern insulator with $C=1$. This provides a parent Chern band for the FQH state at fractional filling. Importantly, there is no competing  $C=0$ state that undermines the global stability of the $C=1$ parent state.
    \item \underline{Excitation gap of parent Chern band:} Excitations above the lowest Landau level are penalized by the cyclotron gap $\hbar \omega_c$. This ensures that the filled Landau level is not prone to local instabilities, and furthermore limits the amount of band-mixing expected in the $\nu=2/3$ FQH state. Actually, the relevant gap is the neutral collective mode gap, which is slightly lower than $\hbar\omega_c$ due to excitonic effects~\cite{kallin1984excitations}. However, since the cyclotron energy scales as $\propto B$ while the Coulomb interaction scales as $\propto \sqrt{B}$, it is always possible to find a magnetic field regime where the interacting correction to the neutral gap is much smaller than the cyclotron gap.
    \item \underline{Effective flatness of parent Chern band:} The lowest Landau level is exactly flat whether it is empty or fully occupied. This facilitates the stabilization of strongly-correlated FQH states (which involve appreciable occupation over the entire Chern band) and staves off more conventional phases such as Fermi liquids that are favored by strong kinetic dispersion. 
\end{enumerate}

\begin{figure*}
\centering
\includegraphics[width=0.95\linewidth]{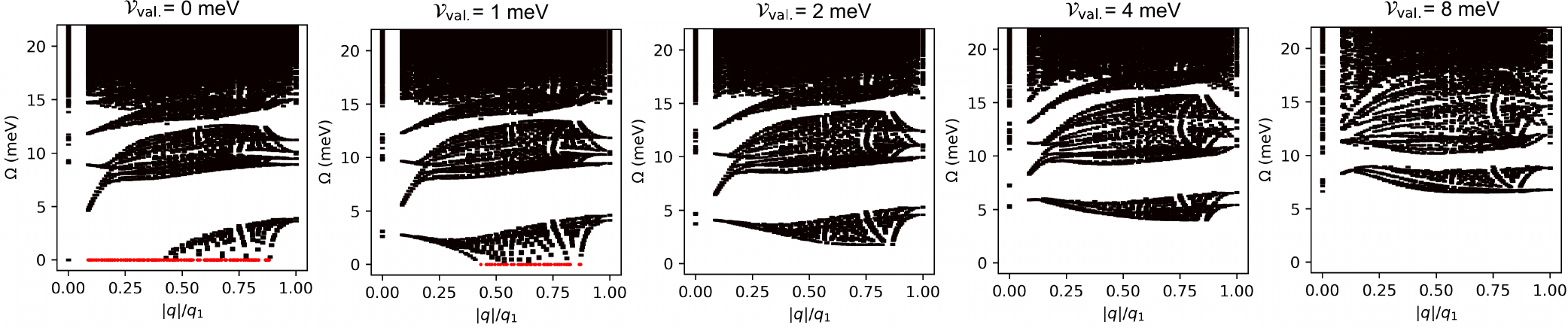}
\caption{TDHF spectrum of the $C=1$ HF state at $\nu=1$ for $V=22\,$meV and varying $\mathcal{V}_\text{val.}$, as a function of momentum transfer $|q|/q_1$. The results for multiple system sizes $L\times L$ with $L=8,9,\ldots,20$ are overlaid. Red dots indicate momenta where there exists a negative of complex TDHF eigenvalue, indicating a local instability. Parameters are $V_{tb}=32\,$meV, $\psi=(\frac{4\pi}{3}+0.12)\,$rad, and $\epsilon_r=5$. $n_c=3$ conduction bands are kept.} 
\label{fig:app_TDHF_sizes_V0.022}
\end{figure*}

\begin{figure*}
\centering
\includegraphics[width=0.95\linewidth]{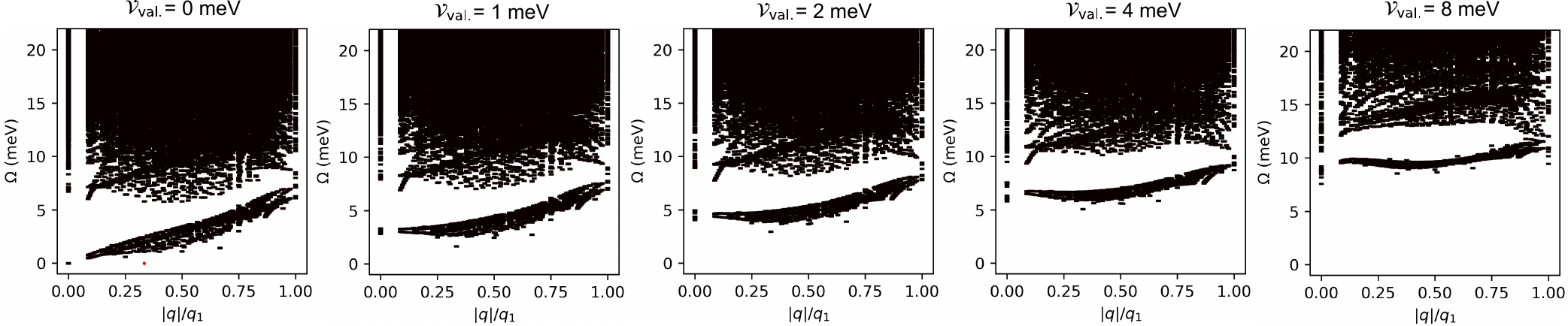}
\caption{TDHF spectrum of the $C=1$ HF state at $\nu=1$ for $V=17\,$meV and varying $\mathcal{V}_\text{val.}$, as a function of momentum transfer $|q|/q_1$. The results for multiple system sizes $L\times L$ with $L=8,9,\ldots,20$ are overlaid. Red dots indicate momenta where there exists a negative of complex TDHF eigenvalue, indicating a local instability. Parameters are $V_{tb}=32\,$meV, $\psi=(\frac{4\pi}{3}+0.12)\,$rad, and $\epsilon_r=5$. $n_c=3$ conduction bands are kept.} 
\label{fig:app_TDHF_sizes_V0.017}
\end{figure*}

We now propose how the above three aspects should be generalized or modified for $\nu=2/3$ FCIs in R$L$G/hBN:
\begin{enumerate}
    \item \underline{Existence of parent Chern band:} The lowest conduction band of non-interacting R$L$G/hBN is typically not a Chern band with the appropriate Chern number, regardless of whether the moir\'e capacitor effect is incorporated. This would seem to preclude the possibility of an FCI given the `conventional wisdom'. However, HF calculations at $\nu=1$ can yield a spin-valley-polarized $C=1$ state. The lowest band of the corresponding HF band structure is then a $C=1$ band which could serve as a parent Chern band for an FCI at fractional filling. (Note that we do not claim that the putative multi-band FCI only has support on the parent Chern band. For a parent Chern band picture to be sensible though, the multi-band FCI should still have significant weight on the parent Chern band.)
    A complication that arises in R$n$G/hBN is that a competing $C=0$ phase at $\nu=1$ can appear. For a robust FCI, the $C=1$ state at $\nu=1$ should be energetically favored compared to the $C=0$ state.

    In Fig.~\ref{fig:app_bandstruct_olap_QG}, we show example HF results for different values of $V$. In particular, while the non-interacting (`kinetic') dispersion exhibits multiple bands in close proximity, especially at the $K_M$ and $K'_M$ points, the lowest HF band is a gapped $C=1$ conduction band. We emphasize that obtaining this $C=1$ HF band requires band inversions at the mBZ corners (see discussion in App.~\ref{app:phasediagram}), as illustrated in `overlap with lowest band'. We also plot the Berry curvature and the trace of the Fubini-Study metric of the $C=1$ HF band. We note that these quantities are heavily concentrated at the mBZ corners where the band inversion has taken place and where the gap to all other bands was the smallest before Hartree-Fock (and negligible in the absence of the moir\'e capacitor effect). On the other hand, the Berry curvature and the trace of the Fubini-Study metric are smallest around $\Gamma_M$.

    \item \underline{Excitation gap of parent Chern band:} 
    While excitations above the lowest Landau level are suppressed by the large single-particle cyclotron gap, the situation for the parent Chern band R$L$G/hBN is markedly different. In the HF band structure of R$L$G/hBN at $\nu=1$, there is a finite HF gap $\Delta_{\text{HF}}$ above the parent Chern band, suggesting some level of protection against inter-band mixing effects. However, since the non-interacting band structure of R$L$G/hBN has the wrong topology and/or is gapless, $\Delta_{\text{HF}}$ is determined wholly by (a fraction of the) interactions. This means that interactions are always  larger, than $\Delta_{\text{HF}}$, meaning that we cannot \emph{a priori} be in a regime where interband effects can be neglected. 

    In fact, the situation is even worse than described above, because $\Delta_\text{HF}$ only describes the energy cost of a well-separated particle-hole pair. The actual neutral gap is suppressed due to collective modes and excitonic effects. For example, in the moir\'e translation-invariant limit, the neutral gap at $\mbf{q}=0$ is (at least) zero due to the Goldstone modes of the Wigner crystal. We estimate the neutral gap by computing the TDHF gap $\Delta_{\text{TDHF}}$ (see App.~\ref{secapp:TDHF_formalism}). While it is not possible for $\Delta_{\text{TDHF}}$ to be larger than the interaction scale, we impose the criterion that $\Delta_\text{TDHF}$ at least be non-negative. This is because $\Delta_{\text{TDHF}}<0$ indicates a local instability of the $\nu=1$ Chern insulator, which compromises the viability of the $C=1$ band as a parent band for the FCI. We anticipate that a larger positive $\Delta_{\text{TDHF}}$ will enhance the robustness of the FCI against potentially deleterious mutliband effects.

    In Figs.~\ref{fig:app_TDHF_sizes_V0.022} and \ref{fig:app_TDHF_sizes_V0.017}, we plot the TDHF collective mode spectrum for $V=22$ and $17\,$meV respectively. The results are shown for various values of $\mathcal{V}_{\text{val.}}$, including the moir\'e-less limit $\mathcal{V}_\text{val.}=0\,$meV. We combine results from multiple system sizes $L\times L$ with $L=8,9,\ldots,20$ to enable better visualization of the low-lying collective modes. For $V=22\,$meV (Fig.~\ref{fig:app_TDHF_sizes_V0.022}), the TDHF reveals local instabilities (red dots) within the lowest collective mode branch for a large region of non-zero momenta for $\mathcal{V}_\text{val.}=0,1\,$meV. As $\mathcal{V}_\text{val.}$ increases, the lowest collective mode branch gets lifted to higher energies, until $\Delta_\text{TDHF}$ and the system becomes locally stable. For a smaller interlayer potential of $V=17\,$meV (Fig.~\ref{fig:app_TDHF_sizes_V0.017}), the TDHF is now stable even in the moir\'e-less limit $\mathcal{V}=17\,$meV, though the HF bandwidth is no longer narrow (see later discussion of Fig.~\ref{fig:app_V_vs_V1_extended}). The two Goldstone phonon modes can be resolved. Note that one of the phonon modes with plasmon character has a large velocity and embeds itself within the higher collective modes already at small momentum transfer $|q|/q_1\lesssim0.1$. A finite $\mathcal{V}_\text{val.}$ gaps the phonons, and increases $\Delta_\text{TDHF}$. 

    \item \underline{Effective flatness of parent Chern band:} As explained in App.~\ref{secapp:HF_formalism}, the bandwidth in an strongly-interacting system is not uniquely defined. For example, the lowest non-interacting conduction band and the lowest HF band at $\nu=1$ can have very different dispersions for R$L$G/hBN. In App.~\ref{secapp:HF_formalism}, we motivated the definition of an effective $\nu=2/3$ bandwidth $W_{2/3}$, which is constructed by rescaling the 
    Hartree and Fock potentials at $\nu=1$ by a factor of $2/3$. We also argued that a $\nu=2/3$ FCI should be favored by a small $W_{2/3}$.

    In Fig.~\ref{fig:app_bandstruct_olap_QG}, the band structure plots (first column) illustrate the qualitative difference between the non-interacting and HF dispersions at $\nu=1$. One key aspect is the substantial lowering of the $\Gamma_M$ point relative to the $K_M,K_M'$ points for the lowest HF band. Since increasing the interlayer potential $V$ 
    pushes the $\Gamma_M$ kinetic energy upwards 
    (since the single-particle wavefunctions there are most layer-polarized), this implies that the region of minimal effective band width $W_{2/3}$ or HF band width lies at a higher $V$ than where the non-interaction dispersion is most narrow. This will be demonstrated in the phase diagrams of App.~\ref{secapp:HD_TDHF_phase_diagrams}.
\end{enumerate}

\subsubsection{Phase diagrams}\label{secapp:HD_TDHF_phase_diagrams}

\begin{figure*}
\centering
\includegraphics[width=0.95\linewidth]{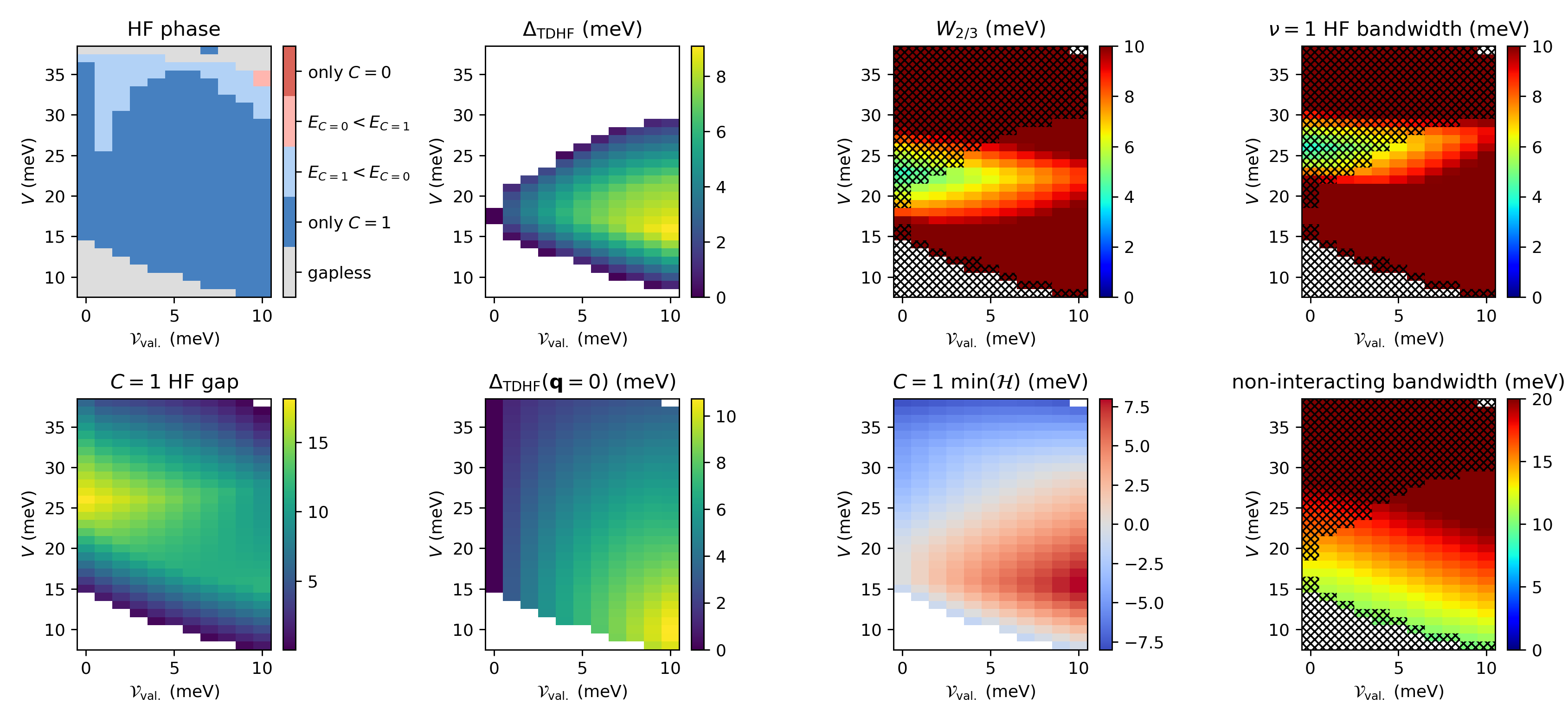}
\caption{HF and TDHF calculations as a function of $V$ and $\mathcal{V}_\text{val.}$.  For the $W_{2/3}$, `$\nu=1$ HF bandwidth' (which refers to the $C=1$ solution), and `non-interacting bandwidth' subplots, the cross-hatching indicates regions where $\Delta_{\text{TDHF}}<0$ (i.e.~the white region in the $\Delta_{\text{TDHF}}$ subplot). The color scales for these subplots are clamped at $10,10,20\,$meV respectively. The `$C=1$ min$(\mathcal{H})$' subplot denotes the lowest eigenvalue of the Thouless stability matrix across all momenta $\mbf{q}$. See text in App.~\ref{secapp:HD_TDHF_phase_diagrams} for further explanation of plots. Parameters are $V_{tb}=32\,$meV, $\psi=(\frac{4\pi}{3}+0.12)\,$rad, and $\epsilon_r=5$. System size is $12\times12$ and $n_c=3$ conduction bands are kept.} 
\label{fig:app_V_vs_V1_extended}
\end{figure*}

\begin{figure*}
\centering
\includegraphics[width=0.95\linewidth]{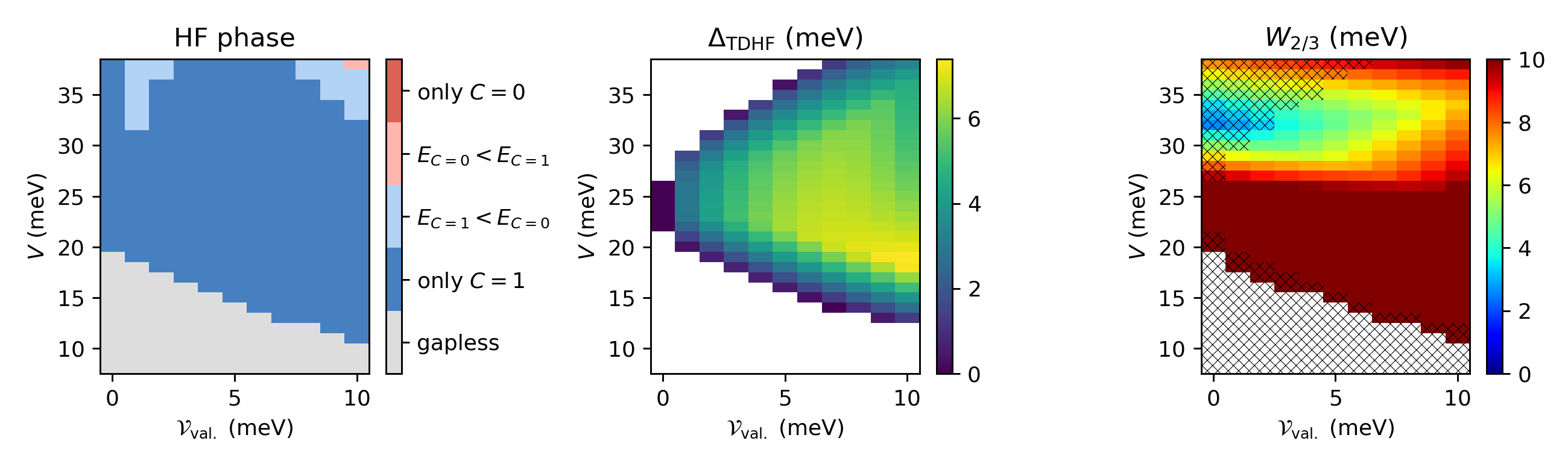}
\caption{HF and TDHF calculations as a function of $V$ and $\mathcal{V}_\text{val.}$. This figure is similar to Fig.~\ref{fig:app_V_vs_V1_extended} except that we set $V_{tb}=0$. In the `HF phase' subplot, `only $C=1$' means that the HF ground state has $C=1$, and no (metastable) $C=0$ state could be obtained. The $\Delta_{\text{TDHF}}$ subplot indicates the lowest collective mode eigenvalue $\Delta_\text{TDHF}$ of the $C=1$ state computed using TDHF. White regions indicate regions where either the $C=1$ state yields negative/complex TDHF eigenvalues, or a $C=1$ HF state could not be obtained. The $W_{2/3}$ subplot indicates the effective $\nu=2/3$ bandwidth of the $C=1$ state. Cross-hatching corresponds to the white region in the $\Delta_{\text{TDHF}}$ subplot. The color scale is clamped at $10\,$meV. System size is $12\times12$ and $n_c=3$ conduction bands are kept. Parameters are $V_{tb}=0\,$meV, $\psi=(\frac{4\pi}{3}+0.12)\,$rad, and $\epsilon_r=5$ (first row).} 
\label{fig:app_V_vs_V_Vtb0.000}
\end{figure*}

\begin{figure*}
\centering
\includegraphics[width=0.95\linewidth]{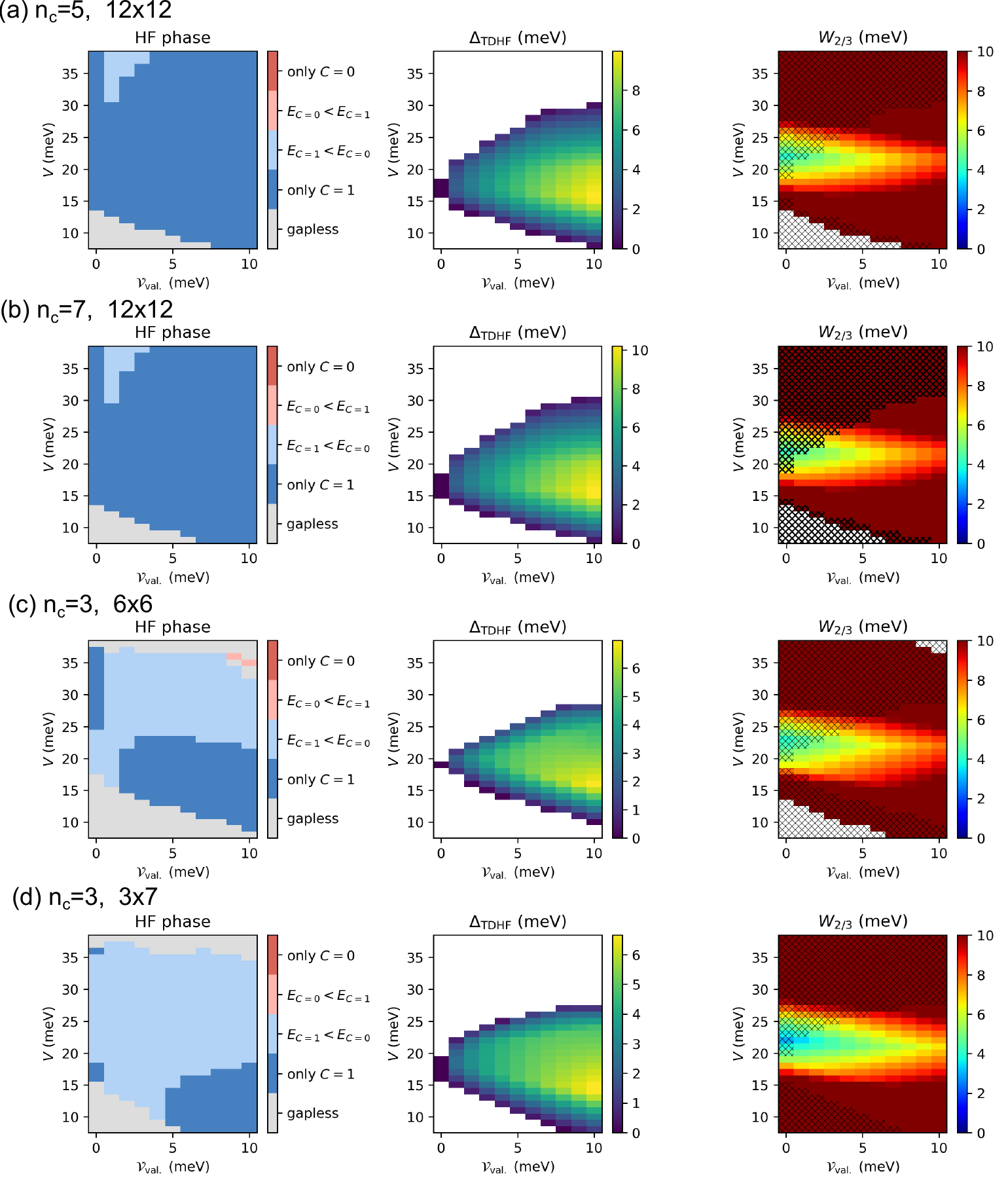}
\caption{HF and TDHF calculations as a function of $V$ and $\mathcal{V}_\text{val.}$. In the `HF phase' subplot, `only $C=1$' means that the HF ground state has $C=1$, and no (metastable) $C=0$ state could be obtained. The $\Delta_{\text{TDHF}}$ subplot indicates the lowest collective mode eigenvalue $\Delta_\text{TDHF}$ of the $C=1$ state computed using TDHF. White regions indicate regions where either the $C=1$ state yields negative/complex TDHF eigenvalues, or a $C=1$ HF state could not be obtained. The $W_{2/3}$ subplot indicates the effective $\nu=2/3$ bandwidth of the $C=1$ state. Cross-hatching corresponds to the white region in the $\Delta_{\text{TDHF}}$ subplot. The color scale is clamped at $10\,$meV. Parameters are $V_{tb}=32\,$meV, $\psi=(\frac{4\pi}{3}+0.12)\,$rad, and $\epsilon_r=5$. 
The system size is $(a)$ $12\times12$ with $n_c=5$ conduction bands, $(b)$ $12\times12$ with $n_c=7$, $(c)$ $6\times6$ and $n_c=3$, and $(d)$ tilted $3\times 7$ with $n_c=3$.} 
\label{fig:app_V_vs_V1_5band}
\end{figure*}

\begin{figure*}
\centering
\includegraphics[width=0.95\linewidth]{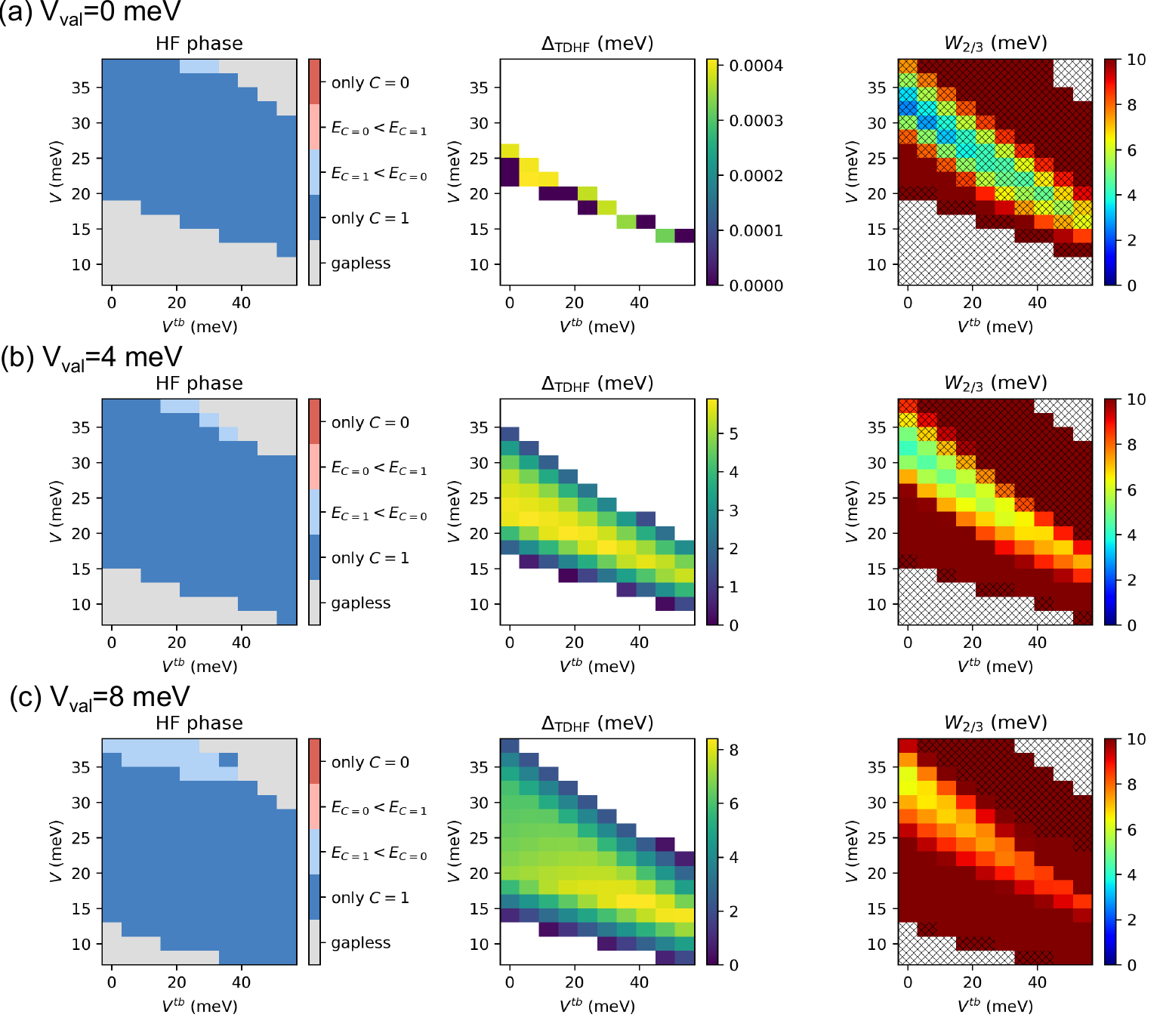}
\caption{HF and TDHF calculations as a function of $V$ and $V_{tb}$. In the `HF phase' subplot, `only $C=1$' means that the HF ground state has $C=1$, and no (metastable) $C=0$ state could be obtained. The $\Delta_{\text{TDHF}}$ subplot indicates the lowest collective mode eigenvalue $\Delta_\text{TDHF}$ of the $C=1$ state computed using TDHF. White regions indicate regions where either the $C=1$ state yields negative/complex TDHF eigenvalues, or a $C=1$ HF state could not be obtained. The $W_{2/3}$ subplot indicates the effective $\nu=2/3$ bandwidth of the $C=1$ state. Cross-hatching corresponds to the white region in the $\Delta_{\text{TDHF}}$ subplot. The color scale is clamped at $10\,$meV. System size is $12\times12$ and $n_c=3$ conduction bands are kept.
Parameters are $\psi=(\frac{4\pi}{3}+0.12)\,$rad, $\epsilon_r=5$, and $\mathcal{V}_\text{val.}=0,4,8\,$meV for $(a)$, $(b)$, $(c)$ respectively. 
} 
\label{fig:app_V_vs_V0_V10.004}
\end{figure*}

\begin{figure*}
\centering
\includegraphics[width=0.95\linewidth]{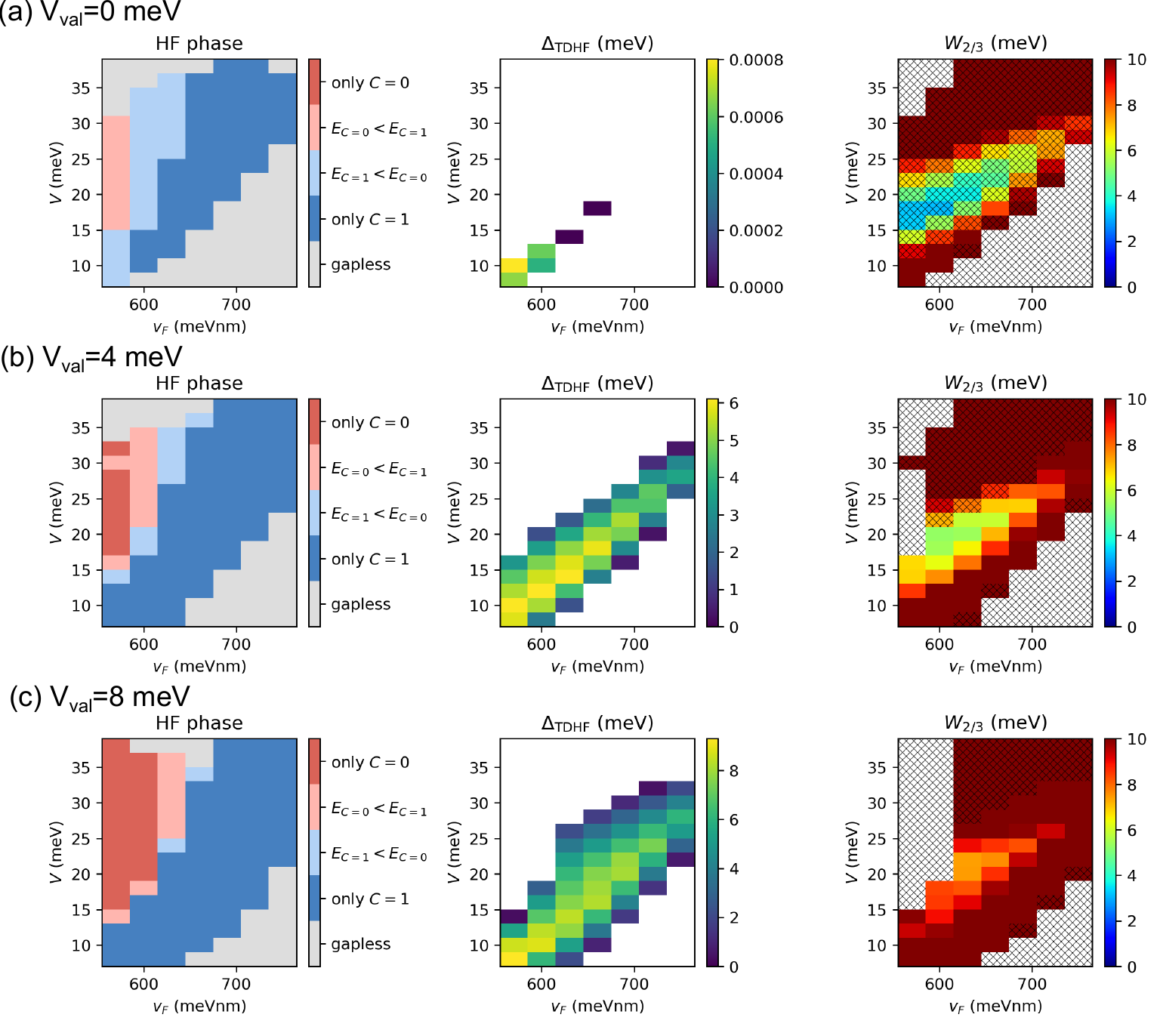}
\caption{HF and TDHF calculations as a function of $V$ and $v_F$. This figure is similar to Fig.~\ref{fig:app_V_vs_V0_V10.004} except that the horizontal axis is varying $v_F$ instead of $V_{tb}$. Here, we use a fixed value of $V_{tb}=32\,$meV, while Fig.~\ref{fig:app_V_vs_V0_V10.004} uses a fixed value of $v_F=660\,$meVnm. In the `HF phase' subplot, `only $C=1$' means that the HF ground state has $C=1$, and no (metastable) $C=0$ state could be obtained. The $\Delta_{\text{TDHF}}$ subplot indicates the lowest collective mode eigenvalue $\Delta_\text{TDHF}$ of the $C=1$ state computed using TDHF. White regions indicate regions where either the $C=1$ state yields negative/complex TDHF eigenvalues, or a $C=1$ HF state could not be obtained. The $W_{2/3}$ subplot indicates the effective $\nu=2/3$ bandwidth of the $C=1$ state. Cross-hatching corresponds to the white region in the $\Delta_{\text{TDHF}}$ subplot. The color scale is clamped at $10\,$meV. Parameters are $\psi=(\frac{4\pi}{3}+0.12)\,$rad, $\epsilon_r=5$, and  $\mathcal{V}_\text{val.}=0, 4, 8$meV for $(a)$, $(b)$, $(c)$ respectively. . System size is $12\times12$ and $n_c=3$ conduction bands are kept.} 
\label{fig:app_V_vs_vF_V10.000}
\end{figure*}

\begin{figure*}
\centering
\includegraphics[width=0.95\linewidth]{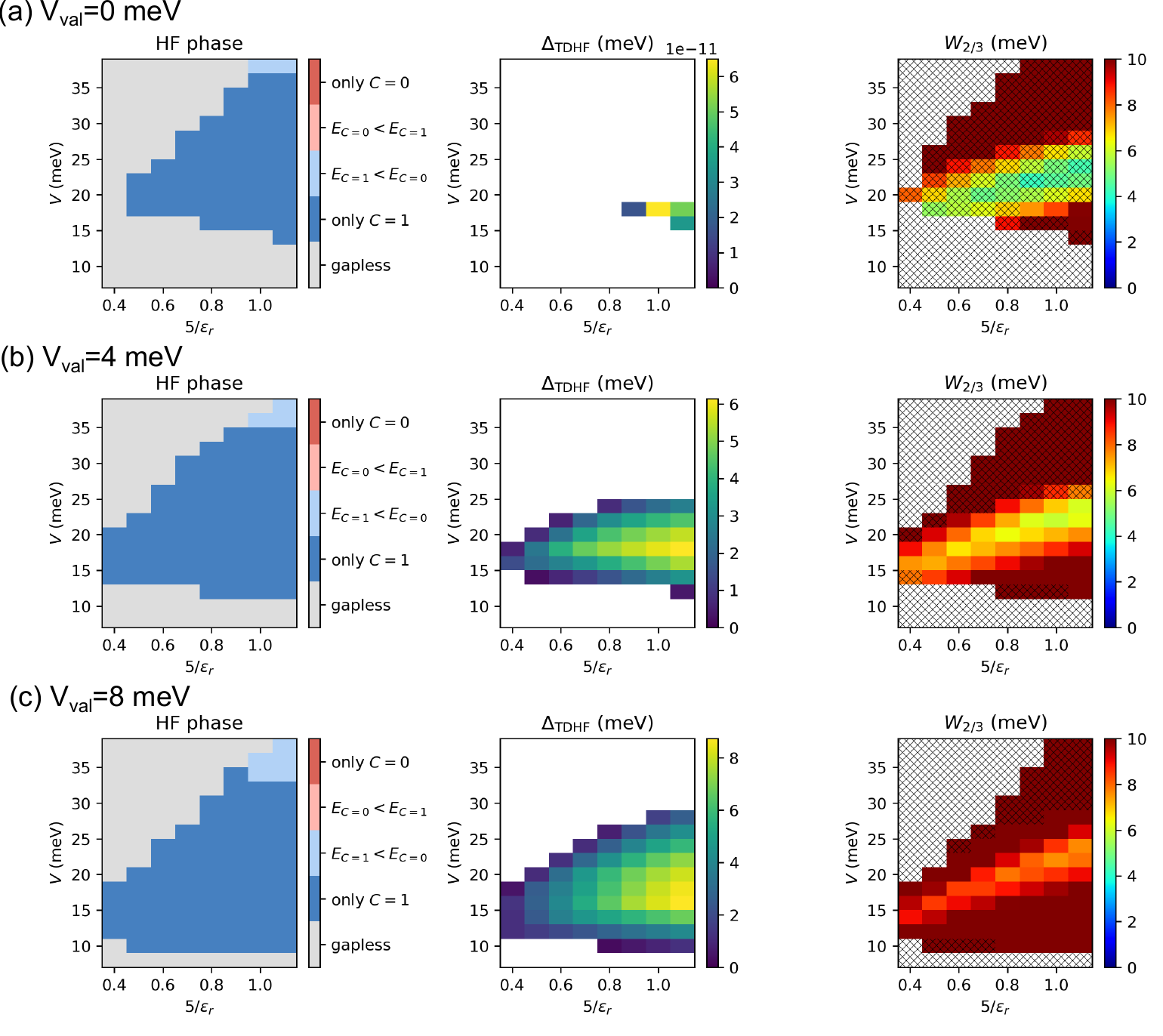}
\caption{HF and TDHF calculations as a function of $V$ and $5/\epsilon_r$. In the `HF phase' subplot, `only $C=1$' means that the HF ground state has $C=1$, and no (metastable) $C=0$ state could be obtained. The $\Delta_{\text{TDHF}}$ subplot indicates the lowest collective mode eigenvalue $\Delta_\text{TDHF}$ of the $C=1$ state computed using TDHF. White regions indicate regions where either the $C=1$ state yields negative/complex TDHF eigenvalues, or a $C=1$ HF state could not be obtained. The $W_{2/3}$ subplot indicates the effective $\nu=2/3$ bandwidth of the $C=1$ state. Cross-hatching corresponds to the white region in the $\Delta_{\text{TDHF}}$ subplot. The color scale is clamped at $10\,$meV. Parameters are $\psi=(\frac{4\pi}{3}+0.12)\,$rad, and $\mathcal{V}_\text{val.}=0, 4, 8\,$meV for $(a)$, $(b)$, $(c)$ respectively. System size is $12\times12$ and $n_c=3$ conduction bands are kept.}
\label{fig:app_V_vs_epsr_V10.004}
\end{figure*}

\begin{figure*}
\centering
\includegraphics[width=0.95\linewidth]{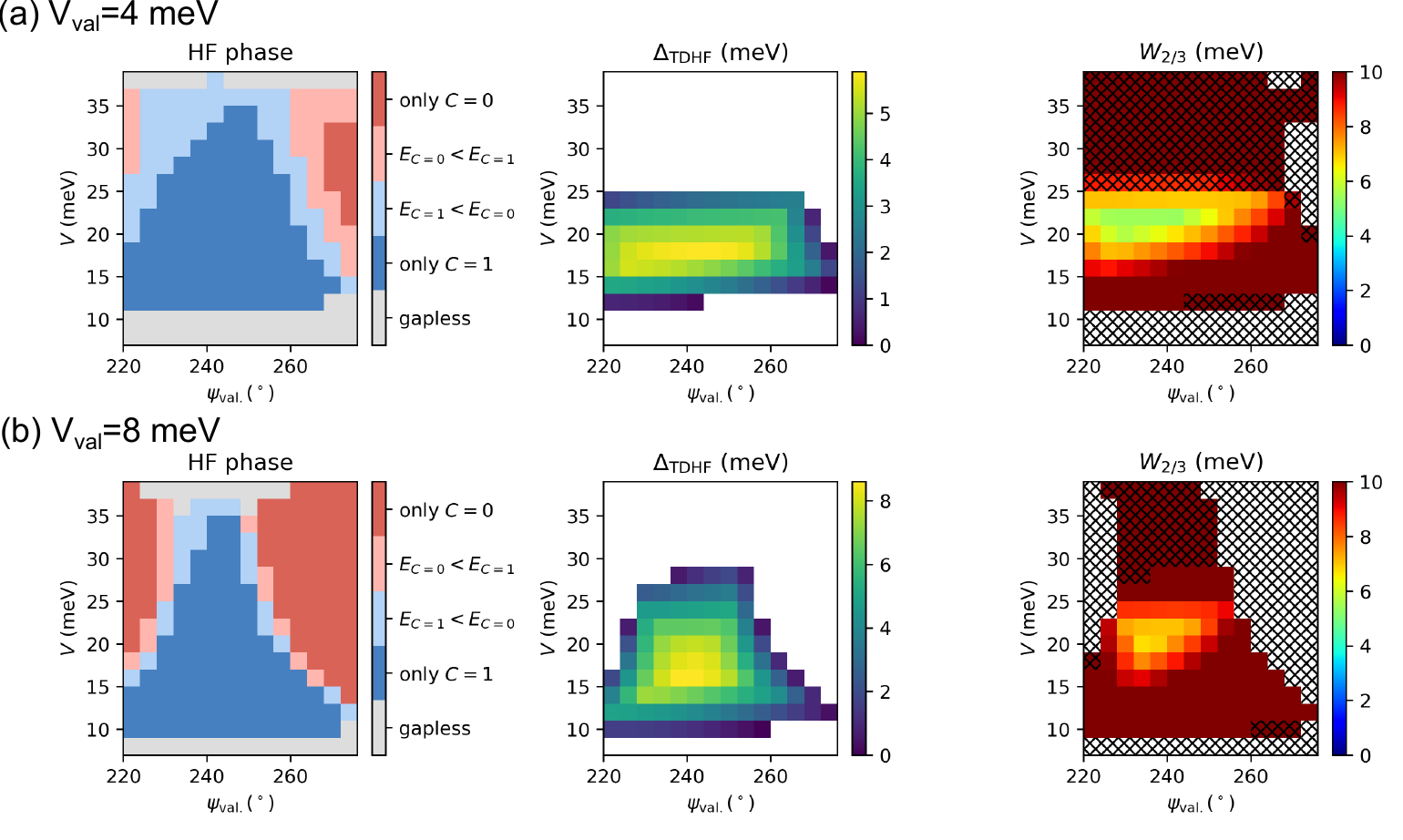}
\caption{HF and TDHF calculations as a function of $V$ and $\psi_{\text{val.}}$. In the `HF phase' subplot, `only $C=1$' means that the HF ground state has $C=1$, and no (metastable) $C=0$ state could be obtained. The $\Delta_{\text{TDHF}}$ subplot indicates the lowest collective mode eigenvalue $\Delta_\text{TDHF}$ of the $C=1$ state computed using TDHF. White regions indicate regions where either the $C=1$ state yields negative/complex TDHF eigenvalues, or a $C=1$ HF state could not be obtained. The $W_{2/3}$ subplot indicates the effective $\nu=2/3$ bandwidth of the $C=1$ state. Cross-hatching corresponds to the white region in the $\Delta_{\text{TDHF}}$ subplot. The color scale is clamped at $10\,$meV. Parameters are $\epsilon_r=5$, and $\mathcal{V}_\text{val.}=4, 8$meV for $(a),(b)$ respectively. System size is $12\times12$ and $n_c=3$ conduction bands are kept.} 
\label{fig:app_V_vs_psi_V10.004}
\end{figure*}

We first summarize
the space of parameters explored in the $\nu=1$ calculations of Figs.~\ref{fig:app_V_vs_V1_extended}-\ref{fig:app_V_vs_psi_V10.004}. Fig.~\ref{fig:app_V_vs_V1_extended} shows HF and TDHF phase diagrams as a function of $V$ and $\mathcal{V}_\text{val.}$, where we present a variety of quantities. In subsequent plots, we focus on the HF ground state phase diagram, $\Delta_{\text{TDHF}}$ and $W_{2/3}$. In Figs.~\ref{fig:app_V_vs_V_Vtb0.000} and \ref{fig:app_V_vs_V1_5band}, we examine the robustness of the results to the number of projected conduction bands, the system size, as well as switching off $V_{tb}$ (recall that $V_{tb}$, introduced in Eq.~\ref{eq:Vxifinal}, captures the alignment-independent spatially-uniform potential generated by the hBN on both sides of R$n$G). In Fig.~\ref{fig:app_V_vs_V0_V10.004}, we present phase diagrams as a function of $V$ and $V_{tb}$ for $\mathcal{V}_{\text{val.}}=0,4,8\,$meV. Fig.~\ref{fig:app_V_vs_vF_V10.000} and Fig.~\ref{fig:app_V_vs_epsr_V10.004} are the same except they are a function of $V$ and $v_F$, and $V$ vs $\epsilon_r$, respectively. Finally, Fig.~\ref{fig:app_V_vs_psi_V10.004} presents phase diagrams as a function of $V$ and $\psi_{\text{val.}}$ for $\mathcal{V}_{\text{val.}}=4,8\,$meV. We now describe the content of these figures in detail:
\begin{itemize}
    \item In Fig.~\ref{fig:app_V_vs_V1_extended}, we show the results of HF and TDHF calculations as a function of $V$ and $\mathcal{V}_\text{val.}$ for $V_{tb}=32\,$meV and $\psi=(\frac{4\pi}{3}+0.12)\,$rad on a $12\times 12$ system with three conduction bands. The `HF phase' subplot reveals that the majority of the parameter space corresponds to a $C=1$ ground state. In fact, there is only a small region where a metastable $C=0$ solution is obtained (labeled by $E_{C=0}<E_{C=1}$). For sufficiently small or large interlayer potential $V$, the system becomes gapless. For example, in the moir\'e-less limit $\mathcal{V}_\text{val.}=0$, the system is gapless if $V<15\,$meV or $V>36\,$meV.
    
    The subplots labeled $W_{2/3}$, `$\nu=1$ HF bandwidth' (which refers to the $C=1$ solution), and `non-interacting bandwidth' chart the different definitions of the bandwidth. We observe that the minimum of the non-interacting bandwidth occurs at a significantly lower $V$ than the other two definitions ($W_{2/3}$ and `$\nu=1$ HF bandwidth').
    This occurs because the Fock self-energy in HF significantly lowers the $\Gamma_M$ point relative to the mBZ corners, as shown in Fig.~\ref{fig:app_bandstruct_olap_QG}. Therefore, the non-interacting band structure needs to be in the `inverted' regime with a peak at $\Gamma_M$, in order to yield a narrow HF band. The minimum of the $\nu=1$ HF bandwidth occurs at a slightly higher $V$ than the minimum of $W_{2/3}$, because the Hartree and Fock self-energy in the latter has been reduced by a factor of 1/3. 
    
    In the bandwidth subplots, the cross-hatching indicates regions where the $C=1$ state does not exist, or is locally unstable as diagnosed by a negative eigenvalue of the Thouless stability matrix (see `$C=1$ min$(\mathcal{H})$' subplot) and negative/complex TDHF eigenvalues (see white region in the $\Delta_{\text{TDHF}}$ subplot). As shown in the $\Delta_\text{TDHF}$ subplot, the $C=1$ state at $\mathcal{V}_{\text{val.}}=0\,$meV is only locally stable for a narrow range of $V$ at low interlayer potentials. We note that this range of $V$ does not coincide with the region where the effective bandwidth $W_{2/3}$ is small. Recall that $\mathcal{V}_{\text{val.}}=0\,$meV corresponds to the moir\'e-less limit, where the $C=1$ HF state is an anomalous Hall crystal (AHC). Our results suggest that a narrow AHC Chern band is incompatible with local stability.

    On the other hand, for finite $\mathcal{V}_{\text{val.}}$, there is a sizable region where $W_{2/3}$ is small and $\Delta_{\text{TDHF}}$ is positive.

    We observe that the HF gap of the $C=1$ state does not correlate with $\Delta_{\text{TDHF}}$. Hence, inspecting the HF band structure alone does not give insight on the latter. Furthermore, restricting to the TDHF gap at $\mbf{q}=0$ (see subplot labeled $\Delta_{\text{TDHF}}(\mbf{q}=0)$) does not capture the local instabilities that occur at finite $\mbf{q}$. 
    
    In the subsequent figures, we only show the subplots for the HF phase diagram, $\Delta_{\text{TDHF}}$, and $W_{2/3}$, because these are the relevant quantities for the $\nu=2/3$ FCI stability criteria outlined in App.~\ref{secapp:discussion_criteria}.

    \item Fig.~\ref{fig:app_V_vs_V_Vtb0.000} is analogous to Fig.~\ref{fig:app_V_vs_V1_extended}, except we use $V_{tb}=0\,$meV. The region where $W_{2/3}$ is small and $\Delta_\text{TDHF}\geq0$ shift together to larger $V$ (see also Fig.~\ref{fig:app_V_vs_V0_V10.004} which shows results as a function of $V$ and $V_{tb}$). Apart from this shift, the results are qualitatively the same as for $V_{tb}=32\,$meV. This suggests that, at least at the level of our proposed FCI criteria, $V_{tb}$ does not play a critical role. As indeed demonstrated later in the ED calculations of App.~\ref{app:ED_23} (see e.g. Fig.~\ref{fig_EDoldparascan}$(b)$), the region of stable FCIs mainly exhibits a simple shift towards larger $V$ as $V_{tb}$ is reduced.

    \item Compared to Fig.~\ref{fig:app_V_vs_V1_extended}, the first and second rows of Fig.~\ref{fig:app_V_vs_V1_5band} are computed using $n_c=5$ and $n_c=7$ conduction bands respectively. The results are quantitatively similar, especially regarding the region where $\Delta_{\text{TDHF}}\geq0$ and $W_{2/3}$ is small. This suggests that projection to 3 bands is reliable for the parameters here.

    \item In the third and fourth rows of Figs.~\ref{fig:app_V_vs_V1_5band}, we project to 3 conduction bands, but consider a $6\times 6$ and tilted $3\times 7$ system respectively. Details of the tilted $3\times 7$ geometry can be found in App.~\ref{app:ED_23}. We find a good qualitative agreement between calculations on the $6\times 6$, tilted $3\times 7$, and $12\times 12$ (Fig.~\ref{fig:app_V_vs_V1_extended}) systems, suggesting that finite-size effects are not severe. We note that the smallest $3\times 7$ system slightly underestimates $W_{2/3}$, which is attributed to the coarser sampling of the mBZ. Furthermore, the $3\times 7$ system can obtain the metastable $C=0$ state over a larger parameter region compared to the larger system sizes. This is likely because the $C=0$ and $C=1$ states have a finite energy difference in the thermodynamic limit, which translates to a lower total energy difference for smaller systems. 

    \item Fig.~\ref{fig:app_V_vs_V0_V10.004} shows the HF phase diagram, $\Delta_\text{TDHF}$, and $W_{2/3}$ as a function of $V$ and $V_{tb}$ for $\mathcal{V}_\text{val.}=0,4,8\,$meV. For all values of $\mathcal{V}_\text{val.}$, the region of small $W_{2/3}$ consistently occupies a diagonal stripe that points down and to the right in the $V-V_{tb}$ plane. In the moir\'e-less limit $\mathcal{V}_\text{val.}=0\,$meV, the region of local stability ($\Delta_\text{TDHF}\geq 0$) also occupies a stripe, but this does not overlap with the region of narrow effective bandwidth. As $\mathcal{V}_\text{val.}$ increases, the region of local stability expands and overlaps with the region where $W_{2/3}$ is small.

    \item Fig.~\ref{fig:app_V_vs_vF_V10.000} shows the HF phase diagram, $\Delta_\text{TDHF}$, and $W_{2/3}$ as a function of $V$ and $v_F$ for $\mathcal{V}_\text{val.}=0,4,8\,$meV. We observe that larger $v_F$ relatively favors the $C=1$ state over the $C=0$ state. This trend can be motivated from the moir\'e-less limit where the Berry curvature of the lowest R$L$G conduction band near the Dirac point scales as $\sim v_F^2$. A larger Berry curvature relatively favors the Chern insulator, as has been demonstrated in previous theoretical studies~\cite{dong2024stabilityahc,bernevig2025berrytrashcanmodelinteracting,crepel2024efficient,soejima2024anomalous}. We are unable to find a parameter region for $\mathcal{V}_\text{val.}=0$ where the Chern band is relatively flat and locally stable.

    \item Fig.~\ref{fig:app_V_vs_epsr_V10.004} shows the HF phase diagram, $\Delta_\text{TDHF}$, and $W_{2/3}$ as a function of $V$ and $5/\epsilon_r$ (proportional to the interaction strength) for $\mathcal{V}_\text{val.}=0,4,8\,$meV. As the interaction strength gets weaker, the window of interlayer potential corresponding to the $C=1$ HF phase gets narrower. Again, we are unable to find a parameter region for $\mathcal{V}_\text{val.}=0$ where the Chern band is relatively flat and locally stable. 

    \item Fig.~\ref{fig:app_V_vs_psi_V10.004} shows the HF phase diagram, $\Delta_\text{TDHF}$, and $W_{2/3}$ as a function of $V$ and $V_{tb}$ for $\mathcal{V}_\text{val.}=4,8\,$meV. We do not show results for $\mathcal{V}_{\text{val}.}=0\,$meV since they are independent of $\psi_\text{val.}$ in the moir\'e-less limit. We note that the results are periodic under $\psi_{\text{val.}}\rightarrow\psi_{\text{val.}}+120^\circ$. At finite $\mathcal{V}_\text{val.}$, the $C=1$ HF phase is the ground state for a window of $\psi_\text{val.}$ centered around $\simeq 240^\circ$. As $\mathcal{V}_\text{val.}$ increases, this window gets narrower as the $C=0$ phase occupies a larger parameter region.
\end{itemize}

\clearpage
\newpage

\section{ED results}
\label{app:ED_23}

In this appendix, we summarize the multi-band ED techniques introduced in Ref.~\cite{PhysRevB.112.075110}. We then present additional ED spectra and phase diagrams with different parameters than those used in the Main Text.

\subsection{Construction and truncation of the Multi-Band ED Hilbert Space}

\begin{figure}
\centering
\includegraphics[width=0.9\columnwidth]{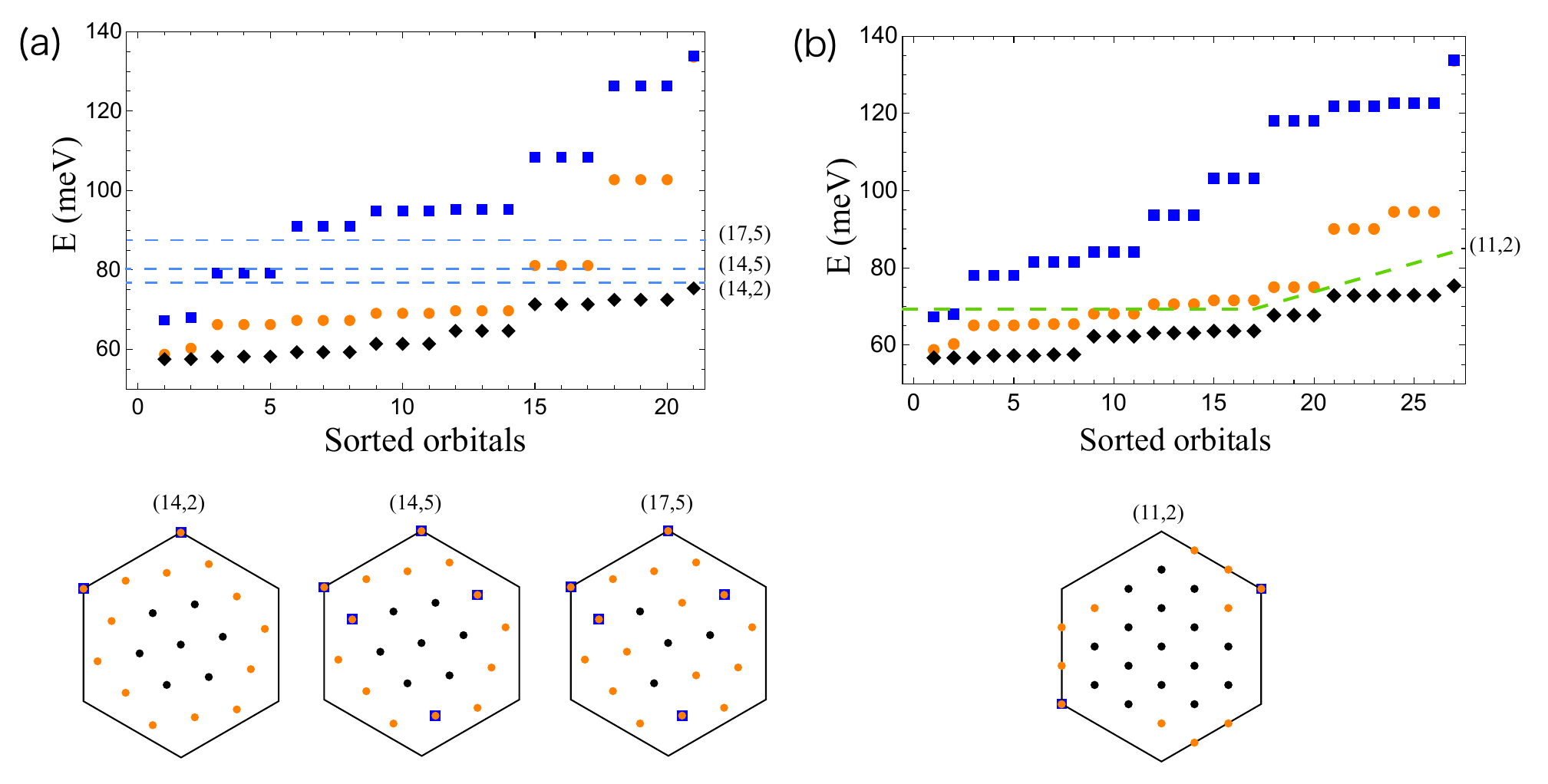}
\caption{ (a) Single-particle levels of $H^{\text{con. sp.}}$ sorted in order of energy for the $3\times 7$ system with $(N_1,N_2,\tilde n_{11},\tilde n_{12},\tilde n_{21},\tilde n_{22})=(3,7,1,1,1,2)$. The three bands are indicated by different colors with black, orange, blue representing band 1,2,3 respectively. Each truncation $(n_2^{\text{orb}},n_3^{\text{orb}})$ contains the states below the corresponding dashed line. The bottom row shows the $k$ points in the mBZ with the various orbital truncations. $k$ points where one/two/three bands are kept are shown in black/orange/blue. (b) is the same as (a) but in a $9\times 3$ system with $(N_1,N_2,\tilde n_{11},\tilde n_{12},\tilde n_{21},\tilde n_{22})=(9,3,1,-2,0,1)$.   }
\label{fig_EDorbital}
\end{figure}

We perform three-band ED computations on the Hamiltonian $H^{\text{con. sp.}} + H^{\text{con. int.}}$ (see \App{app:HF}) and include the lowest 3 bands in HF basis to capture the effect of band-mixing. The momentum mesh for ED is chosen as follows. Let $\bsl b_{M1}$ and $\bsl b_{M2}$ be the shortest moir\'e reciprocal lattice vectors, with $\bsl b_{M2}=R(\frac{2\pi}{6})\bsl b_{M1}$, where $R(\theta)$ is a counterclockwise rotation matrix. The momentum points in the mBZ of an $N_1\times N_2$ system are given by
\bea
\bsl k=\frac{k_1}{N_1}\bsl f_1+\frac{k_2}{N_2}\bsl f_2, 
\eea
where $k_i=0,1,...,N_i-1$. The reciprocal lattice vectors $\bsl f_1$ and $\bsl f_2$ are defined as
\bea
\bsl f_1=\tilde n_{11}\bsl b_{M1}+\tilde n_{12}\bsl b_{M2}, \quad
\bsl f_2=\tilde n_{21}\bsl b_{M1}+\tilde n_{22}\bsl b_{M2}, 
\eea
where $\tilde n_{ij}$ are integers and $\det[\bar{n}_{ij}] = 1$ so that the $\tilde{n}_{ij}$ represent an $SL(2,Z)$ transformation. The momentum mesh is hence specified by $N_1, N_2, \tilde n_{ij}$. 

In a multi-band system, the band degree of freedom exponentially increases the dimension of the Hilbert space. To make the ED computation manageable, we apply a truncation to the Hilbert space by limiting the particle number and allowed single-particle states in the higher bands~\cite{PhysRevB.112.075110,zlqg-sj86}. We represent the truncation of the Hilbert space by the four truncation parameters:
\bea
(n^{\text{orb}}_2,n^{\text{orb}}_3;\ n_2,n_3).
\label{trunconvention}
\eea
Here $n^{\text{orb}}_i$ is the number of single-particle states (orbitals) that are kept in band $i=2,3$, and $n_2,n_3$ are maximum allowed particle number in band 2 and band 3 respectively.  This truncation scheme is useful when the higher bands are dispersive, and $k$-points in some region of the mBZ are much higher in energy than others, and so can safely be removed. This is the case in R5G/hBN as shown in \Fig{fig_EDorbital}. In this case, we choose to truncate $k$-points whose bare kinetic energy in $\bar{H}_{con}$ is above a given threshold. (Note that this preserves $C_3$ symmetry.)

Next, we filter the many-body basis to remove basis states which have more than $n_i$ particles in band $i$. No truncation in orbital or particle number is imposed on the lowest band $i=1$. The case with $n_2=n_3=0$ represents the single-band limit in which the ED computation is fully within the lowest band. Hilbert space dimensions for different truncation levels may be found in Ref.~\cite{PhysRevB.112.075110}. 

The above truncation procedure can be performed in any orthonormal band basis, e.g. the HF basis or the non-interacting band basis. In our case, we choose to work in the basis of the HF Hamiltonian at $\nu=1$ so that the full many-body Hilbert space of the $C=1$ band (the lowest band labeled by $i=1$) is kept. Note that without any truncation, the choice of basis does not affect the ED calculation. Hence a converged ED calculation in the HF basis will be unbiased, whereas a 1-band ED calculation in the HF basis, which neglects interband fluctuations, may be biased towards FCIs~\cite{PhysRevB.112.075110}. 

The $3\times 7$ momentum mesh used in Fig.~\ref{fig_EDbandmax} in the Main Text has $(N_1,N_2,\tilde n_{11},\tilde n_{12},\tilde n_{21},\tilde n_{22})=(3,7,1,1,1,2)$. At 2/3 filling in this lattice, the three FCI momenta are at $\Gamma_M,K_M,K'_M$ with $k_1+k_2N_1=0,1,2$. 

\subsection{ED Spectra at filling $\nu=1$}

\begin{figure}
\centering
\includegraphics[width=\columnwidth]{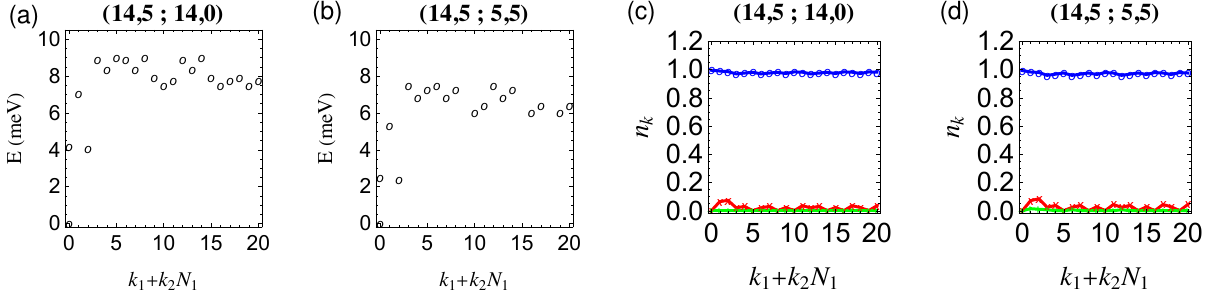}
\caption{ ED energy spectrum (a,b) and occupation number (c,d) for $3\times 7$ system at $\nu=1$ under different truncations (see Eq.\eqref{trunconvention}) with $V=22$ meV, $V_{tb}=32$ meV, $\mathcal{V}_{\text{val.}}=8$ meV, $\psi=(\frac{4\pi}{3}+0.12)\,$rad, and $\epsilon_r=5$. In the occupation number plots, the blue, red and green curves represent the occupation number calculated from the lowest many-body state in the three HF bands 1,2,3 respectively. The occupation is predominantly within the lowest HF Chern band.   }
\label{fig_EDnu1}
\end{figure}

\begin{figure}
\centering
\includegraphics[width=0.6\columnwidth]{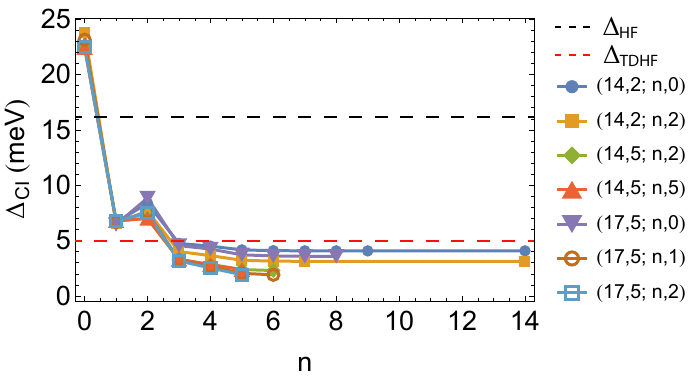}
\caption{ Energy gap above the CI ground state at $\nu=1$ for $3\times 7$ system calculated with different truncation parameters for $V=22$ meV, $V_{tb}=32$ meV, $\mathcal{V}_{\text{val.}}=8$ meV, $\psi=(\frac{4\pi}{3}+0.12)\,$rad, and $\epsilon_r=5$. We have considered truncations that lead to Hilbert space dimension up to $7\times 10^8$. Some intermediate truncation $n$ values are dropped whenever the convergence with largest $n=14$ has already been reached. The black and red horizontal dashed lines represent the HF indirect gap $\Delta_{\text{HF}}$ and the TDHF gap $\Delta_{\text{TDHF}}$ respectively, on the same system size. }
\label{fig_EDnu1bm}
\end{figure}

We first examine the robustness of the parent state at filling $\nu=1$ with $V=22\,$meV, $V_{tb}=32\,$meV, $\mathcal{V}_{\text{val.}}=8$ meV, $\psi_\text{val.}=(\frac{4\pi}{3}+0.12)\,$rad and interaction parameters $\epsilon_r=5$ and $d_\text{sc}=10\,$nm. The HF calculation in Fig.~\ref{fig_maintext_bandstruct_colleigs} shows the lowest HF band has $C=1$ with a 5 meV TDHF gap. In Fig.~\ref{fig_EDnu1} we show the results of multi-band ED calculation in this $3\times 7$ system at $\nu=1$ as a test of HF/TDHF.  Fig.~\ref{fig_EDnu1}(a,c) shows the ED spectrum and one-body density matrix $\rho(\mbf {k})_{mn}=\braket{\tilde{\gamma}^\dag_{\mbf{k}m}\tilde{\gamma}_{\mbf{k}n}}$ for a calculation keeping 14 orbitals and allowing up to 14 particles in the second band (but no particles in the third band). We observe an isolated many-body ground state at $k_1=k_2=0$ separated by a 4meV gap from excited states, and hence is in good agreement with the HF/TDHF. The 1-body density matrix shows that the the $C=1$ HF band is at least $90\%$ occupied at all $k$ points, with maximum occupation less than $10\%$ in the second band at the $K_M,K'_M$ points at the edge of the mBZ. In Fig.~\ref{fig_EDnu1}(b,d), we allow an additional 5 momentum points in the third band, and set $n_2=5$ and $n_3=5$. The ED spectra and density matrix are nearly unchanged. We further investigate the dependence of the gap above the CI ground state on band mixing under different truncation parameters in Fig.~\ref{fig_EDnu1bm}, where we have included truncations that lead to Hilbert space dimension lower than $7\times 10^8$. In the largest Hilbert space of dimension $6.85\times 10^8$ with truncation $(14,5;5,5)$ that we have calculated, the overlap between the multi-band ED ground state $|\Psi_{\text{ED}}\rangle$ and the $C=1$ HF ground state $|\Psi_{\text{HF}}\rangle$ is $|\langle \Psi_{\text{HF}}|\Psi_{\text{ED}}\rangle|^2=0.71$. The gap converges as band mixing increases, indicating a stable CI ground state. This analysis validates our HF/TDHF approach at integer filling.

\subsection{ED Spectra at filling $\nu=2/3$}

\begin{figure}
\centering
\includegraphics[width=\columnwidth]{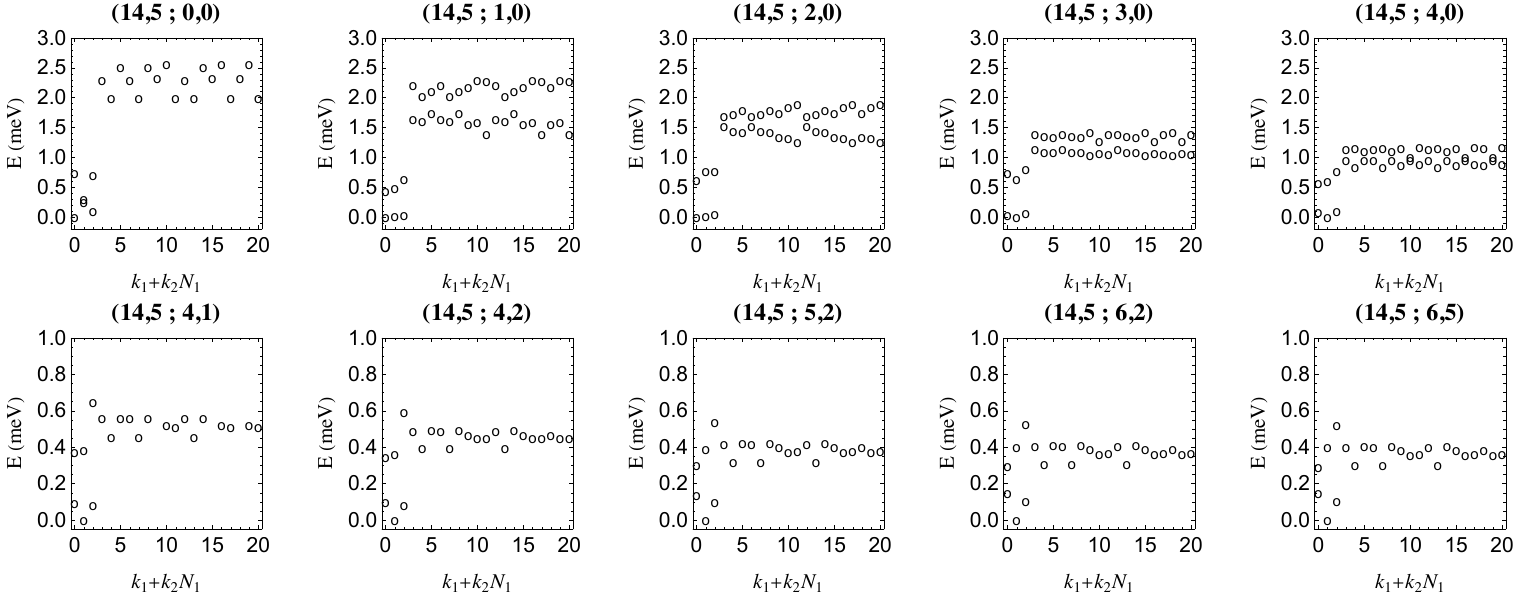}
\caption{ ED energy spectrum for $3\times 7$ at 2/3 filing under different truncations (see Eq.\eqref{trunconvention}) with $V=22$ meV, $V_{tb}=32$ meV, $\mathcal{V}_{\text{val.}}=8$ meV, $\psi=(\frac{4\pi}{3}+0.12)\,$rad, and $\epsilon_r=5$. The FCI momenta are at $k_1+k_2N_1=0,1,2$. The largest Hilbert space dimension in ED reaches $10^9$ and the FCI gap remains finite under large band mixing. }
\label{fig_ED2v3energy}
\end{figure}

\begin{figure}
\centering
\includegraphics[width=\columnwidth]{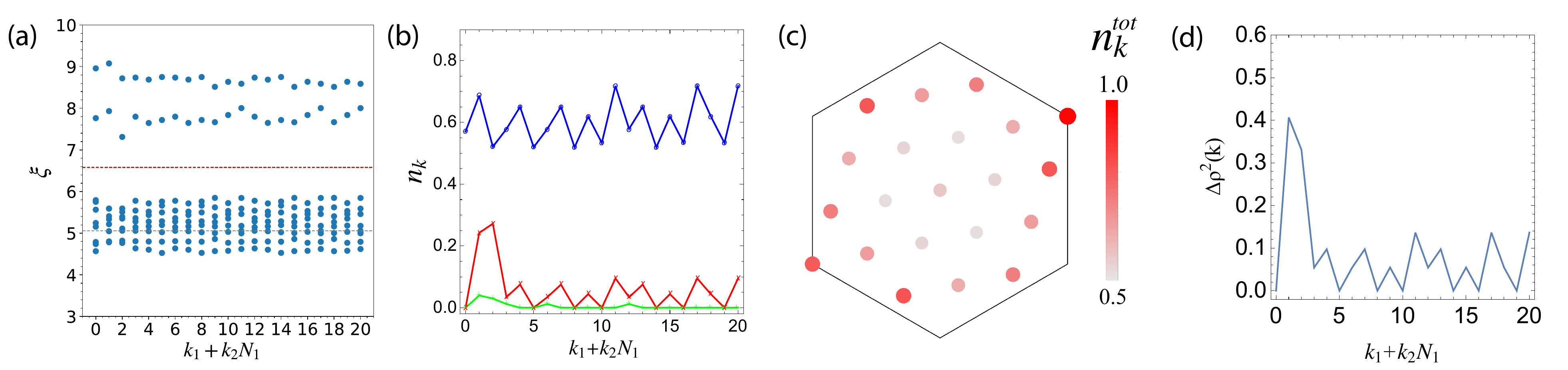}
\caption{ Physical quantities calculated for the ED ground states in $3\times 7$ system at 2/3 filing under truncation $(n^{\text{orb}}_2,n^{\text{orb}}_3;\ n_2,n_3)=(14,5; 6,5)$ with $V=22$ meV, $V_{tb}=32$ meV, $\mathcal{V}_{\text{val.}}=8$ meV, $\psi=(\frac{4\pi}{3}+0.12)\,$rad, and $\epsilon_r=5$. The FCI momenta are at $k_1+k_2N_1=0,1,2$. (a) PES calculated with $N_A=2$. The red line indicates the FCI counting with 168 states below the gap. (b) Occupation number for the same parameters. The blue, red and green symbols represent the occupation number calculated from the lowest many-body state in the three HF bands 1,2,3 respectively, which are the diagonal elements of the one-body density matrix $\rho(\mbf{k})$. The lines are plotted by connecting eigenvalues of $\rho(\mbf{k})$. The diagonal elements and the eigenvalues of $\rho(\mbf{k})$ are almost identical. (c) Total occupation number at different momentum points for the same parameters. (d) Deviation from a one-band FCI measured via $\Delta\rho^2(\mbf{k})$. }
\label{fig_ED2v3PESnk}
\end{figure}

Next we show the ED energy spectrum at $\nu=2/3$ in the same system in Fig.~\ref{fig_ED2v3energy} at various truncation parameters. First, for these parameters, the $n_2,n_3 = 0$ truncation (1-band ED) shows 6 low-energy states with no clear gap. When $n_2 =1,\dots,4$, we see a gap open and 3 states at the correct FCI momentum become nearly degenerate (we will discuss more evidence that these states are an FCI in \Fig{fig_ED2v3PESnk} momentarily). The spectra appear to converge at $n_2 =4$ in the 14,5 truncation shown. Then we consider the effect of non-zero $n_3$. For all the $n_2,n_3 >0$ shown in Fig.~\ref{fig_ED2v3energy}, we observe increased splitting of the FCI manifold and a decrease in the gap. This shows that $n_3$ makes the FCI more sensitive to finite size effects and decreases the gap. However, we see from Fig.~\ref{fig_ED2v3energy} that these effects appear to converge at our largest calculation $(n_2,n_3) = (6,5)$ with Hilbert space dimension $0.958\times 10^9$, and are consistent with a finite-size FCI spectrum.

We now discuss additional diagnostics to establish that this state is indeed an FCI (as opposed to, for instance, a charge density wave (CDW)). To do so, we compute the particle entanglement spectrum (PES)~\cite{PhysRevLett.101.010504-2008,PhysRevLett.106.100405,regnaultbernevig} of the three topological ground states. Recall that at filling $\nu=2/3$, the PES calculated directly from the ED ground states does not obey the counting of the $1/3$ Laughlin state~\cite{zlqg-sj86} since particle-hole symmetry acts non-trivially on the density matrices. Hence we follow Refs.~\cite{zlqg-sj86,PhysRevLett.133.206502_2024,PhysRevB.110.115146-2024} to first project the ground states to the lowest band and perform a particle-hole (PH) transformation to obtain the corresponding many-body state at $1/3$ filling. We compute the PES in these PH-transformed states. The computation is done by dividing the particles into two subsets with $N_A$ and $N_B$ particles, and then we compute the eigenvalues $e^{-\xi_i}$ of the reduced density matrix $\rho_A=\text{Tr}_B\left[\frac{1}{N_{\text{GS}}}\mathcal{P}_{\text{GS}}\right]$, where $\mathcal{P}_{\text{GS}}=\sum_{a\in \{\text{GS}\}} | \Psi_a \rangle \langle \Psi_a |$ is the projection to the PH-transformed ground state manifold $\{\text{GS}\}$ with $N_{\text{GS}}=3$. In the FCI phase, the PES spectrum will be gapped, and the total number of states below the gap for the $3\times 7$ system with $N_A=2$ is 168~\cite{regnaultbernevig}. For a CDW phase, the number of states below the PES gap with $N_A=2$ will be 63~\cite{BernevigNicolas2012thintorus}, and the CDW can be further identified by the peaks in the structure factor $S(\bsl q)=\frac{1}{N_1 N_2}(\langle \hat \rho_{\bsl q} \hat \rho_{-\bsl q}\rangle-\langle \hat \rho_{\bsl q} \rangle \langle \hat \rho_{-\bsl q} \rangle)$~\cite{Wilhelm2021FCITBG} at $K_M$ and $K'_M$ momenta.

As shown in Fig.~\ref{fig_ED2v3PESnk}(a), the PES at large band-mixing indeed has a gap indicated by the red line with 168 states below it, which confirms the FCI nature of the ground state. We also plot the occupation number in Fig.~\ref{fig_ED2v3PESnk}(b,c), which are the diagonal elements of the one-body density matrix $\rho(\mbf{k})_{mn}=\frac{1}{N_{\text{GS}}}\sum_{a\in \text{\{GS\}}}\braket{a|\tilde{\gamma}^\dag_{\mbf{k}m}\tilde{\gamma}_{\mbf{k}m}|a}$. Contrary to the usual setting of an FCI in a single isolated band, the FCI here has significant occupation of the higher bands at the $K_M, K'_M$ points ($k_1 + N_1 k_2 = 1,2$) where the gap is the smallest, demonstrating the importance of band-mixing for accurate calculations of the phase diagram. We further plot the quantity $\Delta\rho^2(\mbf{k})\equiv \text{Tr}[\rho(\mbf{k})]^2-\text{Tr}[\rho(\mbf{k})^2]$ in Fig.~\ref{fig_ED2v3PESnk}(d) as a measure of the deviation of the given FCI state from a one-band FCI. If there exists a one-body basis in which the FCI ground state is fully made of particles in a single band, then $\Delta\rho^2(\mbf{k})$ will be zero. Therefore the value of $\Delta\rho^2(\mbf{k})$ quantifies how much a given FCI state deviates from a one-band FCI.

\begin{figure}
\centering
\includegraphics[width=0.8\columnwidth]{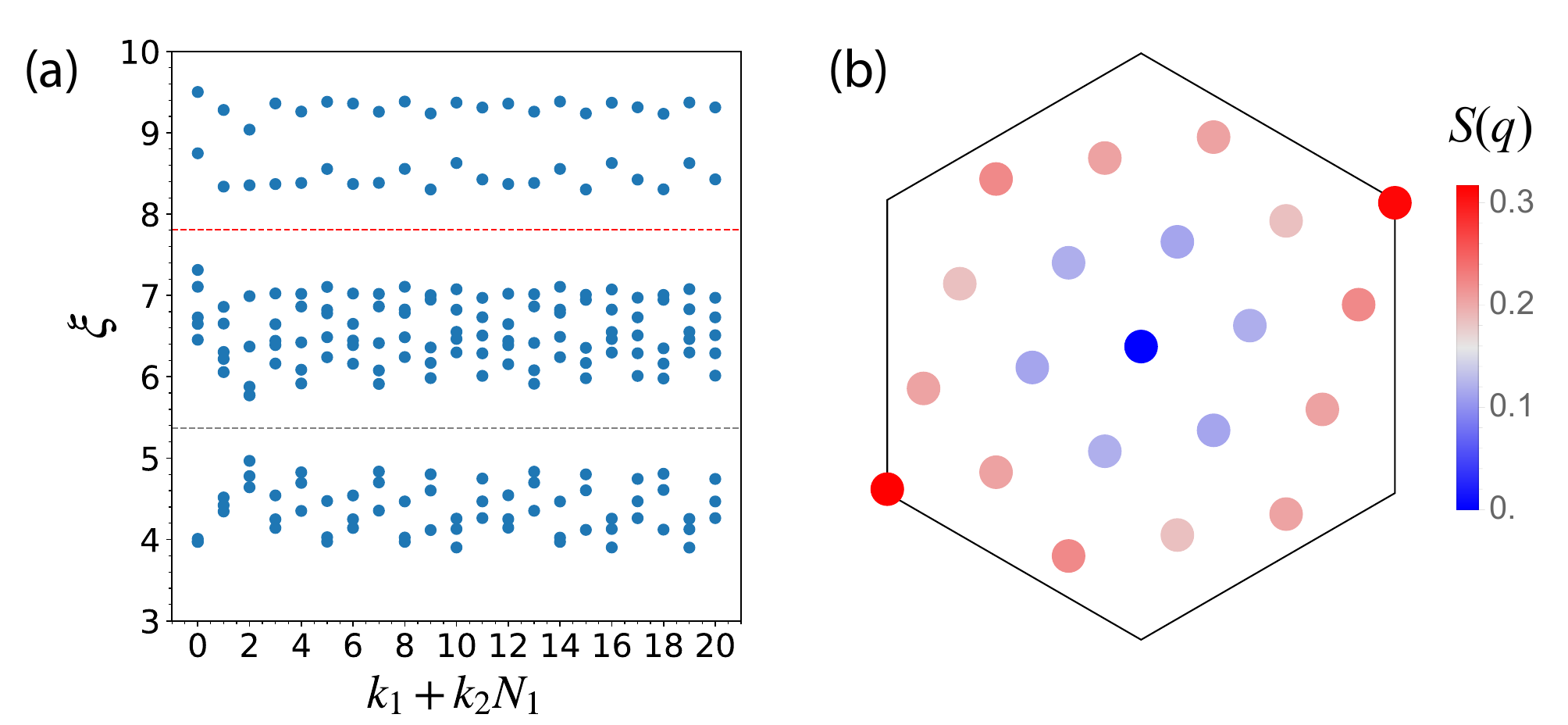}
\caption{ (a) PES calculated with $N_A=2$ for $3\times 7$ at 2/3 filing under truncation $(n^{\text{orb}}_2,n^{\text{orb}}_3;\ n_2,n_3)=(14,5; 0,0)$, namely in the absence of band-mixing, with $V=22$ meV, $V_{tb}=32$ meV, $\mathcal{V}_{\text{val.}}=8$ meV, $\psi=(\frac{4\pi}{3}+0.12)\,$rad, and $\epsilon_r=5$. The PES is calculated using the lowest three states at momenta $k_1+k_2N_1=0,1,2$ (one state per momentum). The red line indicates the FCI counting with 168 states below the gap, and the black line labels the CDW counting with 63 states below the gap. (b) The structure factor $S(\bm q)$ calculated from the same three states under the same parameters. The peaks at $K_M,\ K'_M$ points are consistent with CDW.  }
\label{fig_PES_3b7_oneband}
\end{figure}

For the one-band calculation without band mixing, by contrast, the gap between the third and fourth lowest states is small, as seen in Fig.~\ref{fig_ED2v3energy} for the $(14,5;0,0)$ truncation. We show the PES and the structure factor for the one-band calculation in Fig.~\ref{fig_PES_3b7_oneband}. A direct comparison between the PES in Figs.~\ref{fig_ED2v3PESnk}(a) and \ref{fig_PES_3b7_oneband}(a) further highlights the role of band mixing. With large band mixing, the PES exhibits a clear entanglement gap at the FCI counting of 168 states. In the one-band limit, however, the PES is less conclusive: in addition to the gap at the FCI counting, it also shows a gap at the CDW counting of 63 states. In fact, most of the weight of the state is in this 63 states, unlike for an FCI where the weight is in the 168 entanglement states. Together with the peaks in the structure factor at $K_M$ and $K'_M$ in Fig.~\ref{fig_PES_3b7_oneband}(b), this indicates that the one-band calculation is a CDW state close to the FCI phase boundary, while band mixing suppresses the competing CDW tendency and yields a cleaner FCI entanglement spectrum.

\begin{figure}
\centering
\includegraphics[width=0.9\columnwidth]{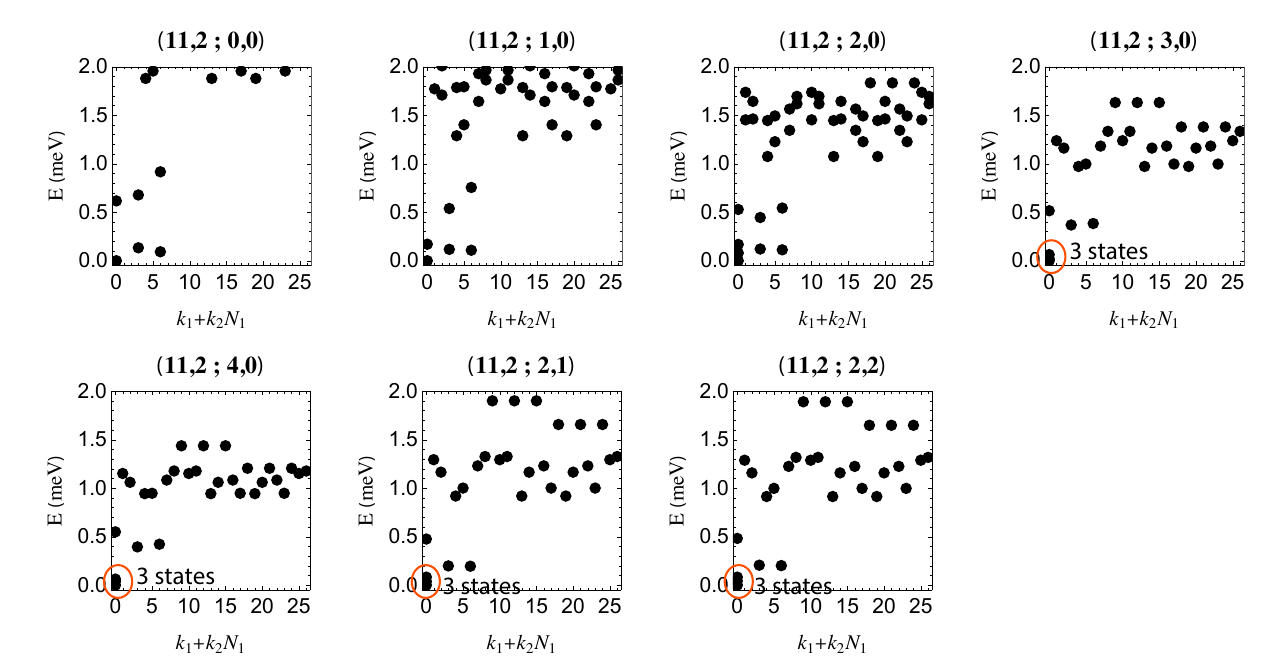}
\caption{ ED energy spectrum for $9\times 3$ system at 2/3 filing under different truncations with $V=22$ meV, $V_{tb}=32$ meV, $\mathcal{V}_{\text{val.}}=8$ meV, $\psi=(\frac{4\pi}{3}+0.12)\,$rad, and $\epsilon_r=5$. In the one-band limit the ground state is CDW with $k_1+k_2N_1=0,3,6$, whereas under finite band mixing the ground state becomes FCI with three states at zero momentum indicated by the red circles, indicating the multi-band nature of the FCI state.   }
\label{fig_comp_CDWFCI}
\end{figure}

The FCI ground states in the $21$ site system in Fig.~\ref{fig_ED2v3energy} at large band-mixing $(14,5;6,5)$ has a $0.18$ meV finite size splitting. To further investigate this splitting, we turn to a larger $9\times 3$ system with $(N_1,N_2,\tilde n_{11},\tilde n_{12},\tilde n_{21},\tilde n_{22})=(9,3,1,-2,0,1)$. In this mesh, the 3 FCI ground states are all at $k_1=k_2=0$, whereas the possible competing CDW states are at $k_1+k_2N_1=0,3,6$. Therefore from the ED energy spectrum we can distinguish FCI and CDW ground states directly. Our results are shown in Fig.~\ref{fig_comp_CDWFCI}. Note that on this larger system, it is not possible to go to large $n_2,n_3$, and we are restricted to the 11,2 orbital truncation (see \Fig{fig_EDorbital}b) with at most $(n_2,n_3) = (3,0)$ or $(2,2)$. Interestingly, Fig.~\ref{fig_comp_CDWFCI} shows that the ground state in the 1-band ED ($(n_2,n_3) = (0,0)$) is a CDW. This is consistent with the 6 low-energy states in 1-band ED on 21 sites in Figs.~\ref{fig_ED2v3energy} and \ref{fig_PES_3b7_oneband}, indicating close competition between the FCI and CDW on small sizes. As $n_2$ is increased, we see a clear transition out of the CDW and into an FCI with a small spread of $0.1$meV even in the presence of level repulsion. This persists with $n_3 > 0$, although the gap decreases slightly. This is further evidence that our FCI is stable and persists to larger sizes.

\subsection{Phase Diagram}

\begin{figure}
\centering
\includegraphics[width=0.8\columnwidth]{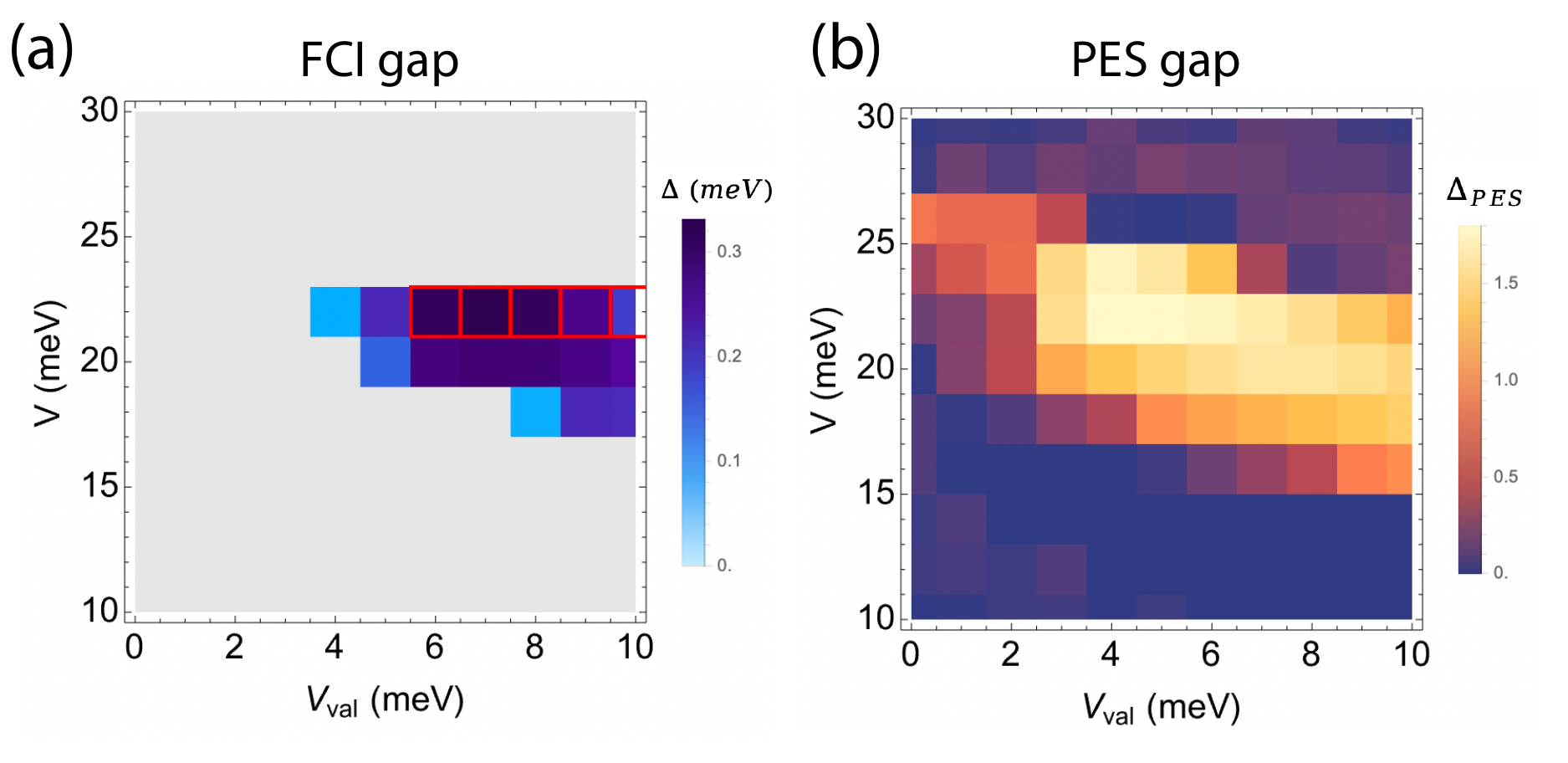}
\caption{$(a)$ FCI gap as a function of $V$ and $\mathcal{V}_{\text{val.}}$ calculated from ED in $3\times 7$ system at 2/3 filling under truncation $(n^{\text{orb}}_2,n^{\text{orb}}_3;\ n_2,n_3)=(14,2; 4,1)$ with parameters $V_{tb}=32$ meV, $\psi=(\frac{4\pi}{3}+0.12)\,$rad, and $\epsilon_r=5$. The parameter regions where the momenta of the three lowest energy states do not agree with FCI momenta are marked gray. The red box indicates the parameters where the FCI gap $E_4-E_3$ is larger than the spread of FCI states $E_3-E_1$, where $E_i$ is the sorted energy with $i=1$ labeling the lowest energy. $(b)$ The gap in PES with $N_A=2$ at the FCI counting with the same parameters. The three lowest states at the FCI momenta are used when calculating the PES.   }
\label{fig_EDVV1}
\end{figure}

\begin{figure}
\centering
\includegraphics[width=0.8\columnwidth]{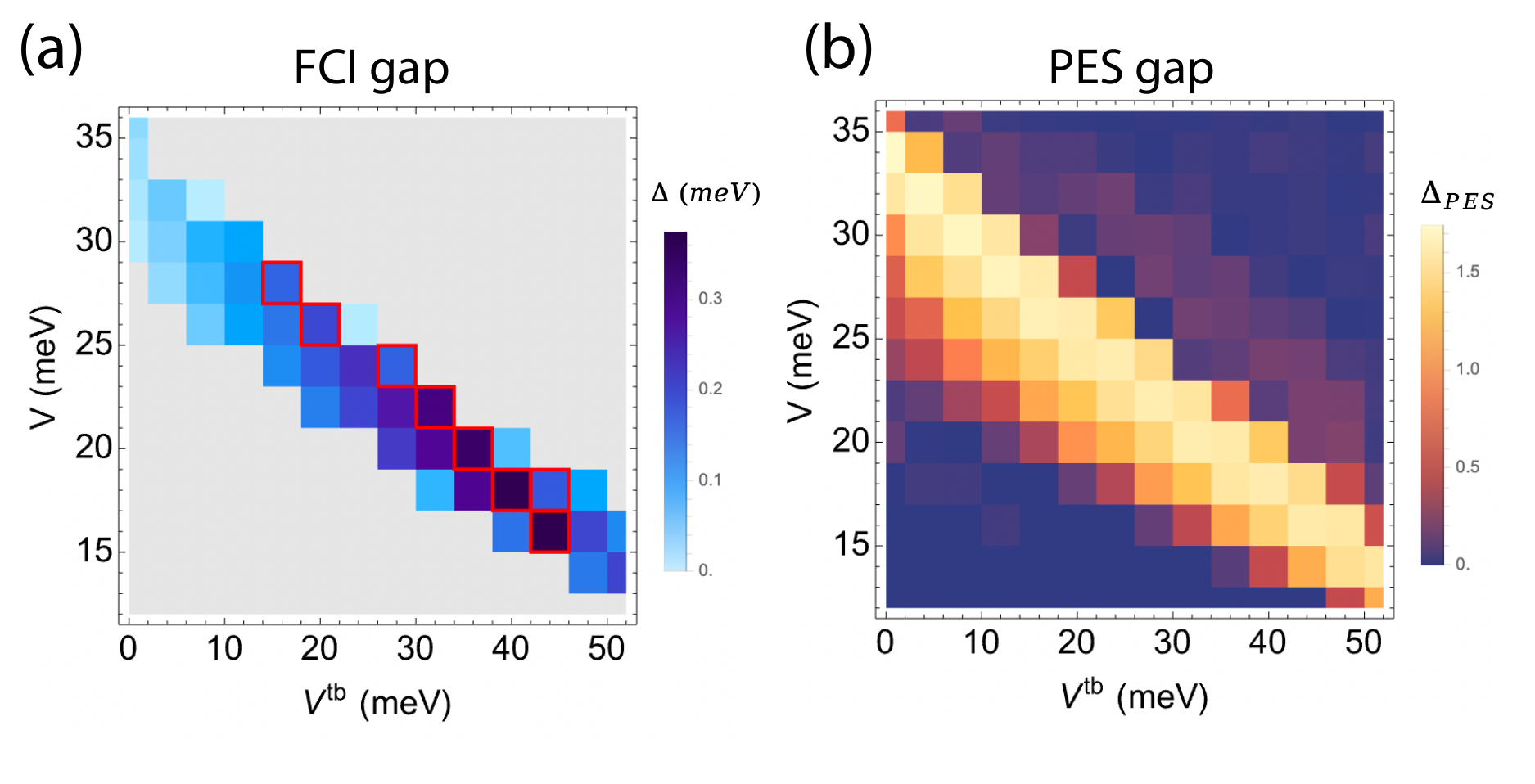}
\caption{ $(a)$ FCI gap as a function of $V$ and $V_{tb}$ calculated from ED in $3\times 7$ system at 2/3 filling under truncation $(n^{\text{orb}}_2,n^{\text{orb}}_3;\ n_2,n_3)=(14,2; 4,1)$ with parameters $\mathcal{V}_{\text{val.}}=8$ meV, $\psi=(\frac{4\pi}{3}+0.12)\,$rad, and $\epsilon_r=5$. The parameter regions where the momenta of the three lowest energy states do not agree with FCI momenta are marked gray. The red box indicates the parameters where the FCI gap $E_4-E_3$ is larger than the spread of FCI states $E_3-E_1$, where $E_i$ is the sorted energy with $i=1$ labeling the lowest energy. $(b)$ The gap in PES with $N_A=2$ at the FCI counting with the same parameters. The three lowest states at the FCI momenta are used when calculating the PES.   }
\label{fig_EDVV0}
\end{figure}

Having established the emergence of FCI for specific parameters, we further examine the dependence of the FCI energy gap and PES gap with different parameters. In Fig.~\ref{fig_EDVV1} and Fig.~\ref{fig_EDVV0} we calculate the ED phase diagram in the $3\times 7$ system at $\nu=2/3$ with truncation (14,2; 4,1) as a function of $V,\mathcal{V}_{\text{val.}}$ and $V,V_{tb}$ respectively. The plots show that the FCI exists in an extended parameter region. 

First, we show the $V,\mathcal{V}_{\text{val.}}$ phase diagram in \Fig{fig_EDVV1} for 14,2 orbital truncation with $n_2=4,n_3=1$. This slightly smaller truncation allows us to incorporate multi-band effects while maintaining a manageable Hilbert space dimension of $6.8\times 10^7$. Our results are consistent with those in the Main Text, showing FCIs appearing at larger values of $\mathcal{V}_{\text{val.}}$, accompanied by a PES gap.

Next, we show the $V,V_{tb}$ phase diagram in \Fig{fig_EDVV0} for the 14,2 orbtial truncation with $n_2=4,n_3=1$. We see that the FCI exists in a diagonal stripe in the space of $V,V_{tb}$, consistent with the generalized FCI stability criteria (see e.g.~$\Delta_{\text{TDHF}}$ and $W_{2/3}$ in \Fig{fig:app_V_vs_V0_V10.004}). Hence, while the precise value of $V_{tb}$ is not known accurately, there is a large range where a critical value of $V$ (as tunable in experiment) will host an FCI. 

\begin{figure}
\centering
\includegraphics[width=0.8\columnwidth]{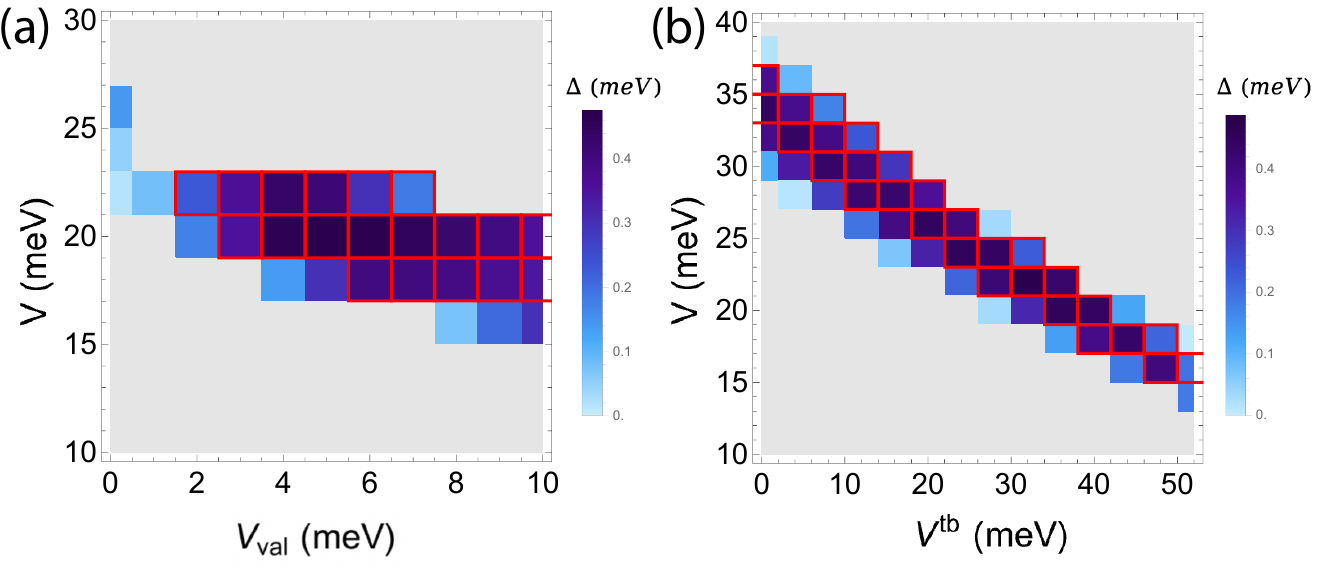}
\caption{ Left: FCI gap as a function of $V$ and $\mathcal{V}_{\text{val.}}$ calculated from ED in $3\times 7$ system at 2/3 filling under truncation $(n^{\text{orb}}_2,n^{\text{orb}}_3;\ n_2,n_3)=(14,0; 5,0)$ with parameters $V_{tb}=36$ meV, $\psi=(\frac{4\pi}{3}+0.12)\,$rad, and $\epsilon_r=5$. The parameter regions where the momenta of the three lowest energy states do not agree with FCI momenta are marked gray. The red box indicates the parameters where the FCI gap $E_4-E_3$ is larger than the spread of FCI states $E_3-E_1$, where $E_i$ is the sorted energy with $i=1$ labeling the lowest energy. $(b)$ FCI gap as a function of $V$ and $V_{tb}$ calculated from ED in $3\times 7$ system at 2/3 filling under truncation $(n^{\text{orb}}_2,n^{\text{orb}}_3;\ n_2,n_3)=(14,0; 5,0)$ with parameters $\mathcal{V}_{\text{val.}}=4$ meV, $\psi=(\frac{4\pi}{3}+0.12)\,$rad, and $\epsilon_r=5$.}
\label{fig_EDoldparascan}
\end{figure}

\begin{figure}
\centering
\includegraphics[width=\columnwidth]{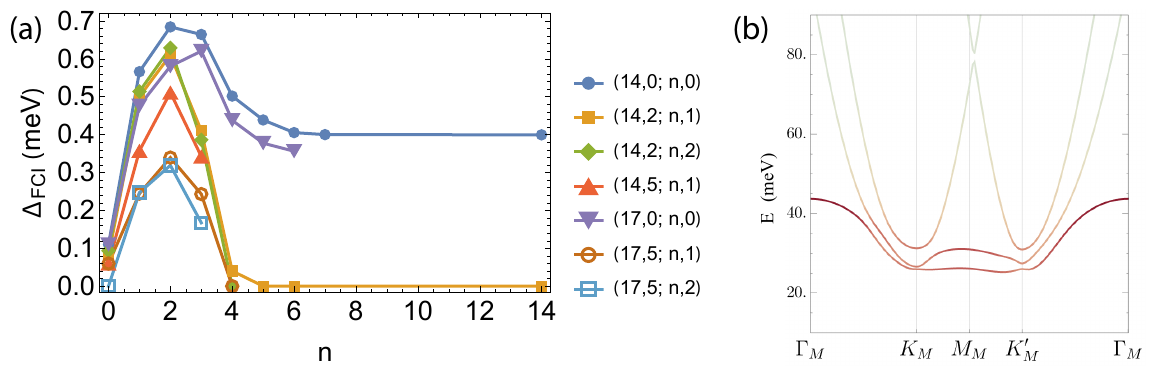}
\caption{ $(a)$ FCI energy gap calculated with different truncation parameters for $V=22$ meV, $V_{tb}=36$ meV, $\mathcal{V}_{\text{val.}}=4$ meV, $\psi=(\frac{4\pi}{3}+0.12)\,$rad, and $\epsilon_r=5$. When $\mathcal{V}_{\text{val.}}$ is small, the FCI gap collapses when there are sufficient particles populating the third band. $(b)$ The non-interacting band structure with the same parameters.  }
\label{fig_ED22364bm}
\end{figure}

Finally, we check that similar behavior is present in 2-band calculations with the truncation $(n^{\text{orb}}_2,n^{\text{orb}}_3;n_2,n_3) = (14,0;5,0)$ as shown in \Fig{fig_EDoldparascan}. However, we caution that not including the third band is likely to overestimate the FCI region, as shown in \Fig{fig_ED22364bm}(a) where we perform detailed 3-band convergence checks similar to \Fig{fig_EDbandmax} of the Main Text but at $\mathcal{V}_{\text{val.}} = 4$ meV (note that the Main Text uses $8$ meV). We observe that at $n_3 = 0$, for both the 14,5 and 17,5 truncations, a nonzero FCI gap appears and converges at $n_2 = 2,\dots, 14$. However, at $n_3 = 1$ on both these truncations, the FCI gap appears to close by $n_2 = 4$. Similar behavior appears for the 14,2 truncation at $n_3 = 2$. We can understand the larger (detrimental) effect of the third band at $\mathcal{V}_{\text{val.}} = 4$ meV compared to $8$ meV (Main Text, where the FCI gap is nonzero) quite simply from the band structure. As shown in \Fig{fig_ED22364bm}(b), the third band is closer in energy to the lower 2 bands. This was discussed in \Fig{fig:backgroundpd} and shown analytically: the splitting at the $K_M,K'_M$ points between the lowest two states (which degenerate at $\psi_{\text{val.}} = 2\pi n/3$) and the third is proportional to $\mathcal{V}_{\text{val}.}$. This is why a larger value of $\mathcal{V}_{\text{val}.}$ lessens the effect of the third band, leaving the FCI intact. Hence, the moir\'e capacitor effect must be sufficiently strong to achieve an FCI. One other way to view this, as pointed out in Ref. \cite{PhysRevB.112.075110}, is that a would-be FCI crystal in the $\mathcal{V}_{\text{val}.} \to 0$ limit would spontaneously break translations so that Goldstone modes would appear in the ED spectrum. This means the gap \emph{must go to zero} in the thermodynamic limit and would be accompanied by an Anderson tower of states. Our calculations here show that a conventional gapped FCI state is stabilized by the moir\'e present in experiments. 

\end{document}